\documentclass[superscriptaddress, nofootinbib, amsmath, amssymb]{revtex4}
\usepackage{graphicx}
\usepackage{wasysym}
\usepackage[colorlinks=true,citecolor=blue,linkcolor=blue,urlcolor=blue]{hyperref}
\usepackage{color}
\usepackage[thinlines]{easytable}
\usepackage{multirow}
\usepackage{array}
\begin{document}
\title{- Gravitational Waves -\\
A Review on the Conceptual Foundations of Gravitational Radiation}
\author{Alain Dirkes}
\email{dirkes@fias.uni-frankfurt.de}
\affiliation{Frankfurt Institute for Advanced Studies (FIAS), Goethe University Frankfurt,\\ Ruth-Moufang Str.1, Frankfurt am Main, D-60438, Germany}
\begin{abstract}
In this manuscript we review the theoretical foundations of gravitational waves in the framework of Albert Einstein's theory of general relativity. Following Einstein's early efforts we first derive the linearised Einstein field equations and work out the corresponding gravitational wave equation. Moreover we present the gravitational potentials in the far away wave zone field point approximation obtained from the relaxed Einstein field equations. We close this review by taking a closer look on the radiative losses of gravitating $n$-body systems and present some aspects of the current interferometric gravitational waves detectors. Each section posses a separate appendix-contribution where further computational details are displayed. To conclude we summarize the main results and present a brief outlook in terms of current ongoing efforts to build a spaced based gravitational wave observatory. 
\end{abstract}
\maketitle

\section{Introduction:}
\label{Introduction}
Only one year after the final formulation of the theory of general relativity, Einstein predicted the existence of gravitational radiation, when he saw that the linearised weak-field equations admit solutions in the form of gravitational waves travelling at the speed of light. Gravitational radiation has been detected indirectly since the mid seventies of the past century in the context of binary systems \cite{Taylor1, Burgay1, Stairs1, Stairs2,Wex1}. Precisely one century after Einstein's theoretical prediction, an international collaboration of scientists (LIGO Scientific Collaboration and Virgo Collaboration) reported the first direct observation of gravitational waves \cite{LIGO1,LIGO2,LIGO3}. The wave signal GW150914 was detected independently by the two LIGO detectors and its basic features point to the coalescence of two stellar black holes. More recently the same collaboration was even able to perform a second (GW151226) and a third GW170104) direct gravitational wave measurement \cite{LIGO4,LIGO5}. Moreover it should be mentioned that gravitational radiation itself is an excellent probe to study Einstein's theory in the fully dynamical, strong field regime \cite{YunesSiemens,Will1}. In this review article we outline the theoretical foundations of gravitational radiation in the context of Albert Einstein's theory of general relativity. Following Einstein's early efforts \cite{Einstein2} we will derive in section \ref{Linearised} the linearised Einstein field equations and work out the corresponding gravitational wave equation. In section \ref{PotentialsFarAway} we present the gravitational potentials in the far away wave zone approximation obtained from the relaxed Einstein field equations \cite{PoissonWill,PatiWill1,PatiWill2,WillWiseman}. We close this review in section \ref{RadiativeLosses} by taking a closer look to the radiative losses of gravitating $n$-body systems. Here we also present some aspects of the current gravitational waves dectetors \cite{Maggiore1,Schutz1,Schutz2}. It should be noticed that each of the sections mentioned previously posses a separate subsection outlined in the appendix \ref{AppendixGRad} where further computational details are displayed. In the conclusion \ref{Conclusion} we briefly summarize the main results presented in this review article and present an outlook in terms of current ongoing efforts to build a spaced based gravitational wave observatory. 

\section{The linearised Einstein field equations}
\label{Linearised}
Long before Albert Einstein published his famous theory of general relativity the relation between matter and the gravitational field had been concisely summarized by the elegant Poisson equation. This law however is purely phenomenological whereas Einstein's theory provides, via the concept of spacetime curvature, a much deeper understanding of the true nature of gravity. It is well known \cite{Weinberg2,Inverno,PoissonWill} that Einstein's famous field equations are obtained from the variation of the total action, $S=S_g+S_m$, with respect to $g_{\alpha\beta}$, the metric tensor,
\begin{equation}
\label{EFE-24.03.17}
G_{\alpha\beta}\,=\,\frac{8\pi G}{c^4} T_{\alpha\beta},
\end{equation} 
where we aim to remind that $G_{\alpha\beta}$ is the Einstein tensor and $T_{\alpha\beta}$ is the energy-momentum tensor of the matter source. The action composed by a gravitational contribution , $S_g=\frac{c^3}{16\pi G}\int d^4x \ \sqrt{-g}\ R$ and a matter piece $S_m\,=\,\int d^4x\ \sqrt{-g}\ \mathcal{L}_m$ respectively. Here $-g$ is the metric determinant and $\mathcal{L}_m$ is the Lagrangian density for all matter and field variables. The Einstein field equations follow from the requirement that $\delta(S_g+S_m)=0$ under an arbitrary variation of the metric tensor. In particular the Einstein tensor results from the functional variation (with respect to the metric) of the gravitational action and the energy-momentum tensor can be defined as being proportional to the functional variation of the matter (energy) action. It should be noticed that the Bianchi identities \cite{Weinberg2,Inverno} give rise to the identities, $\nabla_\beta G^{\alpha\beta}=0$, known as the contracted Bianchi identities. It is straightforward to observe that by introducing this set of equations inside equation \eqref{EFE-24.03.17} we immediately obtain, $\nabla_\beta T^{\alpha\beta}=0$, the energy-momentum conservation of the matter (energy) source term. We recall that general relativity is invariant under the group of all possible coordinate transformations, $x^\alpha\rightarrow x'^\alpha(x)$, where $x'^\alpha$ is an arbitrary smooth function of $x^\alpha$. More precisely $x'^\alpha$ has to be invertible and differentiable with a differentiable inverse \cite{MisnerThroneWheeler,LandauLifshitz,Maggiore1, Buonanno1, Weinberg2, Carroll1}. The field equations given above are diffeomorpism invariant. In this context we recall that we have the following transformation rule for the metric tensor,
\begin{equation}
\label{gtrafo-06.04.17}
g_{\alpha\beta}(x)\rightarrow g'_{\alpha\beta}(x')\,=\,\frac{\partial x^\rho}{\partial x'^\alpha}\frac{\partial x^\sigma}{\partial x'^\beta}\ g_{\rho\sigma}(x).
\end{equation}
This symmetry is usually referred to as the symmetry of general relativity. As a first step towards a deeper understanding of the phenomenon of gravitational radiation, we aim to study the expansion of the Einstein field equations around the flat spacetime metric. In this regard we assume that there exists a reference frame in which, on a sufficiently large spacetime region, we can decompose the metric into the flat Minkowski metric plus a small perturbation,
\begin{equation*}
g_{\alpha\beta}\,=\,\eta_{\alpha\beta}+d_{\alpha\beta}.
\end{equation*}
In the context of weak gravitational fields it can be assumed that the module of the four-dimensional gravitational potentials $|d_{\alpha\beta}|\ll 1$ is a small deviation from Minkowski spacetime \cite{LandauLifshitz,MisnerThroneWheeler,Buonanno1,Schutz0A,Schutz1,Schutz2,Schutz3,Riles1}. This particular assumption allows us to ignore anything that is higher than first order in the gravitational potentials and we can immediately conclude that the metric inverse is $g^{\alpha\beta}=\eta^{\alpha\beta}-d^{\alpha\beta}$. It should be mentioned that by choosing a particular reference frame, we break the invariance of GR under general coordinate transformations. A residual gauge symmetry will however remain \cite{LandauLifshitz,MisnerThroneWheeler,PoissonWill,Maggiore1,Buonanno1,Schutz0A,Schutz1}. Let us in this regard consider the following coordinate transformation,
\begin{equation}
\label{trafocoordinates-07.04.17}
x^\alpha\rightarrow x'^\alpha\,=\,x^\alpha+t^\alpha(x),
\end{equation}
where the derivatives of the spacetime translation components are at most of the same order of smallness as the gravitational potentials, $|\partial_\alpha t_\beta|\leq |d_{\alpha\beta}|$. By taking into account the transformation law for the metric tensor, we obtain the transformation rule for the gravitational potentials to lowest order in the translation vector,
\begin{equation}
\label{trafodeviation-07.04.17}
d_{\alpha\beta}(x)\rightarrow d'_{\alpha\beta}(x')\,=\,d_{\alpha\beta}(x)-(\partial_\alpha t_\beta+\partial_\beta t_\alpha).
\end{equation} 
We notice that this slowly varying coordinate transformation is a symmetry of the linearised theory \cite{MisnerThroneWheeler,Maggiore1,Buonanno1}. Moreover we can also perform finite, global ($x$-independent) Lorentz transformations, $x^\alpha\rightarrow \Lambda^\alpha_{\ \beta}\ x^\beta$, where by definition of the Lorentz transformation, the transformation matrices $\Lambda^\alpha_{\ \beta}$ have to satisfy the relation $\Lambda^{\ \rho}_{\alpha } \Lambda_{ \beta}^{\ \sigma}\ \eta_{\rho\sigma}=\eta_{\alpha\beta}$. We observe that the linearised metric is invariant under a generic Lorentz transformation,
\begin{equation}
g_{\alpha\beta}(x)\,=\,\eta_{\alpha\beta}+d_{\alpha\beta}\rightarrow g'_{\alpha\beta}(x')\,=\,\Lambda^{\ \rho}_{\alpha} \Lambda_\beta^{\ \sigma}\ (\eta_{\rho\sigma}+d_{\rho\sigma})\,=\,\eta_{\alpha\beta}+d'_{\alpha\beta}(x'),
\end{equation} 
where $d'(x')=\Lambda^{\ \rho}_\alpha \Lambda^{\ \sigma}_\beta d_{\rho\sigma}(x)$ is a tensor under Lorentz transformations. Rotations never spoil the condition $|d_{\alpha\beta}|\ll 1$, while for boosts we must limit ourselves to those that do not spoil this condition \cite{Maggiore1}. It is straightforward to see from the general transformation law of the metric that the gravitational potentials are invariant under constant transformations of the form, $x^\alpha\rightarrow x'^\alpha+a^\alpha$, where $a^\alpha$ is not restricted to be infinitesimal but can be finite. From this we deduce that the linearised theory is invariant under finite Poincaré transformations as well as under infinitesimal local transformations. We remind that in the complete theory of general relativity the flat spacetime metric does not play a special role and in this sense the Poincar\'{e} symmetry or the infinitesimal local transformations are only sub-transformations of the general coordinate transformation invariance of the full theory. By making use of the linearised Christoffel symbols (affine connections), $\Gamma^\rho_{\alpha\beta}=\frac{1}{2}\eta^{\rho\lambda}(\partial_\alpha d_{\beta\lambda}+\partial_\beta d_{\lambda\alpha}-\partial_\lambda d_{\alpha\beta})$, we can work out the linearised Riemann tensor, $
R_{\alpha\beta\rho\sigma}=1/2\ [\partial_{\rho}\partial_{\beta} d_{\alpha\sigma}+\partial_\sigma\partial_\alpha d_{\rho\beta}-\partial_\rho \partial_\alpha d_{\sigma\beta}-\partial_\sigma\partial_\beta d_{\alpha\rho}]+\mathcal{O}(d^2)$ and from this we easily obtain the linearised Einstein tensor,
\begin{equation}
\label{LET-06.04.17}
G_{\alpha\beta}\,=\,\frac{1}{2}\big(\partial_\rho \partial_\alpha d^\rho_{\ \beta}+\partial_\beta \partial_\mu d^\mu_{\ \alpha} -\Box d_{\alpha\beta}-\partial_\alpha\partial_\beta d\big)-\frac{1}{2}\big(\partial_\mu \partial_\nu d^{\mu\nu}-\Box d\big)\eta_{\alpha\beta}+\mathcal{O}(d^2).
\end{equation}
A detailed derivation of these results can be found in the appendix-section \ref{AppendixGRad} related to this chapter. It should be noticed that the linearised Einstein tensor is invariant under the transformations outlined in equation \eqref{trafocoordinates-07.04.17} and equation \eqref{trafodeviation-07.04.17}. Moreover it should be mentioned that the form of the Einstein tensor can be simplified even further by using the trace-reversed Minkowski metric deviation, $\bar{d}_{\alpha\beta}=d_{\alpha\beta}-(1/2)\eta_{\alpha\beta} \ d$. This particular deviation is called in this way because of the property, $\bar{d}=-d$, so that we have $d_{\alpha\beta}=\bar{d}_{\alpha\beta}-(1/2)\eta_{\alpha\beta}\bar{d}$. If in addition we impose the condition $\partial^\beta \bar{d}_{\alpha\beta}=0$ on the trace-reversed Minkowski metric deviation, the Einstein field equations reduce to a wave equation,
\begin{equation}
\label{SWave-06.04.17}
\Box \bar{d}_{\alpha\beta}\,=\,-\frac{16 \pi G}{c^4} T_{\alpha\beta},
\end{equation}
for the trace-reversed potentials \cite{LandauLifshitz,MisnerThroneWheeler}. The metric can be worked out from the potentials, $g_{\alpha\beta}=\eta_{\alpha\beta}+\bar{d}_{\alpha\beta}-1/2\ \eta_{\alpha\beta}\bar{d}$ \cite{MisnerThroneWheeler,Maggiore1,Buonanno1,Schutz0A,Schutz1,Schutz2,Schutz3}. Further computational details can be withdrawn from the appendix-section \ref{AppendixGRad} related to this chapter.

\section{Gravitational potentials in the far away wave zone}
\label{PotentialsFarAway}
\begin{figure}
\begin{center}
\includegraphics[width=0.2\textwidth]{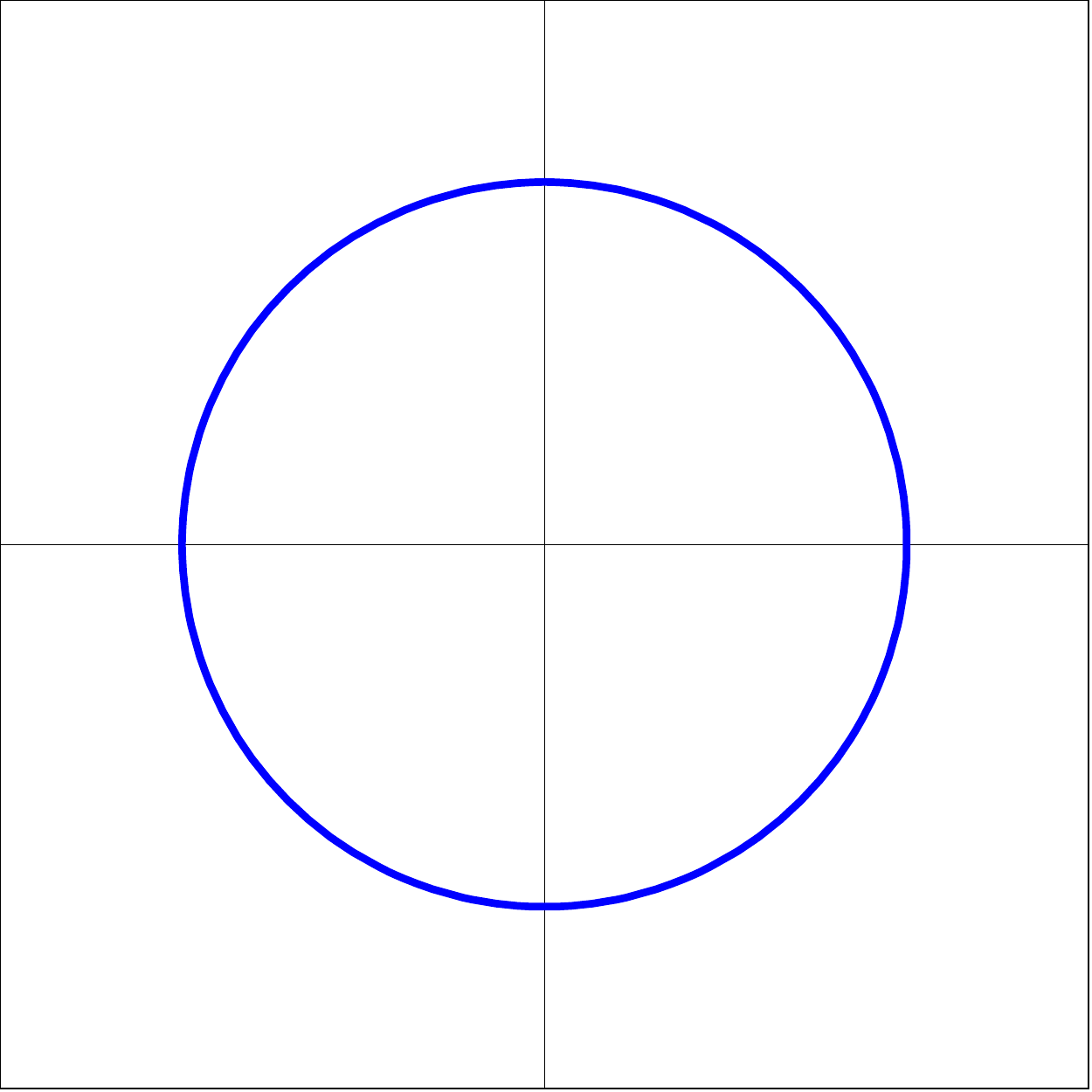}\hspace{2.5cm} \includegraphics[width=0.2\textwidth]{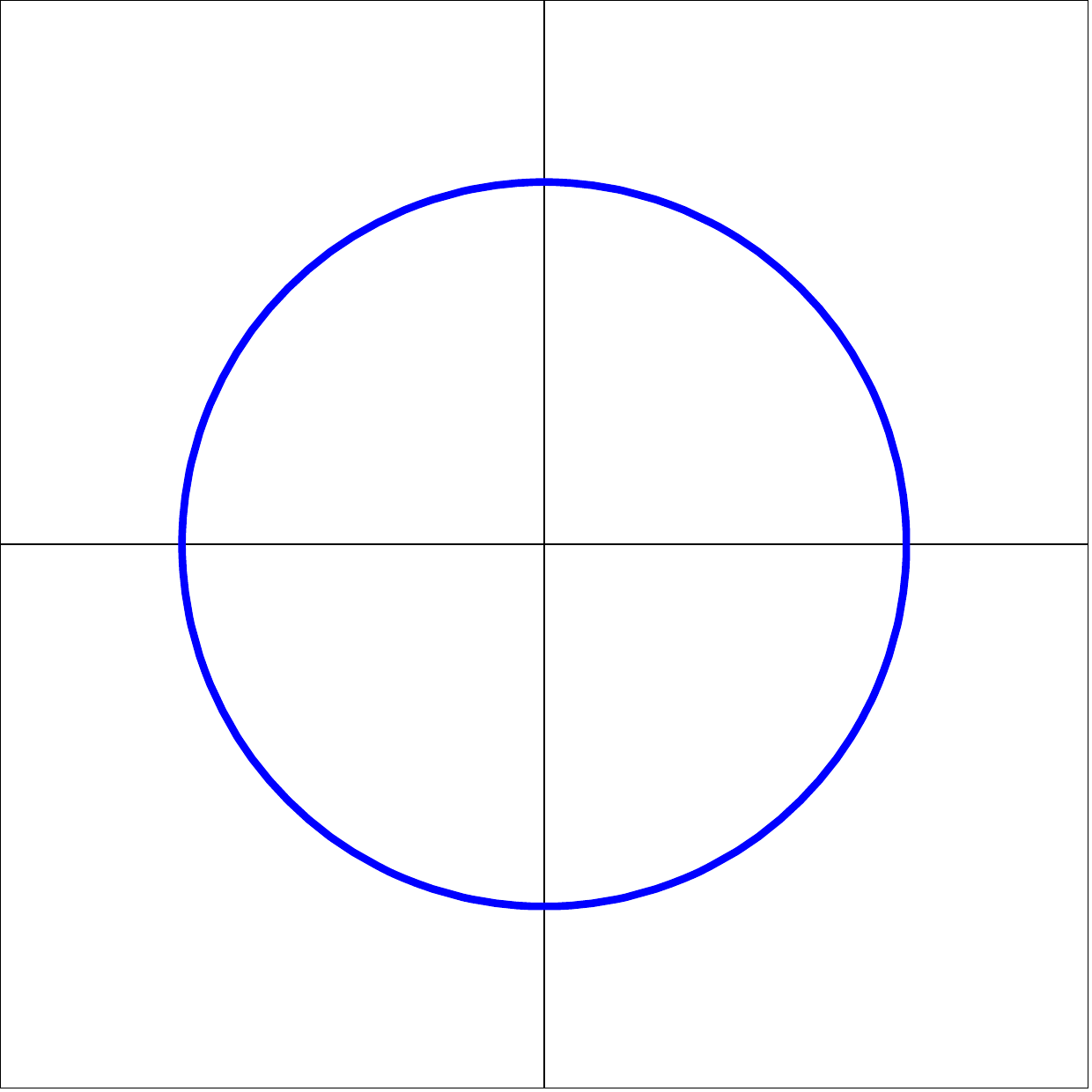}\\ \includegraphics[width=0.2\textwidth]{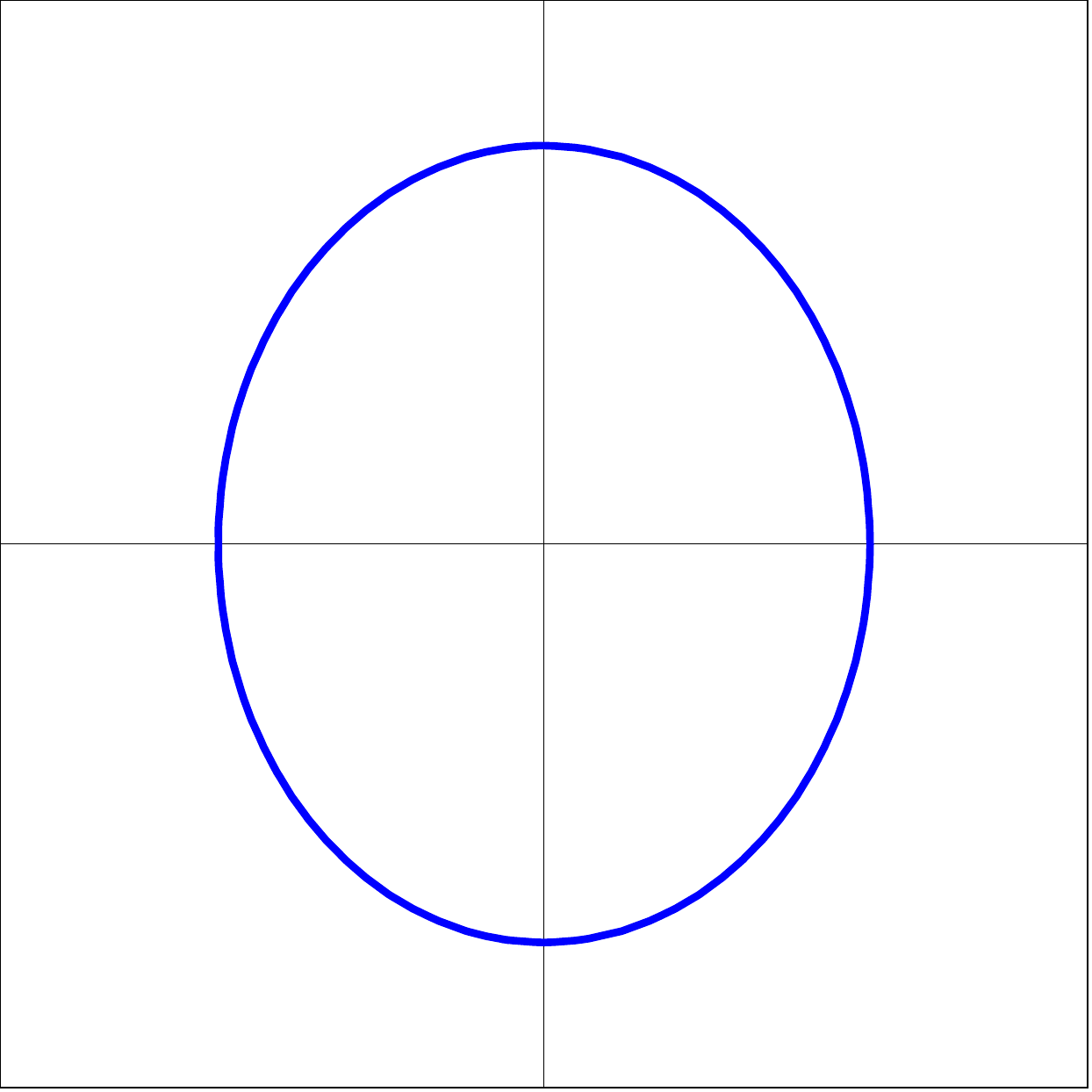}\hspace{2.5cm} \includegraphics[width=0.2\textwidth]{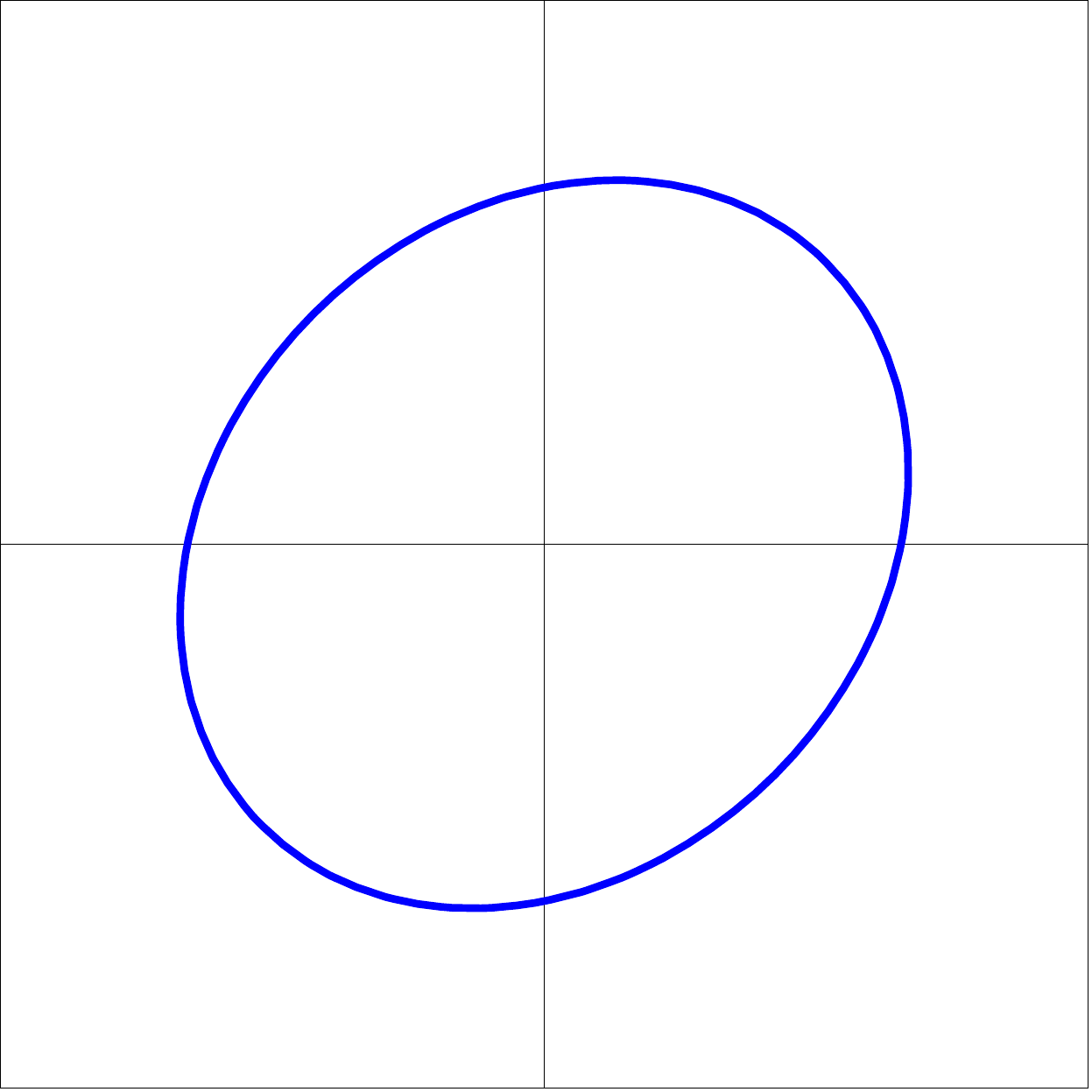}\\ \includegraphics[width=0.2\textwidth]{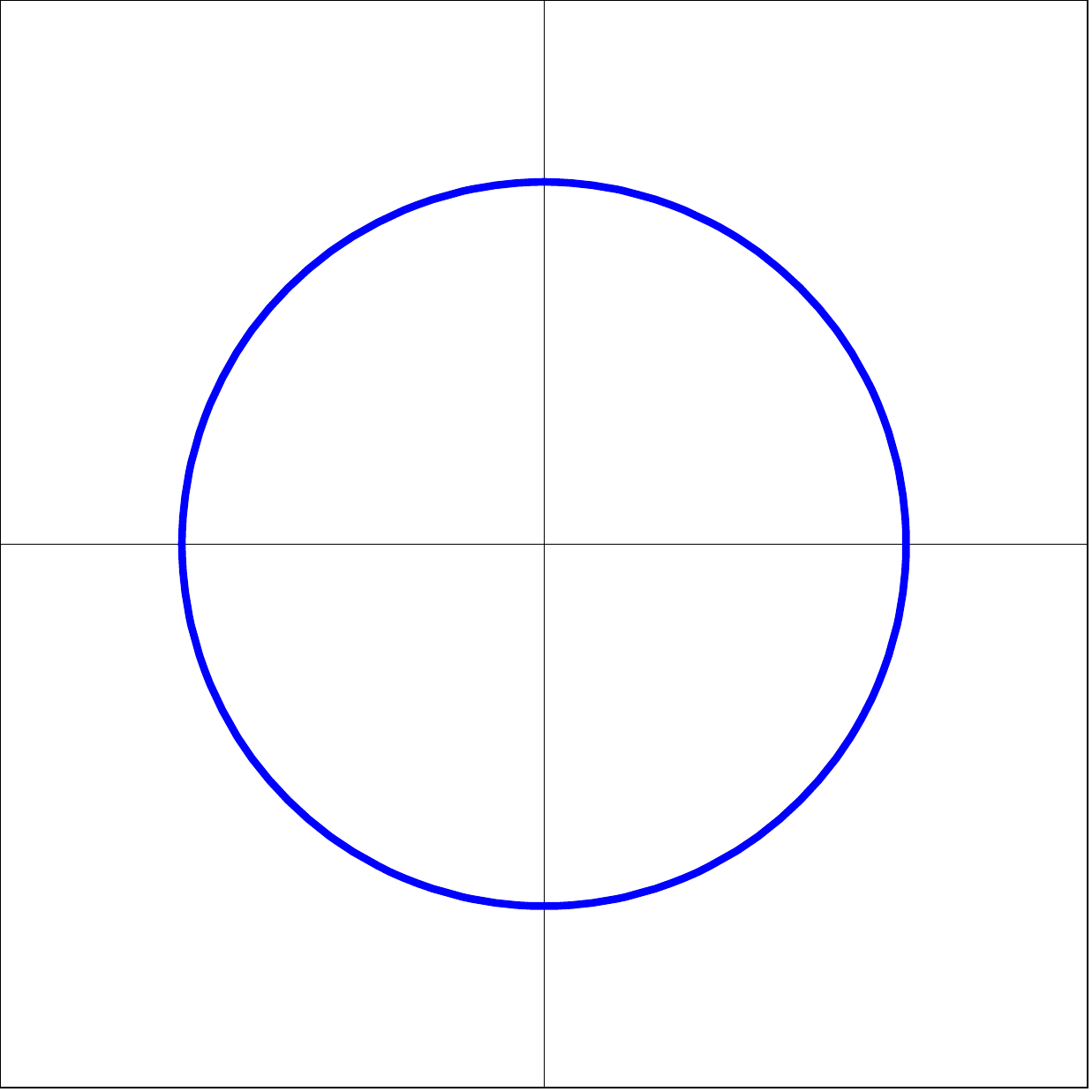}\hspace{2.5cm} \includegraphics[width=0.2\textwidth]{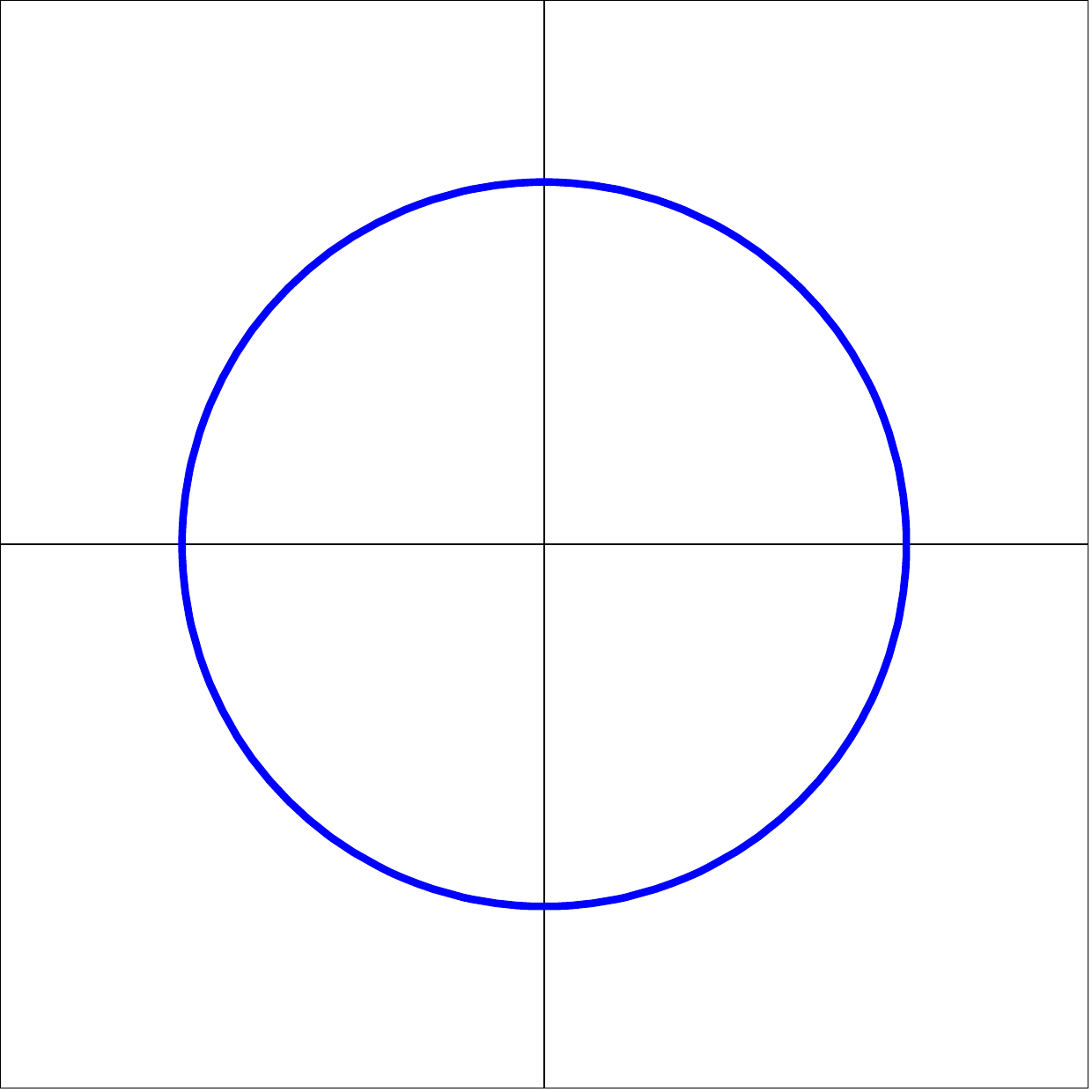}\\ \includegraphics[width=0.2\textwidth]{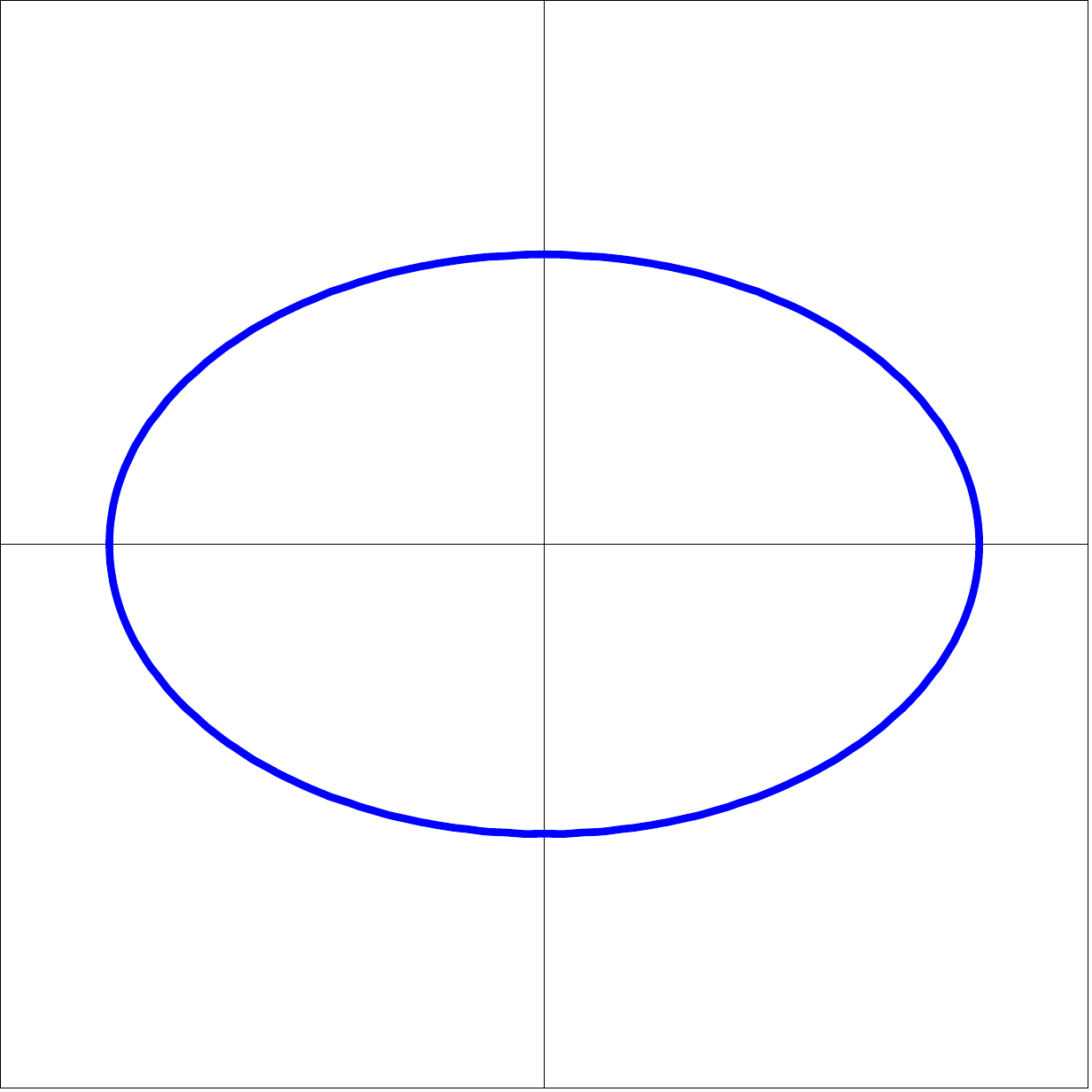}\hspace{2.5cm} \includegraphics[width=0.2\textwidth]{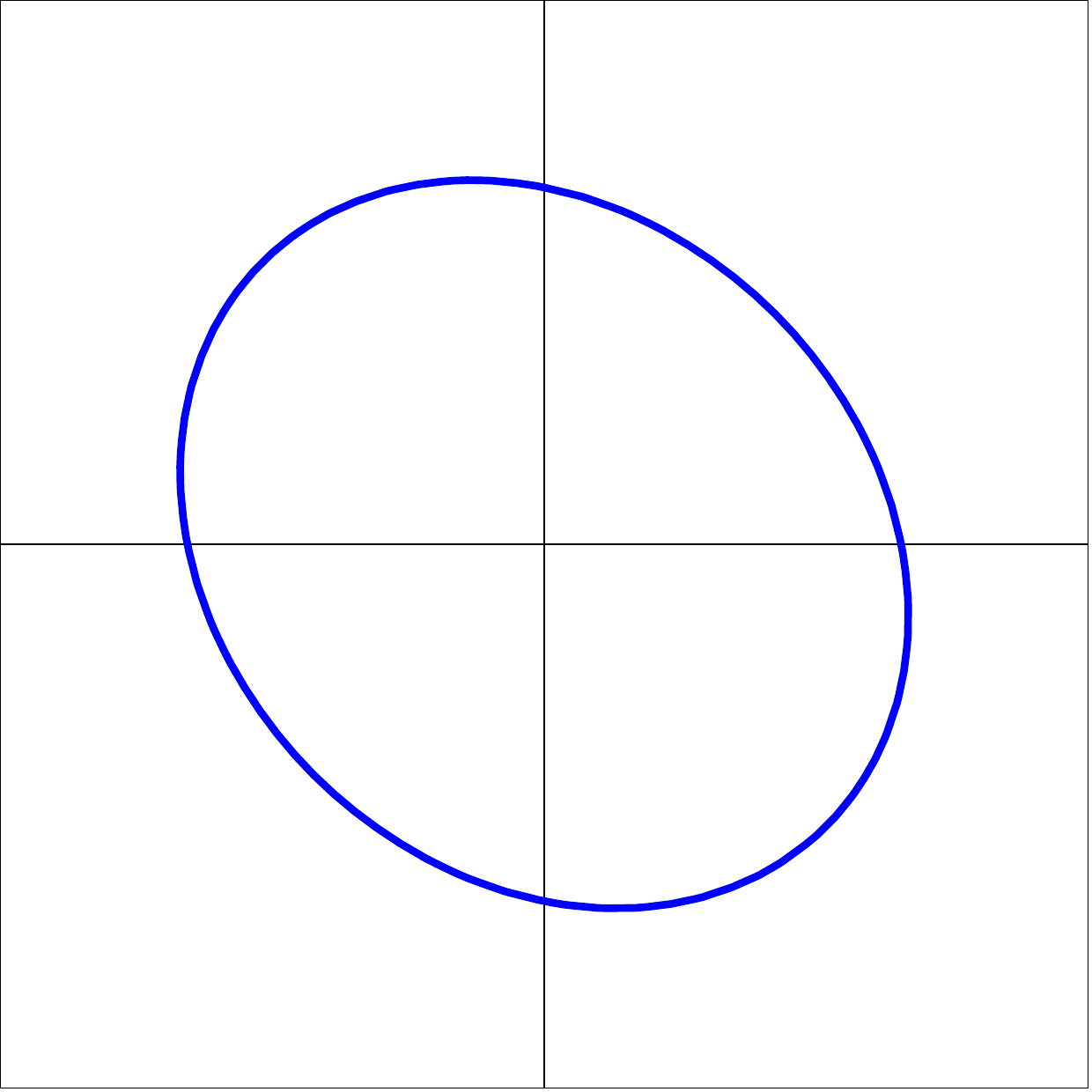}\\ \includegraphics[width=0.2\textwidth]{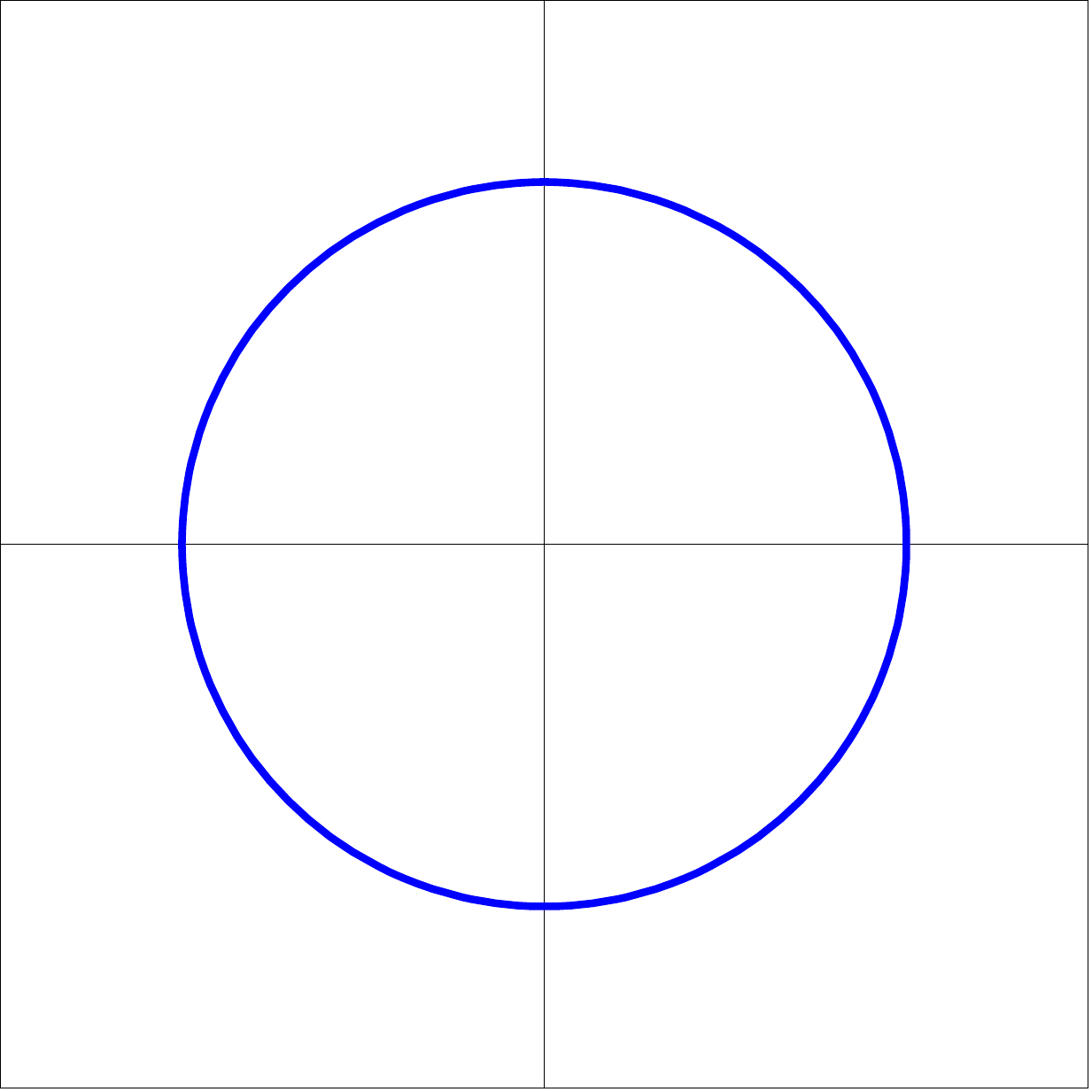}\hspace{2.5cm} \includegraphics[width=0.2\textwidth]{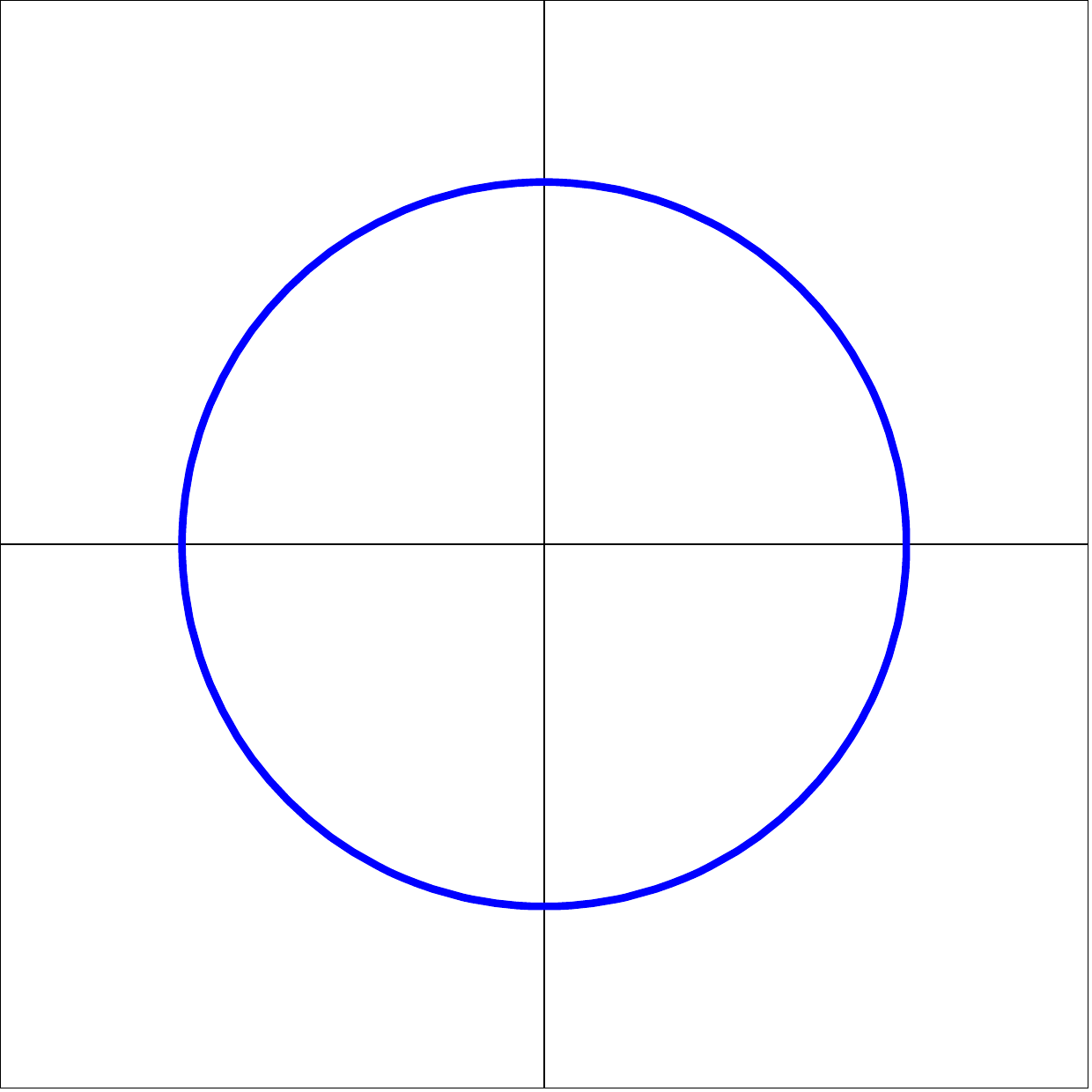}
\end{center}
\caption[Gravitational waves' polarisations: $+$-mode and $\times$-mode]{\label{PolarisationsPlusCross}This figure illustrates the deformation of a ring of freely falling test masses due to the $+$-mode (left cycle) and the $\times$-mode (right cycle) of a gravitational wave that propagates in the direction perpendicular to the ring of test masses. An entire deformation-cycle for the two polarisations is shown from the top to the bottom.}
\end{figure}

It is possible to solve the relaxed Einstein equations, $\Box h^{\alpha\beta}=-\frac{16 \pi G}{c^4}\ \tau^{\alpha\beta}$, for the gravitational potentials in the context of a far away wave zone field point with a near zone matter source term \cite{PoissonWill,Maggiore1,WillWiseman,PatiWill1,PatiWill2,WagonerWill1976}. This set of equations is the nonlinear generalization of the linearised Einstein field equations outlined in equation \eqref{SWave-06.04.17}. Here $\tau^{\alpha\beta}=\tau^{\alpha\beta}_m+\tau^{\alpha\beta}_{LL}+\tau^{\alpha\beta}_H$ is defined as the effective energy-momentum pseudotensor composed respectively by the matter contribution, the Landau-Lifshitz contribution and the harmonic gauge contributions, $\tau_m^{\alpha\beta}=(-g) T^{\alpha\beta}$, $\tau^{\alpha\beta}_{LL}=(-g)t^{\alpha\beta}_{LL}$, $\tau^{\alpha\beta}_H\,=\,(-g)t^{\alpha\beta}_H=\frac{c^4}{16\pi G} \big(\partial_\mu h^{\alpha\nu}\partial_\nu h^{\beta\mu}-h^{\mu\nu}\partial_{\mu\nu}h^{\alpha\beta}\big)$ \cite{PoissonWill,Maggiore1,WillWiseman,PatiWill1,PatiWill2}. It should be noticed that relaxed Einsetin equations are an exact formulation of the Einstein field equations presented an that all the nonlinearity stored within the Einstein tensor $G_{\alpha\beta}$ in equation \ref{EFE-24.03.17} reappears in the Landau-Lifshitz pseudotensor $t^{\alpha\beta}_{LL}$ \cite{LandauLifshitz,PoissonWill,PatiWill1,WillWiseman}. While the near zone is defined by $R=|\textbf{x}|<\lambda_c$, the position of the far away wave zone field point is the region in space where we the field point is situated even further away from the source, $|\textbf{x}|\gg \lambda_c$. Here $\lambda_c$ is the characteristic wavelength of the source term related to the source's characteristic time scale $\lambda_c=t_c c$. In this particular context, the gravitational potentials can be expressed in terms of the radiative multipole moments $Q^{ab\cdots}_{GR}$ and we eventually obtain,
\begin{equation}
\label{FAWZFP-28.03.17}
\begin{split}
h^{ab}\,=&\, \frac{2G}{c^4 R} \frac{\partial^2}{\partial \tau^2} \Big[Q^{ab}_{GR}+Q^{abc}_{GR} \ N_c+Q^{abcd}_{GR} \ N_c N_d+\cdots\Big],
\end{split}
\end{equation}
where $N_a=x^a/R$ is the the $a$-th component of the radial unit vector $\textbf{N}=\textbf{x}/R$ \cite{Alain1,Alain0A,Alain0B,Alain2,PoissonWill,PatiWill1}. The gravitational potentials can be truncated at their leading order contribution proportional to $|\textbf{x}|^{-1}$. In this situation context the gravitational potentials become,
\begin{equation}
\label{PotentialsFAWZ-05.04.17}
h^{00}\,=\,\frac{4GM}{c^2 R}+\frac{G}{c^4 r} X(\tau,\textbf{N}),\quad \
h^{0a}\,=\,\frac{G}{c^4 R}Y^a(\tau,\textbf{N}),\quad \
h^{ab}\,=\, \frac{G}{c^4 R} Z^{ab}(\tau,\textbf{N}),
\end{equation}
where $M$ is the total gravitational mass \cite{PoissonWill,PatiWill1,Maggiore1}. For reasons of clarity the superscript $GR$ was omitted here. The functions $X$, $Y^a$ and $Z^{ab}$ depend on the retarded-time variable $\tau=t-\frac{R}{c}$ as well as on the unit vector $\textbf{N}=\frac{\textbf{x}}{R}$. We recall that $R=|\textbf{x}|$ is the distance between the source and the field point and it should not be confused with the relative separation of two bodies. It should be noticed that at the present stage we do not need the precise form of the gravitational potentials \cite{PoissonWill,Maggiore1}. It is straightforward to verify that the set of solutions  outlined under equation \eqref{PotentialsFAWZ-05.04.17} obey the wave equation as long as the effective energy-momentum tensor falls off at least as fast as $R^{-2}$. For later convenience we aim to introduce a differentiation rule that applies in the far away wave zone, $\partial_a h^{\alpha\beta}=-\frac{N_a}{c}\partial_\tau h^{\alpha\beta}$. It turns out to be useful to decompose the the vector $Y^a$ as well as the tensor $Z^{ab}$ into longitudinal and transverse components,
\begin{equation}
\begin{split}
Y^a\,=\, Y N^a+Y^a_T,\quad \
Z^{ab}\,=\,\frac{\delta^{ab}}{3} A+\big(N^aN^b-\frac{\delta^{ab}}{3}\big)B+N^aZ^b_T+N^bZ_T^a+Z^{ab}_{TT},
\end{split}
\end{equation}
where the longitudinal direction is identified with $\textbf{N}$ \cite{MisnerThroneWheeler,PoissonWill,Buonanno1,Schutz1}. Here $Y N^a$ is representing the longitudinal part of $Y^a$ and $Y^a_T$ is its transverse component which has to satisfy, $N_a Y^a_T=0$. The three components of the vector $Y^a$ are devided into one longitudinal component $Y$ and two transverse components contained in $Y^a_T$. In a very similar way the tensor $Z^{ab}$ can be decomposed into a trace part $\frac{\delta^{ab}}{3}A$, a longitudinal-tracefree part, $\big(N^aN^b-\frac{\delta^{ab}}{3}\big)B$, a longitudinal transverse part, $N^aZ^b_T+N^bZ^a_T$, as well as a transverse-tracefree component $Z^{ab}_{TT}$. We impose the constraints,
\begin{equation}
\label{constraints-06.04.17}
N_a Z^a_T\,=\,0, \quad \quad N_a Z^{ab}_{TT}\,=\,0\,=\, \delta_{ab} Z^{ab}_{TT}.
\end{equation}
The six independent components of $Z^{ab}$ are contained in two scalars $A$ and $B$, in two components of the transverse vector $Z^a_T$ as well as in two components of the transverse-tracefree tensor $Z^{ab}_{TT}$. The latter is usually referred to as the transverse-tracefree component, or TT component of $Z^{ab}$ and we will see that the radiative parts of the gravitational potentials are contained entirely within the $Z^{ab}_{TT}$ piece. In what follows we will analyse in how far the harmonic gauge condition will affect the general form of the gravitational potentials outlined in equation \eqref{PotentialsFAWZ-05.04.17}. By introducing the gravitational potentials decomposed into irreducible components inside the differentiation rule outlined above we obtain for the potentials,
\begin{equation}
\label{IntermediatPot-07.04.17}
\begin{split}
h^{00}\,=&\,\frac{4 G M}{c^2 R}+\frac{G}{c^4 R}\frac{A+2B}{3},\quad \
h^{a0}\,=\,\frac{G}{c^4 R}\Big[\frac{A+2B}{3} N^a+Y^a_T\Big],\\
h^{ab}\,=&\, \frac{G}{c^4 R}\Big[\delta^{ab} A+\big(N^aN^b-\frac{\delta^{ab}}{3}\big) B+N^aY^b_T+N^b Y^a_T+Z_{TT}^{ab}\Big].
\end{split}
\end{equation}
It should be noticed that making use of the harmonic gauge conditions we reduced the initial ten degrees of freedom to six independent quantities which are contained within the functions $A$, $B$, $Y^a_T$ and $Z^{ab}_{TT}$. Additional computational details about the derivation of the potentials can be found in the appendix-section \ref{AppendixGRad} related to this section. It should be noticed that it is possible, for a far away wave zone field point, to specialize the harmonic gauge even further in order to eliminate four additional redundant quantities \cite{MisnerThroneWheeler,Schutz1,Maggiore1,PoissonWill,Buonanno1,Schutz2,Schutz3,Riles1}. We previously saw that when the spacetime metric is decomposed into a Minkowski contribution and a small deviation $d_{\alpha\beta}$, where we remind that the latter transforms under infinitesimal local translations in the following way, $d_{\alpha\beta}\rightarrow d'_{\alpha\beta}=d_{\alpha\beta}-\partial_\alpha t_\beta-\partial_\beta t_\alpha$. We remind that here we assumed that the metric deviation and the first derivative of the local translation are of the same order of magnitude, $|\partial_\alpha t_\beta|\leq|d_{\alpha\beta}|$. In order to relate the metric-deviation to the gravitational potentials we use the post-Minkowskian metric expansion \cite{Alain1,Alain0A,PoissonWill,PatiWill1} and we find, $d_{\alpha\beta}=h_{\alpha\beta}-h/2\ \eta_{\alpha\beta}$. An equivalent relation is given by, $h_{\alpha\beta}=d_{\alpha\beta}-d/2\ \eta_{\alpha\beta}$, where we used $d=-h$ and we observe that the gravitational potentials can be related to the trace-reversed metric-deviation introduced in the previous subsection.
From this, together with equation \eqref{trafodeviation-07.04.17}, we deduce that the gravitational potentials transform under the coordinate transformation, outlined in equation \eqref{trafocoordinates-07.04.17} in the following way, $h^{\alpha\beta}\rightarrow h'^{\alpha\beta}=h^{\alpha\beta}-\partial^\alpha t^\beta-\partial^\beta t^\alpha+\eta^{\alpha\beta}\partial_\sigma t^\sigma$, where we remind that $h_{\alpha\beta}=\eta_{\alpha\sigma} \eta_{\beta\rho} h^{\sigma\rho}$ and $h=\eta_{\sigma\rho}h^{\sigma\rho}$. The transformation law for the derivative of the gravitational potentials is,
\begin{equation}
\label{HG-07.04.17}
\partial_\beta h^{\alpha\beta}\rightarrow \partial'_\beta h'^{\alpha\beta}\,=\,\partial_\beta h^{\alpha\beta}-\partial_\beta \partial^\alpha t^\beta-\partial_\beta \partial^\beta t^\alpha +\eta^{\alpha\beta}\partial_\beta \partial_\nu t^\nu\,=\,\partial_\beta h^{\alpha\beta}-\Box t^\alpha.
\end{equation} 
We conclude that the harmonic gauge conditions for the gravitational potentials will be preserved whenever the components of the vector field are harmonic, $\Box t^\alpha =0$ \cite{Maggiore1,Buonanno1,Schutz1}. Further computational details can be withdrawn from the appendix-section \ref{AppendixGRad} related to this chapter. This choice of gauge can always be enforced and we decompose the small local translations into a temporal and spatial component,
\begin{equation}
\label{Vectors-07.04.17}
t^0\,=\, \frac{G}{c^3 R}\gamma(\tau,\textbf{N})+\mathcal{O}(R^{-2}),\quad \ t^a\,=\, \frac{G}{c^3 R} \kappa^a(\tau,\textbf{n})+\mathcal{O}(R^{-2}),
\end{equation}
where $\gamma$ and $\kappa^a$ are arbitrary functions depending on the retarded time $\tau$ and unit-vector $\boldsymbol{N}=\textbf{x}/R$. The vector can be decomposed in terms of its irreducible components, $\kappa^a=\kappa N^a+\kappa_T^a$ with the orthogonal relation, $N_a \kappa^a_T=0$. This allows us to work out the gauge transformation behaviour of the irreducible components of the gravitational potentials which, as we previously saw, contain the six independent degrees of freedom,
\begin{equation}
\label{Transformation-07.04.17}
\begin{split}
&A\ \rightarrow\  A'\,=\,A+3 \dot{\gamma}-\dot{\kappa},\quad \ Y^a_T\ \rightarrow\  Y'^a_T\,=\, Y^a_T+\dot{\kappa}^a_T,\\
&B\ \rightarrow\  B'\,=\,B+2\dot{\kappa},\quad \hspace{0.8cm} Z^{ab}_{TT}\ \rightarrow\  Z'^{ab}_{TT}\,=\, Z^{ab}_{TT},
\end{split}
\end{equation}
where we remind that the overdot stands for the differentiation with respect to the retarded time $\tau$. We observe that the transverse-tracefree component of the tensor $Z^{ab}$ is invariant under local translations of the form mentioned in equation \eqref{trafocoordinates-07.04.17}. Moreover it should be noticed that the free parameter-functions can be chosen in such a way that the components $A$, $B$ and $Y^a_T$ cancel out. In this regard we eventually arrive at the most concise formulation of the gravitational potentials for a far away wave zone field point,
\begin{equation}
\label{PotentialsTT-11.04.17}
h^{00}\,=\,\frac{4 G M}{c^2 R},\quad \ h^{0a}\,=\,0,\quad \ h_{TT}^{ab}\,=\,\frac{G}{c^4 R} Z^{ab}_{TT}(\tau,\textbf{N}).
\end{equation}
Additional computational details can be found in the appendix-section \ref{AppendixGRad} related to the present chapter. This particular form for the gravitational potentials is usually referred to as the transverse-tracefree gauge, or TT-gauge for short, which results from a specialization of the harmonic gauge condition \cite{MisnerThroneWheeler,PoissonWill,Schutz1,Schutz2,Schutz3,Maggiore1,Buonanno1,Riles1}. By virtue of the conditions imposed on the TT-components of the gravitational potentials, outlined in equation \eqref{constraints-06.04.17}, we infer that the number of time-dependent quantities has been reduced to two. From this we conclude that the radiative degrees of freedom of the gravitational field must be included within these two independent components. In what follows we will analyse in how far the radiative degrees of freedom contained within $Z^{ab}_{TT}$ will affect a hypothetical detector composed by two test masses that are moving freely in the far away wave zone. We assume that the masses are separated by a spacetime vector $\epsilon^\alpha$ and that they move with a four velocity $u^\alpha$. Furthermore we suppose that the separation between the masses is small, $\sqrt{\epsilon^b\epsilon_b}\ll \lambda_c$, compared to the characteristic wavelength of the radiation. It should be noticed that this defines a short gravitational-wave detector such as the LIGO (Laser Interferometer Gravitational-Wave Observatory) instrument \cite{LIGO1,LIGO2,LIGO3}. The behaviour of the separation vector between the two test masses is governed by the geodesic deviation equation and reads, $\frac{D^2\epsilon^\alpha}{D\lambda^2}=-R^\alpha_{\ \rho\kappa \beta}\ \epsilon^\kappa u^\beta u^\rho$, where $\frac{DV^\alpha}{D\lambda}=\frac{d V^\alpha}{d\lambda}+\Gamma^\alpha_{\nu\rho} V^\nu \frac{dx^\rho}{d\lambda}$ is the covariant derivative of a generic vector field $V^\alpha(x)$ along the curve $x^\alpha(\lambda)$. The equation of geodesic deviation states that there is a relative acceleration between geodesics when the spacetime is curved, which is the case whenever the Riemann tensor is non-vanishing. By assuming that the test masses are slowly moving, $u^\alpha\approx (c,\boldsymbol{0})$ and $\epsilon^\alpha\approx (0,\boldsymbol{\epsilon})$, the geodesic deviation equation reduces to the approximate form,
\begin{equation}
\label{GeoForce-10.04.17}
\frac{d^2\epsilon^a}{dt^2}\,=\,-c^2 R_{0a0b}\ \epsilon^b\,=\,\frac{G}{2c^4 R}\ddot{Z}_{ab}^{TT} \ \epsilon^b\,=\,\frac{1}{2}\ddot{h}_{TT}^{ab}\epsilon_b,
\end{equation}
which involves ordinary differentiation with respect to $t$ as well as the spatial components of the separation vector \cite{MisnerThroneWheeler,PoissonWill,Maggiore1,Weinberg2}. Further computational details regarding the derivation of this result can be found in the appendix-section \ref{AppendixGRad} related to this chapter. It should be noticed that the physical interpretation of this equation is rather simple because in the proper detector frame it claims that the effect of gravitational radiation exerted on a test body with a certain mass $M$ can be described in terms of a Newtonian force, $F^a=(M/2)\ \ddot{h}_{TT}^{ab}\epsilon_b$. It should be observed that in this particular context the response of the detector to the passage of gravitational radiation can be understood by employing a purely Newtonian language \cite{Maggiore1,Buonanno1}. We see that the gravitational wave detector is driven by the traceless transverse piece of the gravitational potentials. They contain the radiative degrees of freedom, while the remaining contributions, which can be eliminated by a coordinate transformation (equation \eqref{Transformation-07.04.17}), contain no radiative information. Hence we will call the $h^{ab}_{TT}$ contributions of the gravitational potentials the gravitational wave field \cite{PoissonWill,Maggiore1}. The radiative contributions of the gravitational potentials can be efficiently extracted by making use of the transverse traceless projector, 
\begin{equation}
Z^{ab}_{TT}\,=\,(TT)^{ab}_{\ \ kl}\ Z^{kl}\,=\,\big(P^a_{\ k}P^{b}_{\ l}-\frac{1}{2}P^{ab}P_{kl}\big)\ Z^{kl}.
\end{equation}
We see that the TT-operator is essentially composed by the transverse projector, $P^a_{\ b}=\delta^a_{\ b}-N^aN_b$, which removes the longitudinal components of vectors and tensors \cite{WillWiseman,Maggiore1}. This can be illustrated by having a closer look at the vector $Y^a=YN^a+Y^a_T$, where we remind that $N_aY^a_T=0$. It is straightforward to observe that when applying the transverse projector on this vector we precisely extract its transverse component, $P^a_{\ b}Y^b=Y^a_T$. We aim to provide some additional 
properties of this important quantity, $P^a_{\ b}P^b_{\ c}=P^a_{\ c}$, $P^a_{\ b} N^b=0$, $P^a_{\ a}=2$, which will allow us to immediately conclude that,
\begin{equation}
(TT)^{ab}_{\ \ kl}\ N^k\,=\,0,\quad \ (TT)^{ab}_{\ \ kl}\ \delta^{kl}\,=\,0,\quad \ (TT)^{ab}_{\ \ kl}\ Z^{kl}_{TT}\,=\,Z^{ab}_{TT}. 
\end{equation}
For a general symmetric tensor $Z^{ab}=\frac{\delta^{ab}}{3} A+\big(N^aN^b-\frac{\delta^{ab}}{3}\big)B+N^aZ^b_T+N^bZ_T^a+Z^{ab}_{TT}$ it can be observed without much computational pain that we have, $(TT)^{ab}_{\ \ kl}\ Z^{kl}=Z^{ab}_{TT}$. Additional computational details can be found in the appendix-section \ref{AppendixGRad} related to this section. In order to perform these formal operations it will turn out to be convenient to introduce a vectorial basis in the transverse subspace parametrised by the polar angles,
\begin{equation}
\label{PolarAngles-10.04.17}
\begin{split}
\textbf{k}\,=&\,[-\sin\varphi,\cos\varphi,0],\\
\textbf{n}\,=&\,[\sin\theta\cos\varphi,\sin\theta\sin\varphi,\cos\theta],\\
\textbf{m}\,=&\,[\cos\theta\cos\varphi,\cos\theta\sin\varphi,-\sin\theta].
\end{split}
\end{equation}
The vector $\textbf{m}$ points in the direction of increasing colatitude on the surface of a sphere and $\textbf{k}$ points in the direction of increasing longitude. Both vectors are orthogonal to each other and span the transverse subspace orthogonal to the third vector which is normal to the sphere. In this particular basis we get the transverse projector to be $P^{ab}=m^am^b+k^ak^b$ and the completeness relations are $\delta^{ab}=n^an^b+m^am^b+k^ak^b$. Any symmetric , transverse and tracefree tensor can be decomposed in a tensorial basis that is exclusively constructed from the vectors $\textbf{m}$ and $\textbf{k}$ \cite{Schutz1,PoissonWill,Riles1}. Such a tensor possesses two independent components, $Z^{ab}_{TT}=Z_+(m^am^b-k^ak^b)+Z_\times(m^ak^b+k^am^b)$, called the two polarisations of the tensor. From this we obtain,
\begin{equation}
\label{Polar-10.04.17}
\begin{split}
Z_+\,=&\,\frac{1}{2}(m_am_b-k_ak_b)\ Z_{TT}^{ab}\,=\,\frac{1}{2}(m_am_b-k_ak_b)\ Z^{ab},\\
Z_\times\,=&\,\frac{1}{2}(m_ak_b+k_am_b)\ Z_{TT}^{ab}\,=\,\frac{1}{2}(m_ak_b+k_am_b)\ Z^{ab}.
\end{split}
\end{equation}
This result is obtained by virtue of the fact that the tensorial operators acting on $Z_{TT}^{ab}$ are already transverse and tracefree. By inserting the vectors of the orthogonal basis mentioned in equation \eqref{PolarAngles-10.04.17} we can rephrase the two polarisations, given in equation \eqref{Polar-10.04.17}, in terms of the polar angles (appendix-section \ref{AppendixGRad}). By using the vectors of the orthogonal basis inside the relation, $Z^{ab}_{TT}=Z_+(m^am^b-k^ak^b)+Z_\times(m^ak^b+k^am^b)$ mentioned already previously, it is possible to construct the different entries of the $Z^{ab}_{TT}$ matrix,
\begin{equation}
\label{Components-11.04.17}
\begin{split}
Z^{xx}_{TT}\,=&\,-\frac{1}{2}\ \big[\sin^2\theta-(1+\cos^2\theta)\cos2\varphi\big]\ Z_+-\cos\theta\ \sin2\varphi\ Z_\times,\\
Z^{xy}_{TT}\,=&\,+\frac{1}{2}\ (1+\cos^2\theta)\sin2\varphi\ Z_++\cos\theta\ \cos2\varphi\ Z_\times,\\
Z^{xz}_{TT}\,=&\,-\frac{1}{2}\ \sin 2\theta\ \cos\varphi\ Z_++\sin\theta\sin\varphi\ Z_\times,\\
Z^{yy}_{TT}\,=&\,-\frac{1}{2}\ [\sin^2\theta+(1+\cos^2\theta)\cos2\varphi]\ Z_++\cos\theta\sin2\varphi\ Z_\times,\\
Z^{yz}_{TT}\,=&\,-\frac{1}{2}\ \sin2\theta\ \sin\varphi\ Z_+-\sin\theta\cos\varphi\ Z_\times,\\
Z^{zz}_{TT}\,=&\,+\frac{1}{2}\ (1-\cos2\theta)\ Z_+.
\end{split}
\end{equation}
In order to develop a better understanding of how the gravitational radiation interacts with matter we consider the case of a gravitational wave propagating along the z direction. Moreover we will consider an initially circular ring, lying in the $x-y$ plane, of freely moving particles in an inertial frame \cite{Schutz0C,PoissonWill,Riles1}. In this particular case $\theta=0$ and we can choose $\phi=0$ and we obtain,
\begin{equation}
\label{Tensor-11.04.17}
Z^{ab}_{TT}\,=\,
\begin{pmatrix}
\ Z_+ & Z_\times & 0\ \\
\ Z_\times & -Z_+ & 0\ \\
\ 0&0&0 \ 
\end{pmatrix}.
\end{equation}
The geodesic deviation equation can be solved by assuming that the transverse tracelss components of the gravitational potentials are small \cite{Maggiore1,Schutz1,Schutz2,Schutz3,Riles1,Schutz0C}. In this regard we obtain the leading order solution, $\epsilon^a(t)=\epsilon^a(0)+(1/2)h^{ab}_{TT}(\tau)\epsilon_b(0)$, where $\epsilon^a(0)$ is the initial separation between two geodesics. We observe that the variations in the displacement vector are essentially driven by transverse tracelss part of the potentials and that they are, to leading order, proportional to their initial separation. More explicitly the geodesic deviations in terms of their respective components are,
\begin{equation}
\label{Displ-11.04.17}
\delta x(\tau)\,=\,\frac{G}{c^4 R}\frac{Z_+x_0+Z_\times y_0}{2},\quad \ \delta y(\tau)\,=\,\frac{G}{c^4 R}\frac{Z_\times x_0-Z_+ y_0}{2}, \quad \ \delta z(\tau)\,=\,0,
\end{equation}
where $\boldsymbol{\epsilon}(0)=[x_0,y_0,z_0]$ is the initial position of the test mass, $\boldsymbol{\epsilon}(\tau)=[x(\tau),y(\tau),z(\tau)]$ is the temporal displacement vector and $\boldsymbol{\delta}=[\delta x=x(\tau)-x_0,\delta y=y(\tau)-y_0,\delta z=z(\tau)-z_0]$ is the displacement of the test mass itself. We see that in the context of the particular choice made previously for a gravitational wave travelling in the $z$-direction ($\theta=0$) there is no displacement for the test masses in this direction. The radiative degrees of freedom are contained exclusively within the $x-y$ plane, which is perpendicular to the $z$-direction. It is instructive to study the time-dependent displacement in the framework of a periodic sinusoidal source term. We assume that the polarisations have the form, $Z_+=h_+\sin(\omega_c \tau)$ and $Z_\times=h_\times\sin(\omega_c\tau)$ with $h_+$ and $h_\times$ being arbitrary time independent parameters and $\omega_c$ being proportional to the characteristic frequency of the matter source term. With this we obtain for the from equation \eqref{PotentialsTT-11.04.17} together with equation \eqref{Tensor-11.04.17} the following relation for the gravitational potentials,
\begin{equation}
\label{Polarisation-24.04.17}
h_{TT}^{ab}\,=\,\frac{G}{c^4 R}
\begin{pmatrix}
\ h_+ & h_\times \ \\
\ h_\times & -h_+ \ \\
\end{pmatrix}
\ \sin(\omega_c \tau),
\end{equation}
For a given detector position (test masses) we observe that the sinusoidal periodicity of the source reappears at the level of the displacements of the test masses \cite{PoissonWill,Maggiore1,Buonanno1,Schutz0A,Schutz1,Schutz2,Schutz3,Riles1}. By inserting these potentials into the leading order solution of the geodesic deviation equation, we observe that the displacement of the test particle ring is composed by the superposition of the two polarisations,
\begin{equation}
\delta x(\tau)\,=\,\frac{G}{c^4 R}\frac{h_+x_0+h_\times y_0}{2}\sin(\omega_c \tau),\quad \ \delta y(\tau)\,=\,\frac{G}{c^4 R}\frac{h_\times x_0-h_+ y_0}{2}\sin(\omega_c \tau).
\end{equation} 
We observe that the further away from the source the ring of test particles is situated the smaller the deviation of the test masses will. When we consider only the $+$-mode or only the $\times$-mode we see that a circle of test masses of unit radius will be deformed in both cases into an ellipse described respectively by the following two relations,
\begin{equation}
\label{Ellipses-11.04.17}
\Big[\frac{x}{1+\Delta_+}\Big]^2+\Big[\frac{x}{1-\Delta_+}\Big]^2\,=\,1, \quad \ \Big[\frac{x-\Delta_\times y}{1-\Delta^2_\times}\Big]^2+\Big[\frac{y-\Delta_\times x}{1-\Delta^2_\times}\Big]^2\,=\,1,
\end{equation}
where $\Delta_+=\frac{G\ Z_+(t)}{2c^4\ R}$ and $\Delta_\times=\frac{G\ Z_\times(t)}{2c^4\ R}$ are two small and time-dependent quantity that vary between their minimum and maximum values. Further computational details can be withdrawn from the appendix-section \ref{AppendixGRad} related to the present chapter. The situation is illustrated by Figure \ref{PolarisationsPlusCross} and we observe that the second ellipse is rotated with respect to the first one by an angle of 45 degrees \cite{Maggiore1,Schutz1,Schutz2,Schutz3}. It is straightforward to observe that when $\Delta_+(t)=\Delta_\times(t)=0$ both curves, displayed in equation \eqref{Ellipses-11.04.17}, reduce to a circle of unit radius.
In what follows we will study the gravitational wave field produced by a binary system of orbiting bodies. In order to describe the orbital motion of this two-body system it turns out to be convenient to use an orbit-adapted coordinate system $(x,y,z)$. The orbital plane coincides with the $x-y$ plane and the $z$-axis points in the direction of the angular-momentum vector (equation \eqref{Angular-03.04.17}). In addition we choose the orbit-adapted coordinate system in such a way that its origin is located at the barycentre of the binary system and that the $x$-axis is aligned with the orbit's major axis, while the $y$-axis is aligned with the minor axis. We recall from that $\textbf{e}_r=[\cos\phi,\sin\phi,0]$ and $\textbf{e}_\phi=[-\sin\phi,\cos\phi,0]$ are the time dependent unit polar vectors expressed in a cartesian basis ($\textbf{e}_x$, $\textbf{e}_y$) satisfying $d\textbf{e}_r/d\phi=\textbf{e}_\phi$ and $d\textbf{e}_\phi/d\phi=-\textbf{e}_r$. Furthermore we remind from equation \eqref{FAWZFP-28.03.17} that to leading order the gravitational potentials for a far away wave zone field point are proportional to the second temporal derivative of the radiative quadrupole moment. In addition we determine the radiative quadrupole moment for a two-body system at the 1.5 post-Newtonian order of accuracy. From this particular result, truncated at the Newtonian order of accuracy, we obtain,
\begin{equation}
\label{BSGP-12.04.17}
\begin{split}
h^{ab}\,=\,\frac{2G}{c^4 R}\ \ddot{I}^{ab}\,=\,\frac{4\eta}{c^4 R}\frac{(Gm)^2}{p}\ & \Big[e^2\sin^2\phi-(1+e\cos\phi)\textbf{e}^a_r\textbf{e}^b_r+e\sin\phi\\ &\ \quad (1+e\cos\phi)
 (\textbf{e}_r^a\textbf{e}_\phi^b+\textbf{e}^a_\phi\textbf{e}_r^b)+(1+e\cos\phi)^2\textbf{e}_\phi^a\textbf{e}_\phi^b\Big],
\end{split}
\end{equation}
where we recall that $I^{ab}$ is the Newtonian quadrupole moment. Moreover we remind that $\eta=m_1m_2/m^2$ is a dimensionless parameter and that $p$ is a quantity of dimension length commonly known as the orbit's {\it semi-latus rectum} \cite{WagonerWill1976,WillWiseman,PoissonWill}. Additional details about the derivation of this result can be found in the appendix-section \ref{AppendixGRad} related to this chapter. It turns out to be useful to introduce, in addition to the original coordinate system ($x$, $y$, $z$) a detector-adapted coordinate system ($X$, $Y$, $Z$) in order to construct the gravitational wave polarisations $h_+$ and $h_\times$. The origin of the new system coincides with the orbit-adapted coordinate system. The $Z$-axis points in the direction of the gravitational wave detector where the polarisations of the incoming gravitational radiation are measured. Moreover the $X-Y$ plane is orthogonal to the $Z$-axis and the $X$-axis is aligned with the line where the orbital plane cuts the reference plane (line of nodes). By convention it points towards the point where the orbit cuts the plane from below (ascending node). 
\begin{figure}[h]
\begin{center}
\includegraphics[width=10.6cm,height=5.2cm]{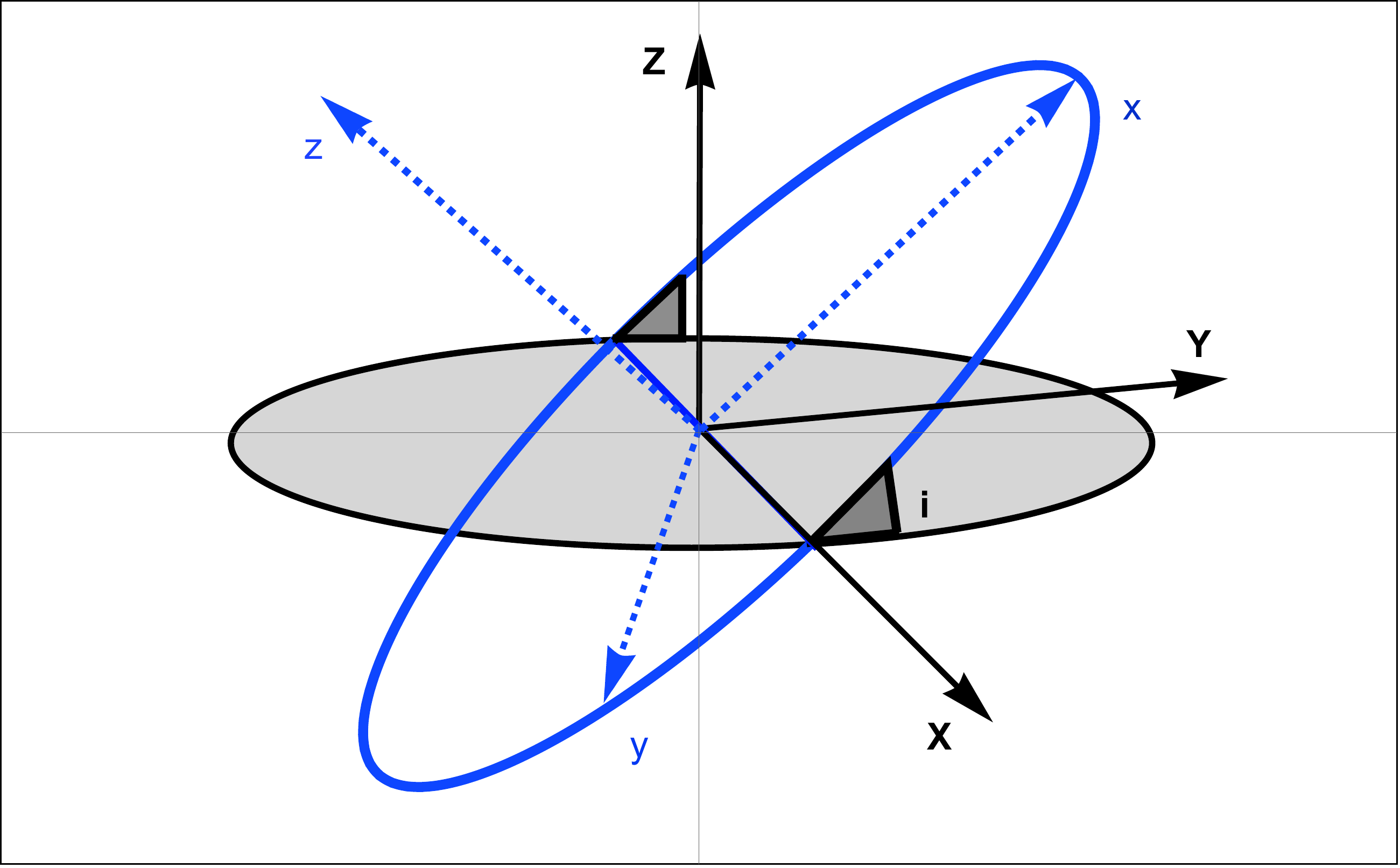}
\end{center}
\caption[Orbital plane - detector adapted plane]{\label{DetAd}This figure illustrates the orbital plane which coincides with the $x-y$ plane (blue dashed coordinate system) and the detector-adapted plane which coincides with the $X-Y$ plane (black coordinate system). The angle $i$ is the inclination angle between the orbital plane and the detector-adapted plane and $\omega$, which for reasons of clarity is not indicated in the figure, is the angle between the $x$-axis of the orbital plane and the $X$-axis of the detector-adapted plane. The line of intersection between the two planes is known as the line of nodes and coincides with the $X$-axis of the detector-adapted plane.}
\end{figure}
The situation is illustrated by Figure \ref{DetAd} and in terms of the original coordinates ($x$,$y$,$z$) the new coordinate directions are given by,
\begin{equation}
\begin{split}
\textbf{e}_X\,=&\,[\cos\omega,-\sin\omega,0],\\
\textbf{e}_Y\,=&\,[\cos i\sin\omega,\cos i \cos\omega,-\sin i],\\
\textbf{e}_Z\,=&\,[\sin i\sin\omega,\sin i\cos\omega,\cos i].
\end{split}
\end{equation}
In the detector adapted frame the inclination angle $i$ measures the inclination of the orbital plane with respect to the $X-Y$ plane, while the longitude of pericenter $\omega$ is the angle between the pericenter ($x$-axis) and the line of nodes ($X$-axis), as measured in the orbital plane \cite{Schutz0C,Schutz1, Buonanno1,Riles1,PoissonWill,Maggiore1}. The vectors forming a basis in the orbital plane become,
\begin{equation}
\begin{split}
\label{polllla-13.04.17}
\textbf{e}_r\,=&\,[\cos(\omega+\phi),\ \cos i\sin(\omega+\phi),\ \sin i \sin(\omega+\phi)],\\
\textbf{e}_\phi\,=&\, [-\sin(\omega+\phi),\cos i\cos(\omega+\phi),\sin i\cos(\omega+\phi)],
\end{split}
\end{equation}
when formulated in the detector adapted coordinate system. It is convenient to use the transverse subspace composed by $\textbf{e}_X$ and $\textbf{e}_Y$ as a basis to describe the gravitational radiation propagating from the binary system towards the detector along the $Z$-axis. In this context we can work out the two gravitational waves polarizations,
\begin{equation}
\label{polla-13.04.17}
h_+\,=\,\frac{1}{2}\ [\textbf{e}^a_X\textbf{e}_X^b-\textbf{e}_Y^a\textbf{e}_Y^b]\ h_{ab},\quad\ 
h_\times\,=\,\frac{1}{2}\ [\textbf{e}^a_X\textbf{e}_Y^b+\textbf{e}_Y^a\textbf{e}_X^b]\ h_{ab},
\end{equation}
and it should be noticed that this new basis only differs in notation ($\textbf{k}=-\textbf{e}_X$, $\textbf{m}=\textbf{e}_Y$, $\textbf{n}=\textbf{e}_Z$) when compared to the vectors displayed in equation \eqref{PolarAngles-10.04.17} and that the old angles are related to the new ones by $\theta=i$ and $\varphi=\frac{\pi}{2}-\omega$. In this particular basis the gravitational wave polarisations become,
\begin{equation}
\label{polla2-20.04.17}
h_+\,=\,A_o\ P_+,\quad \ h_\times\,=\,A_o\ P_\times,
\end{equation}
where $A_o=\frac{2\eta}{c^4 R}\frac{(Gm)^2}{p}$ is the amplitude of the gravitational wave and $P_+$ and $P_\times$ are the two scale free polarisations of the incoming gravitational radiation \cite{WillWiseman,PoissonWill,Schutz1}. Their precise form can be withdrawn from the appendix-section \ref{AppendixGRad} related to this chapter. For the special case of circular orbits ($e=0$), where we remind that $\phi=\Omega \tau$ increases with a constant orbital frequency $\Omega=\sqrt{G m/p^3}$, the two polarisations become,
\begin{equation}
\label{Circular-13.04.17}
P_+\,=\,-(1+\cos^2 i)\ \cos[2(\Omega \tau+\omega)],\quad \ P_\times\,=\,-2\cos i\ \sin[2(\Omega \tau+\omega)]. 
\end{equation}
We observe that because of the quadrupolar nature of the gravitational waves, the latter see their orbital frequency doubled. To conclude this section we display in Table \ref{Amplitudes-20.04.17} different values for the amplitude of the two gravitational wave polarisations outlined in equation \eqref{polla2-20.04.17}.
\vspace{0.15cm}
\begin{table}[h]\begin{center}
\begin{tabular}{c|c|c|c|c}
$A_o$ &Earth-Sun&Double Pulsar&Stellar BH& SMassive BH\\ \hline
\parbox[0pt][3em][c]{0cm}{}
$A_o(R_1)\ $ &\hspace{0.5cm}$\sim10^{-27}\hspace{0.6cm}$& $\sim10^{-22}$&\hspace{0.5cm} $\sim10^{-19}\hspace{0.5cm}$ & $\sim 10^{-10}$\\ 
$A_o(R_2)\ $ &$\sim10^{-17}$ &$\sim10^{-11}$ & $\sim10^{-8}$ & $\sim 10^{+1}$
\end{tabular}
\caption[Gravitational waves' amplitudes: different binary systems]{\label{Amplitudes-20.04.17}In this table we present different orders of magnitude for the values of the amplitude $A_o$ for various binary systems (Earth-Sun, Double Pulsar, stellar black hole binary system and supermassive black hole binary system) as a function of the distance ($R_1\sim 1.8\times 10^{16} \ km$, $R_2\sim 3.844\times 10^{5}\ km$) between the source and a potential detector. We use the semi-major axis and the orbital eccentricity of the Double Pulsar (PSR J0737-3039A/B) binary system \cite{DoublePulsar}. The masses of the stellar black holes are those deduced from the gravitational wave signal GW150914 \cite{LIGO1} and the supermassive black hole binary system have masses of the order of the mass of Sgr $\text{A}^*$ \cite{SGRA}.}\end{center}
\end{table}
We present four different bound two-body systems (Earth-Sun, Double Pulsar, stellar black hole binary system and supermassive black hole binary system) and work out the various orders of magnitude of the amplitude $A_o$ for two different distances between the source and a potential detector. We observe that for the distance $R_1\sim 1.8\times 10^{16} \ km$, which is the approximate spatial interval between the Earth and the Double Pulsar (PSR J0737-3039A/B) the different amplitudes range from $\sim 10^{-27}$ to $\sim 10^{-10}$. For the smaller distance Earth-Moon $R_2\sim 3.844\times 10^{5}\ km$ the amplitudes become much stronger than for the previous case. In both situations we assumed the orbital period and orbital eccentricity to be the ones of the Double Pulsar binary system \cite{Burgay1,DoublePulsar}.

\section{Radiative losses in gravitating systems:}
\label{RadiativeLosses}
\begin{figure}[h]
\begin{center}
\includegraphics[width=11.6cm,height=5.8cm]{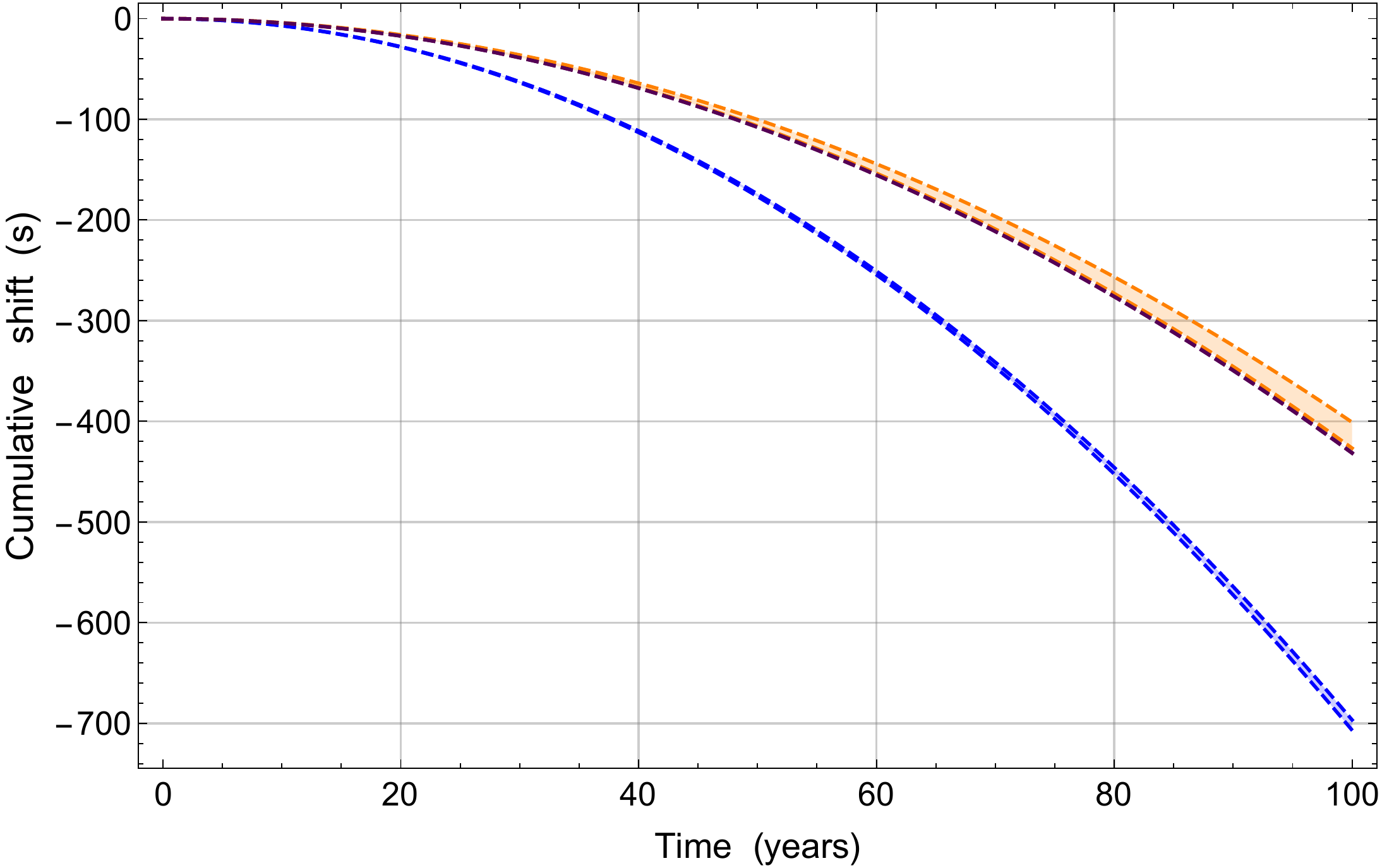}
\end{center}
\caption[Cumulative shift of periastron time: PSR J0737-3039A/B, PSR B1913+16, PSR B1534+12]{\label{PeriastroTimeShift}This figure illustrates the accumulated decay-time of the orbital period, due to the permanent emission of gravitational radiation, for the Double Pulsar (PSR J0737-3039A/B), the Hulse-Taylor binary system (PSR B1913+16) and for the the double neutron star binary (PSR B1534+12) for a hundred years observation time interval. We observe that the cumulated shift of periastron time is strongest for the Double Pulsar (blue lower curves) and almost identical for the Hulse-Taylor binary system (red intermediate curves) and PSR B1534+12 (orange upper curves). The spreading of the three different curves results from the standard errors on the different parameters outlined in Table \ref{B1534} and Table \ref{HulseTaylorPulsar}.}
\end{figure}

In this section we will use the conservation identities outlined in \cite{PoissonWill} in order to reformulate these equations in terms of the transverse traceless gravitational potentials discussed previously. In this regard we know from equation \eqref{conservationrelations2} that the temporal energy variation of a gravitating system is given by,
\begin{equation}
\label{balance-16.04.17}
\frac{dE}{dt}\,=\,-\mathcal{P},
\end{equation}
where $\mathcal{P}=c\oint_\infty (-g) t_{LL}^{0b} dS_b$ is the rate at which gravitational waves remove energy from the system. The surface integrals are evaluated in the limit $R\rightarrow \infty$ and $dS_a=R^2N_ad\Omega$ is an outward-directed surface element. Furthermore we remind that $R=|\textbf{x}|$ is the position of a generic field point, $\textbf{N}=\textbf{x}/R$ is a unit vector pointing from the source towards the field point and $d\Omega$ is an element of solid angle in the direction of the unit vector $\textbf{N}$. In a similar manner we obtain the momentum and angular balance equations respectively,
\begin{equation}
\frac{dP^a}{dt}\,=\,-\mathcal{F}^a,\quad \quad \frac{dJ^{ab}}{dt}\,=\,-\mathcal{A}^{ab},
\end{equation}
where $P^a$ and $J^{ab}$ are respectively the total linear and the total angular moments of the system. Moreover we remind that, $\mathcal{F}^a=\oint_\infty (-g) t^{ab}_{LL}dS_b$, is the rate at which gravitational waves remove linear momentum and angular momentum, $\mathcal{A}^{ab}=\oint_\infty\big[x^a(-g)t^{bp}_{LL}-x^b(-g)t^{ap}_{LL}\big]dS_p$, away from the source. The surface integrals are evaluated in the limit $R\rightarrow \infty$ and we recall that $dS_a=R^2N_ad\Omega$ is an outward-directed surface element. $\textbf{N}=\textbf{x}/R$ is a unit vector and $d\Omega$ is an element of solid angle in the direction of $\textbf{N}$ \cite{Schutz3,MisnerThroneWheeler,LandauLifshitz,Schutz1,WillWiseman}. It should be noticed that the Landau-Lifshitz formulation of the Einstein field equations provides a physically sensible description of how the energy, the linear momentum and the angular momentum change as a result of the emission of gravitational radiation. A different formulation of the gravitational field equations, based on other variables, other coordinate systems and a different choice of pseudotensors may lead to different definitions for these three quantities with different expressions for $\mathcal{P}$, $\mathcal{F}^a$ and $\mathcal{A}^{ab}$. An elegant way to extract the leading order terms from the full Landau-Lifshitz pseudotensor $(-g)\tau^{\alpha\beta}_{LL}$ is to use an approximation technique for a far away wave zone field point. This particular technique introduced for the first time by Misner, Throne and Wheeler in 1973 is called the shortwave approximation \cite{MisnerThroneWheeler}. The latter is broader than the post-Minkowskian approximation introduced previously and fully includes the post-Minkowskian estimation procedure \cite{PoissonWill}. We know that in the far away wave zone the characteristic wavelength of the gravitational radiation is much smaller than the distance between the barycentre of the source and a generic wave field point. We will use this property to formally expand the gravitational potentials in terms of this particular expansion parameter $\epsilon=\lambda_c/R\ll 1$ and we have,
\begin{equation}
h^{\alpha\beta}\,=\,\epsilon \ w^{\alpha\beta}_1(\tau,\textbf{N})+\epsilon^2 \ w^{\alpha\beta}_2(\tau,\textbf{N})+\epsilon^3 \ w^{\alpha\beta}_3(\tau,\textbf{N})+\cdots,
\end{equation}
where the functions $w^{\alpha\beta}_l$ are functions of the retarded time and the angles contained within $\textbf{N}$. It should be noticed that in the context of a post-Minkowskian approximation the functions $w^{\alpha\beta}_l$ are accurate to all orders in the gravitational constant $G$. The retarded time dependence of $w^{\alpha\beta}_l$ originates from the fact that the gravitational potentials must satisfy the wave equation, $\Box h^{\alpha\beta}=-(16\pi G/c^4)\ \tau^{\alpha\beta}$. As the effective Landau-Lifshitz pseudotensor is essentially composed by the gravitational potentials and its spacetime derivatives we wish to work out the leading order derivative term of the potentials in the shortwave approximation and find after some lengthy computations the following result for the spacetime derivative of the potentials,
\begin{equation}
\label{DerivativePot-15.04.17}
\partial_\gamma h^{\alpha\beta}\,=\,-c^{-1}s_\gamma \partial_\tau h^{\alpha\beta}+\mathcal{O}(\epsilon^2),
\end{equation}
where $s_\gamma=(-1,\textbf{N})$ is a spacetime vector which satisfies the null condition $\eta^{\kappa\gamma}s_\kappa s_\gamma=0$. The reader interested in the computational details is referred to the appendix-section \ref{AppendixGRad} related to this chapter. By taking into account the harmonic gauge condition $\partial_\alpha h^{\alpha\beta}=0$, we know from equation \eqref{LLTensorPotentials}, that the Landau-Lifshitz pseudotensor, previously expressed in terms of the gothic inverse metric, reduces to,
\begin{equation}
\label{LL-15.04.17}
\begin{split}
(-g)t^{\alpha\beta}_{LL}\,=&\,\frac{c^4}{16\pi G}\ \Big[g_{\lambda\mu}g^{\nu\rho}\partial_\nu h^{\alpha\lambda}\partial_\rho h^{\beta\mu}+\frac{1}{2} g_{\lambda\mu}g^{\alpha\beta}\partial_\rho h^{\lambda\nu}\partial_\nu h^{\rho\mu}-2g_{\mu\nu} g^{\lambda(\alpha}\partial_\rho h^{\beta)\nu}\partial_\lambda h^{\rho\mu}\\
&\,\hspace{1.7cm}+\frac{1}{8}\big(2g^{\alpha\lambda}g^{\beta \mu}-g^{\alpha\beta}g^{\lambda\mu}\big)\big(2g_{\nu\rho}g_{\sigma\tau}-g_{\rho\sigma} g_{\nu\tau}\big)\partial_\lambda h^{\nu\tau}\partial_\mu h^{\rho\sigma}\Big],
\end{split}
\end{equation}
where for reasons of completeness it should be mentioned that, $-2g_{\mu\nu} g^{\lambda(\alpha}\partial_\rho h^{\beta)\nu}\partial_\lambda h^{\rho\mu}=\\-g_{\mu\nu} g^{\lambda\alpha}\partial_\rho h^{\beta\nu}\partial_\lambda h^{\rho\mu}-g_{\mu\nu} g^{\lambda\beta}\partial_\rho h^{\alpha\nu}\partial_\lambda h^{\rho\mu}$. We will use this result and refine the harmonic gauge to the transverse traceless gauge in which $h^{00}=\mathcal{O}(\epsilon^2)$, $h^{0a}=\mathcal{O}(\epsilon^2)$ and $h^{ab}=h^{ab}_{TT}+\mathcal{O}(\epsilon^2)$ in order to obtain after a rather long computation (appendix),
\begin{equation}
\label{TT-15.04.17}
(-g)t^{\alpha\beta}_{LL}\,=\,\frac{c^2}{32\pi G}\ \big[\dot{h}^{ab}_{TT}\dot{h}^{TT}_{ab}\big]\ s^\alpha s^\beta,
\end{equation}
where we recall that $s^\alpha=(1,\textbf{N})$ and $h^{ab}_{TT}=(TT)_{\ cd}^{ab}\ h^{cd}$ is the transverse tracefree component of the gravitational potentials. The derivation of this result can be withdrawn from the appendix-section \ref{AppendixGRad}. Furthermore it should be noticed that the present relation is of the order $\epsilon^2$ and that it can be used within the relation $\mathcal{P}=c\oint_\infty (-g) t_{LL}^{0b} dS_b$ in order to determine the emitted radiation energy per unit time \cite{WillWiseman,MisnerThroneWheeler,PoissonWill,Maggiore1, Buonanno1}. In addition we remind, from the previous subsection, that it is possible to decompose the transverse traceless gravitational potentials in terms of the two wave polarisations $h_+$ and $h_-$ using the transverse basis ($\textbf{k},\textbf{m}$) outlined in equation \eqref{PolarAngles-10.04.17}. From this we eventually obtain for the energy and the linear-momentum flux rates,
\begin{equation}
\label{Fluxes-16.04.17}
\begin{split}
\mathcal{P}\,=&\,\frac{c^3}{32\pi G}\int d\Omega\ R^2\ \big[\dot{h}^{ab}_{TT}\ \dot{h}^{TT}_{ab}\big]\,=\,\frac{c^3}{16\pi G}\int d\Omega \ R^2\ \big[\dot{h}^2_++\dot{h}_{\times}^2\big],\\
\mathcal{F}^a=&\,\frac{c^2}{32\pi G}\int d\Omega \ R^2\ \big[\dot{h}^{ab}_{TT}\ \dot{h}^{TT}_{ab}\big]\ N^a\,=\,\frac{c^2}{16\pi G}\int d\Omega\ R^2\ \big[\dot{h}_+^2+\dot{h}_\times^2\big]\ N^a.
\end{split}
\end{equation}
It should be noticed that these two expressions are exact within the shortwave approximation and accurate to all orders in a post-Minkowskian expansion of the potentials in powers of the Newtonian constant. In a similar way it is possible to derive a relation that describes the emission of angular momentum from a given source and we display this particular result in equation \eqref{AMf-16.04.17} in the appendix-section \ref{AppendixGRad} related to this chapter. In what follows we will derive from the first expression outlined in equation \eqref{Fluxes-16.04.17} the quadrupole formula, which will provide an easy to use relation to determine the power removed from a given source due to the emitted gravitational radiation. As the name of this famous relation suggests we will formulate the amount of emitted energy per unit time in terms of the quadrupole moments. In this regard we remind that to leading Newtonian order the relation presented in equation \eqref{FAWZFP-28.03.17} is, 
\begin{equation}
\label{N-16.04.17}
h^{ab}\,=\,\frac{2G}{c^4R}\ddot{I}^{ab}\,=\,\frac{2G}{c^4R}[\ddot{I}^{\langle ab\rangle}+(\delta^{ab}/3)\ \ddot{I}^{qq}],
\end{equation}
where we remind from equation \eqref{NewtonPo-12.04.17} that $I^{ab}=\sum_A m_A x^a_Ax^b_A$ is the Newtonian radiative quadrupole moment of an $n$-body system and $I^{ab}=I^{\langle ab\rangle}+(\delta^{ab}/3)\ I^{qq}$ was decomposed in its irreducible components. It should be observed that in the TT-gauge, $h^{ab}_{TT}=(TT)^{ab}_{\ cd}h^{cd}$, the term proportional to $\delta^{ab}$ will not survive the transverse traceless projection operation and we will obtain from equation \eqref{balance-16.04.17} together with equation \eqref{Fluxes-16.04.17} for the emitted energy per unit time,
\begin{equation}
\label{QF-16.04.17}
\frac{dE}{dt}\,=\,-\frac{G}{5c^5} \dddot{I}^{\langle ab\rangle}\dddot{I}_{\langle ab\rangle}.
\end{equation}
This is the famous quadrupole formula which relates the energy radiated off the source under the form of gravitational waves to the quadrupole moment tensor of the matter distribution \cite{LandauLifshitz,MisnerThroneWheeler,WagonerWill1976,Buonanno1,Schutz0A,Schutz1,Schutz2,Schutz3}. Additional computational details for the precise derivation of this important relation can be found in the appendix-section \ref{AppendixGRad}. In what follows we will use this result in order to determine the gravitational wave flux for a binary system composed by two bodies with respective masses $m_1$ and $m_2$. We describe the mutual gravitational attraction between the two bodies at the Newtonian order of accuracy employing the Keplerian laws \cite{PoissonWill,Weinberg2,Inverno}. We obtain, after a rather long calculation outlined in the appendix-section \ref{AppendixGRad}, for the released energy-flux of a binary system evolving on elliptic orbits,
\begin{equation}
\label{AverageEmission-17.04.17}
\frac{d E}{dt}\,=\,-\langle \mathcal{P}\rangle \,=\,-\frac{32}{5c^5}\ \eta^2 \ \frac{G^4m^5}{p^5}\ (1-e^2)^{3/2}\ \Big[1+\frac{73}{24}e^2+\frac{37}{96}e^4\Big],
\end{equation}
where we remind that $p=a(1-e^2)$ is the {\it semi-latus rectum}, $e$ is the eccentricity and $\eta=(m_1m_2)/(m_1+m_2)^2$ is a dimensionless parameter. The $\langle... \rangle$ stands for the average energy released over a complete orbital cycle and more details about this notation can be found in the appendix-section \ref{AppendixGRad} related to this chapter. In this particular context we can determine the temporal variation of the semi-major axis of the fictitious body with reduced mass $\mu=m_1m_2/m$ orbiting the binary system's barycentre,
\begin{equation}
\label{SMAD-17.04.17}
\frac{da}{dt}\,=\,-\frac{64}{5c^5}\eta \ \Big(\frac{Gm}{ a}\Big)^3\ \frac{1+\frac{73}{24}e^2+\frac{37}{96}e^4}{(1-e^2)^{7/2}}\,=\,-\frac{C_a}{a^3},
\end{equation}
where we recall that $\eta=m_1m_2/m^2$ is a dimensionless parameter describing the two-body system. Computational details are presented in the appendix-section \ref{AppendixGRad}. We observe that the rate of decrease of the semi-major axis grows with increasing values of the latter and by integrating the equation above we get, $a(t)=(a^4_o-4 C_a\ t)^{1/4}$, where $a_o$ is the initial value of the semi-major axis. Using equation \eqref{SMAD-17.04.17} and reminding Kepler's third law \cite{PoissonWill,Inverno} it is straightforward to derive a relation that describes the temporal decrease of the orbital period,
\begin{equation}
\label{Periodvar-18.04.17}
\dot{P}\,=\,-3\ C_a\ \Big[\frac{(2\pi)^5}{P^5(Gm)^4}\Big]^{1/3},
\end{equation}
where $C_a$ was defined in equation \eqref{SMAD-17.04.17}. Here again we observe that for increasing values of the orbital period the decrease rate of the latter becomes more important \cite{WagonerWill1976,WillWiseman,PatiWill1,PoissonWill}. It is possible to work out a relation between the orbital period and the orbital energy. From this we can derive an equation which describes their respective temporal variations and in combination with the result displayed in equation \eqref{AverageEmission-17.04.17} we eventually obtain,
\begin{equation}
\label{PeriodVar-18.04.17}
\frac{\dot{P}}{P}\,=\,-\frac{3}{2}\frac{\dot{E}}{E}\,=\,-\frac{3}{2}\ \frac{2a}{Gm\mu}\ \langle\mathcal{P}\rangle\,=\,-\frac{96}{5}\ \frac{G^3\mu m^2}{c^5 a^4}\ \frac{1+\frac{73}{24}e^2+\frac{37}{96}e^4}{(1-e^2)^{7/2}},
\end{equation}
where we remind that $\mu=m_1m_2/m$ is the reduced mass of the fictitious body orbiting the binary system's barycentre and $m=m_1+m_2$ is the sum of the two-body system's respective masses. Additional computational details, on this particular result, are outlined in the appendix-section \ref{AppendixGRad}. Due to the permanent emission of gravitational radiation the total energy of the system decreases. As a consequence the two bodies approach their common center-of mass and we observe from equation \eqref{Periodvar-18.04.17} that the orbital period $P$ decreases too. The correlation between the accumulated orbital phase and the binary system's orbital frequency $\nu=P^{-1}$ is given by the relation $\Phi(T)=\int_0^Tdt \ \nu(t)$, where $T$ is the proper time in the two-body system's frame \cite{Maggiore1}. The $n$-th time of periastron passage $T_n$ takes place whenever we the condition $\Phi(T_n)=2\pi n$ is fulfilled and from this, together with equation \eqref{PeriodVar-18.04.17}, we can deduce a relation which describes the cumulative shift of periastron time as a function of the observation time,
\begin{equation}
\label{Periastronshift-18.04.17}
T_n-P\cdot n\,=\,-\frac{48}{5}\ \frac{G^3\mu m^2}{c^5 a^4}\ \frac{1+\frac{73}{24}e^2+\frac{37}{96}e^4}{(1-e^2)^{7/2}}\ T_n^2.
\end{equation}
It should be noticed that this relation is accurate only to leading order in the periastron passage time and in this regard it neglects terms of the order $\mathcal{O}(T_n^3)$. More details on the derivation of this equation can be withdrawn from the appendix-section \ref{AppendixGRad} related to the presented chapter. In Table \ref{B1534} we present some of the observed and derived parameters of the double neutron star binary system, PSR1534+12, which was discovered by the Arecibo Observatory in 1990 \cite{Wolszczan1}. We will see that this two-body system is quite similar to the famous Hulse-Taylor binary system (PSR B1913+16) in the sense that both systems have a rather short orbital period of the order of $P\sim 0.3-0.4$ day.
\begin{table}[h]
\begin{center}
\scalebox{0.69}{
\begin{tabular}{||c|c||}
\hline\hline
B1534+12&\quad \quad Pulsar \quad\quad\quad  Companion\\ \hline
Pulse Period P ($ms$)&\quad37.9\quad \ \hspace{2.2cm}- \\ \hline
Masses ($M_\odot$)&\quad\quad1.3332(10)\quad\quad\quad\quad 1.3452(10)\\ \hline
Projected semi-major axis $x = a\sin i/c$ (sec)&3.729464(2) \\ \hline
Shapiro delay parameter $s=\sin i$&0.975(7) \\ \hline
Orbital period $P$ ($day$)&0.420737299122(10)\\ \hline
Semi-major axis $a\ (m)$&$1.14673(823)\cdot 10^{9} $ \\ \hline
Eccentricity $e$&$0.2736775(3)$\\ \hline \hline
Gravitational constant $G$ ($\frac{m^3}{kg\ s^2}$)&$6.67408(31)\cdot10^{-11}$\\ \hline
Solar Mass $M_\odot$ ($m$) &$1.98855(25)\cdot 10^{30}$\\ \hline
Speed of light $c$ $(\frac{m}{s})$&299792458\\ \hline \hline
\end{tabular}}
\caption[Binary system: PSR B1534+12]{\label{B1534}In this table we present in the upper part some of the observed and derived parameters of the PSR B1534+12 binary system. In the lower part we further outline some fundamental constants and their respective standard errors. Standard errors are given in parentheses after the values and are in units of the least significant digit(s) \cite{Stairs1, Stairs2,Wolszczan1,Maggiore1}.}
\end{center}
\end{table}  
In Table \ref{HulseTaylorPulsar} we outline some of the observed and derived numerical values of the famous Hulse-Taylor binary system, which we used in order to work out the results presented in Figure \ref{PeriastroTimeShift}.
\begin{table}[h]
\begin{center}
\scalebox{0.69}{
\begin{tabular}{||c|c||}
\hline\hline
B1913+16&\quad \quad Pulsar \quad\quad\quad  Companion\\ \hline
Pulse Period P ($ms$)&\quad59.0\quad \ \hspace{2.2cm}- \\ \hline
Masses ($M_\odot$)&\quad\quad1.4414(2)\quad\quad\quad\quad 1.3867(2)\\ \hline
Projected semi-major axis $x = a\sin i/c$ (sec)&\hspace{0.3cm} 2.341774(1) \\ \hline
Orbital period $P$ ($day$)&0.322997462727(5)\\ \hline
Orbital period derivative $\dot{P}\ (10^{-12})$&$ -2.4211(14)$ \\ \hline
Eccentricity $e$&0.6171338(4)\\ \hline \hline
Gravitational constant $G$ ($\frac{m^3}{kg\ s^2}$)&$6.67408(31)\cdot10^{-11}$\\ \hline
Solar Mass $M_\odot$ ($m$) &$1.98855(25)\cdot 10^{30}$\\ \hline
Speed of light $c$ $(\frac{m}{s})$&299792458\\ \hline \hline
\end{tabular}}
\caption[Hulse-Taylor binary system: PSR B1913+16]{\label{HulseTaylorPulsar}In this table we present in the upper part some of the observed and derived parameters of the Hulse-Taylor binary system (B1913+16). In the lower part we further outline some fundamental constants and their respective standard errors. Standard errors are given in parentheses after the values and are in units of the least significant digit(s) \cite{Taylor1,Taylor2,Taylor3,Maggiore1}.}
\end{center}
\end{table}
R. A. Hulse, J. H. Taylor were awarded the Nobel Prize in 1993, {\it for the discovery of a new type of pulsar, a discovery that has opened up new possibilities for the study of gravitation} and the subsequent experimental study of the pulsar PSR B1913+16 in terms of the shift of periastron time led to the first indirect evidence for the existence of gravitational waves. In this context we illustrate in Figure \ref{PeriastroTimeShift}, with the use of the relation presented in equation \eqref{Periastronshift-18.04.17} and the result outlined in equation \eqref{HulseTaylor-21.04.17} the computed shift of periastron time for the Double Pulsar (PSR J0737-3039A/B) \cite{Burgay1,DoublePulsar}, the Hulse-Taylor binary system (PSR B1913+16) \cite{Taylor1,Taylor2,Taylor3} as well as for the double neutron star binary system (PSR B1534+12) \cite{Stairs1,Stairs2, Wolszczan1}. We observe that the cumulated shift of periastron time is strongest for the Double Pulsar (blue lower curves) and almost identical for the Hulse-Taylor binary system (red intermediate curves) and PSR B1534+12 (orange upper curves). The spreading of the three different curves results from the standard errors on the different parameters outlined in Table \ref{B1534} and Table \ref{HulseTaylorPulsar}. It should be mentioned that due to the detailed study of the Hulse-Taylor binary system over the last forty years the spreading of the curves is rather small and the measurements coincide pretty well with the theoretical predictions \cite{Taylor1,Taylor2,Taylor3, Stairs1, Maggiore1}. It should however be noticed that for the Double Pulsar system, with an orbital period of just $\sim2.4$ hours, relativistic effects are much more important than for the Hulse-Taylor binary system. In this regard PSR J0737-3039A/B is as a two-body system much more appropriate to discover deviations from the standard theory of gravity.
While Figure \ref{PeriastroTimeShift} and the related publications \cite{Taylor1,Taylor2,Taylor3, Stairs1} describe the indirect detection of gravitational waves, we aim to close the present chapter by reviewing some of the basic experimental techniques used in the context of the direct interferometric detection of gravitational radiation.
\begin{figure}[h]
\begin{center}
{\includegraphics[width=3.55cm,height=3.7cm]{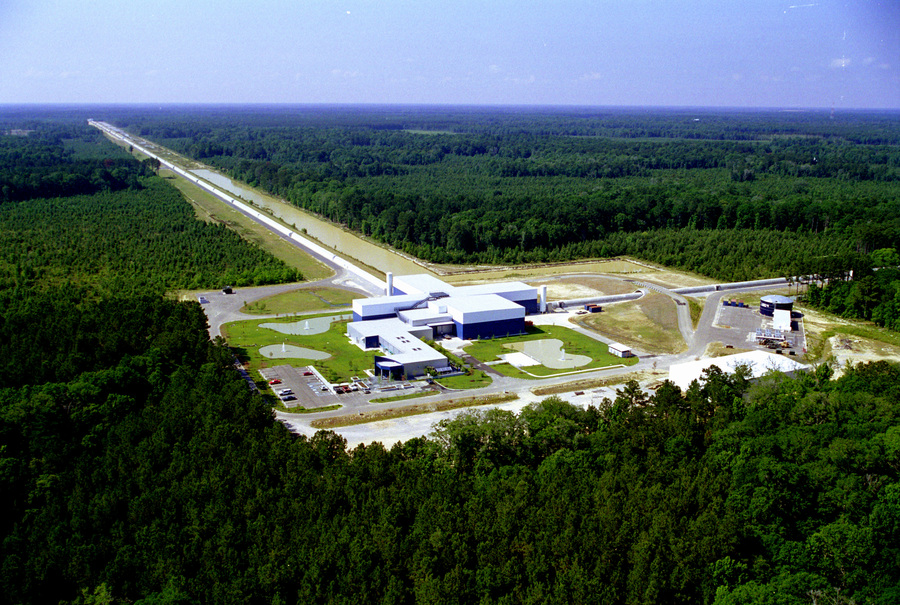}\quad\includegraphics[width=4.45cm,height=3.7cm]{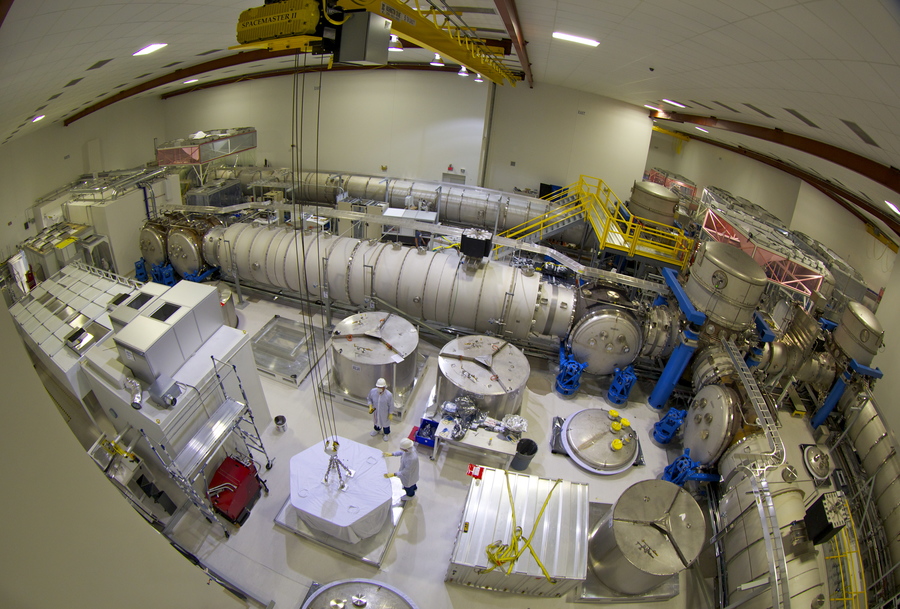}\quad\includegraphics[width=4.45cm,height=3.7cm]{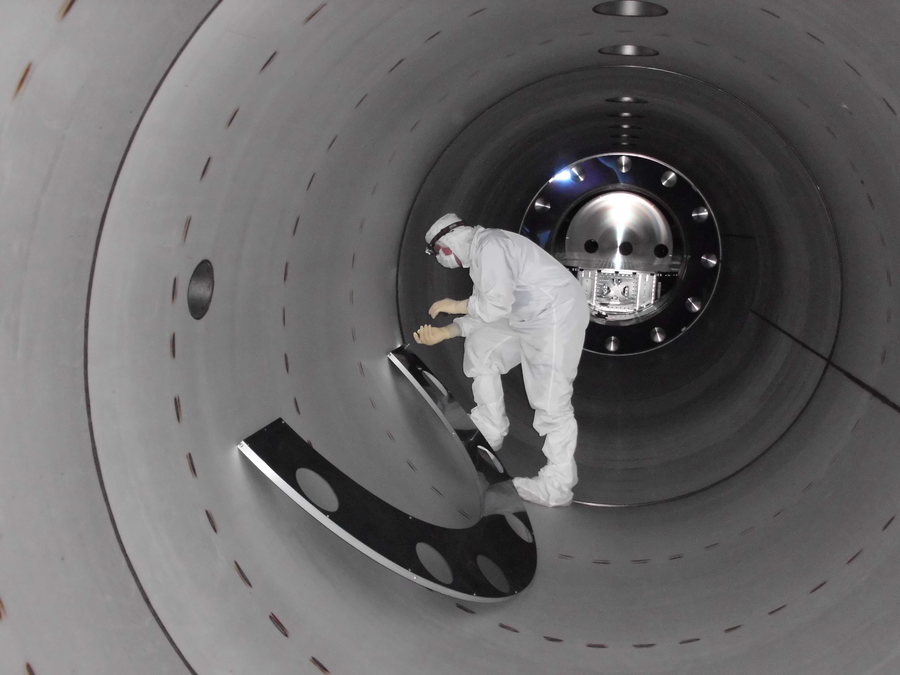}\quad \includegraphics[width=3.55cm,height=3.7cm]{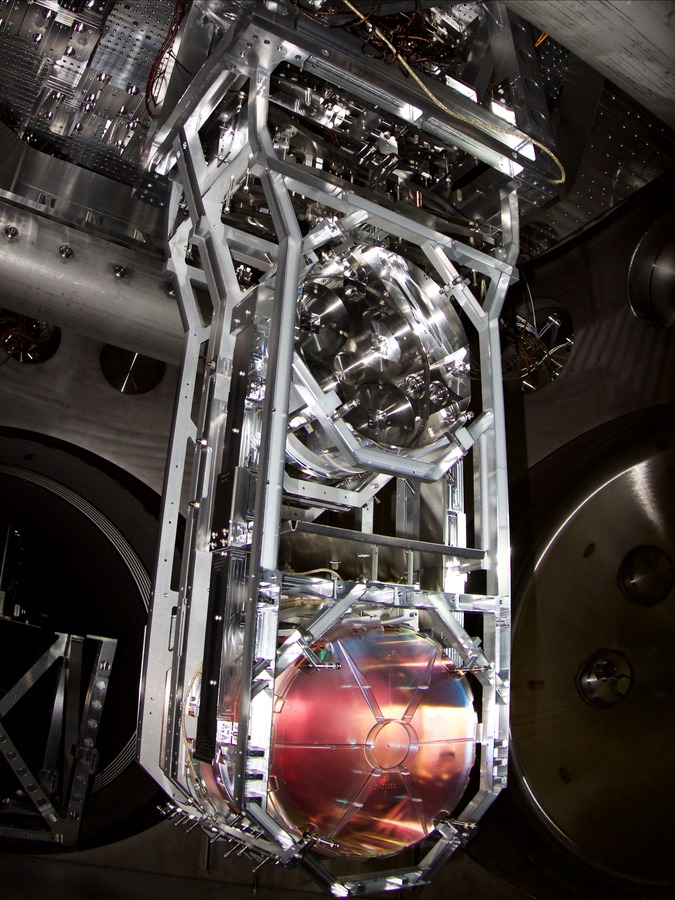}}
\end{center}
\caption[LIGO interferometer: 4 pictures]{\label{LIGOInterferometer}From left to right: The first picture shows the LIGO-detector located in Livingston (State of Louisiana). In the second picture we see the laser and vacuum equipment area at the corner station of the second LIGO-detector located in Hanford (Washington State). This area houses, among other things, the interferometer's beam splitter (Figure \ref{Interferometer-23.04.17}). The third photograph displays a LIGO technician installing a mode cleaner tube baffle used to control stray light. The last photo shows one of LIGO's test masses installed as the 4th element in a 4-element suspension system. These test masses, also called the mirrors, reflect the laser beams along the lengths of the detector arms. The 40 kg test mass is suspended below the metal mass above by 4 silica glass fibers (courtesy \href{http://www.ligo.org/}{LIGO Scientific Collaboration}).}
\end{figure}
We remind that precisely one century after Einstein's theoretical prediction, an international collaboration of scientists (LIGO Scientic Collaboration and Virgo Collaboration) reported the first direct observation of gravitational waves \cite{LIGO1,LIGO2}. The first direct wave signal GW150914 was detected independently by the two LIGO ground based interferometric detectors and its main features point to the coalescence of two stellar black holes. Three months later the same collaboration was able to perform a second direct gravitational wave measurement GW151226 allowing for an even better estimation of the stellar black hole population as well as for more robust constraints on possible general relativity deviations \cite{LIGO3}. Today there are several ground based interferometric detectors located around the world. The most famous are the two interferometers with arms of 4 km length situated in the United States of America. The two detectors, located in Hanford (Washington State) and Livingston (State of Louisiana), are physically separated by a rather large distance in order to have maximally uncorrelated noise sources.
\begin{figure}[h]
\begin{center}
{\includegraphics[width=0.224\textwidth]{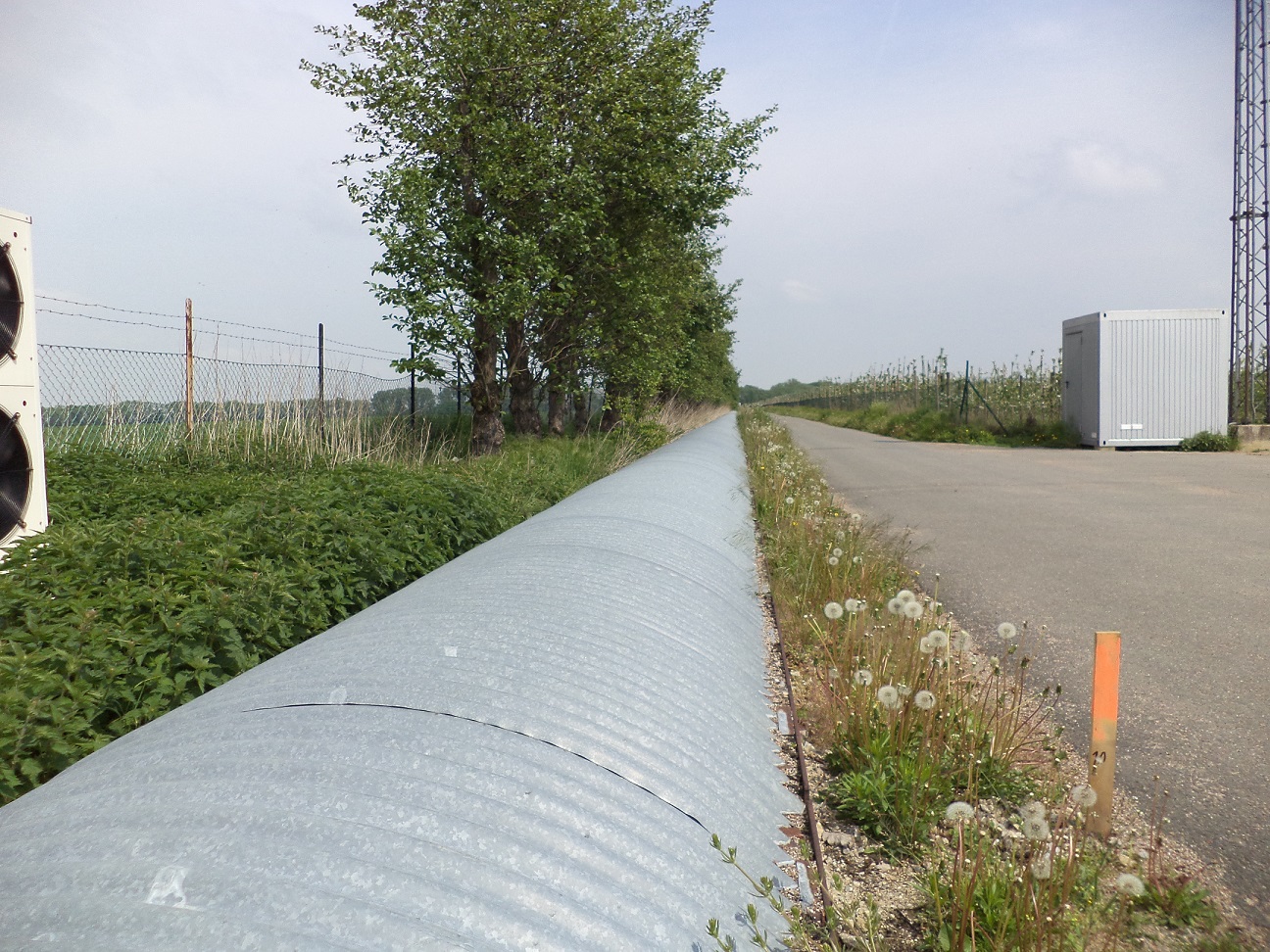} \quad \includegraphics[width=0.224\textwidth]{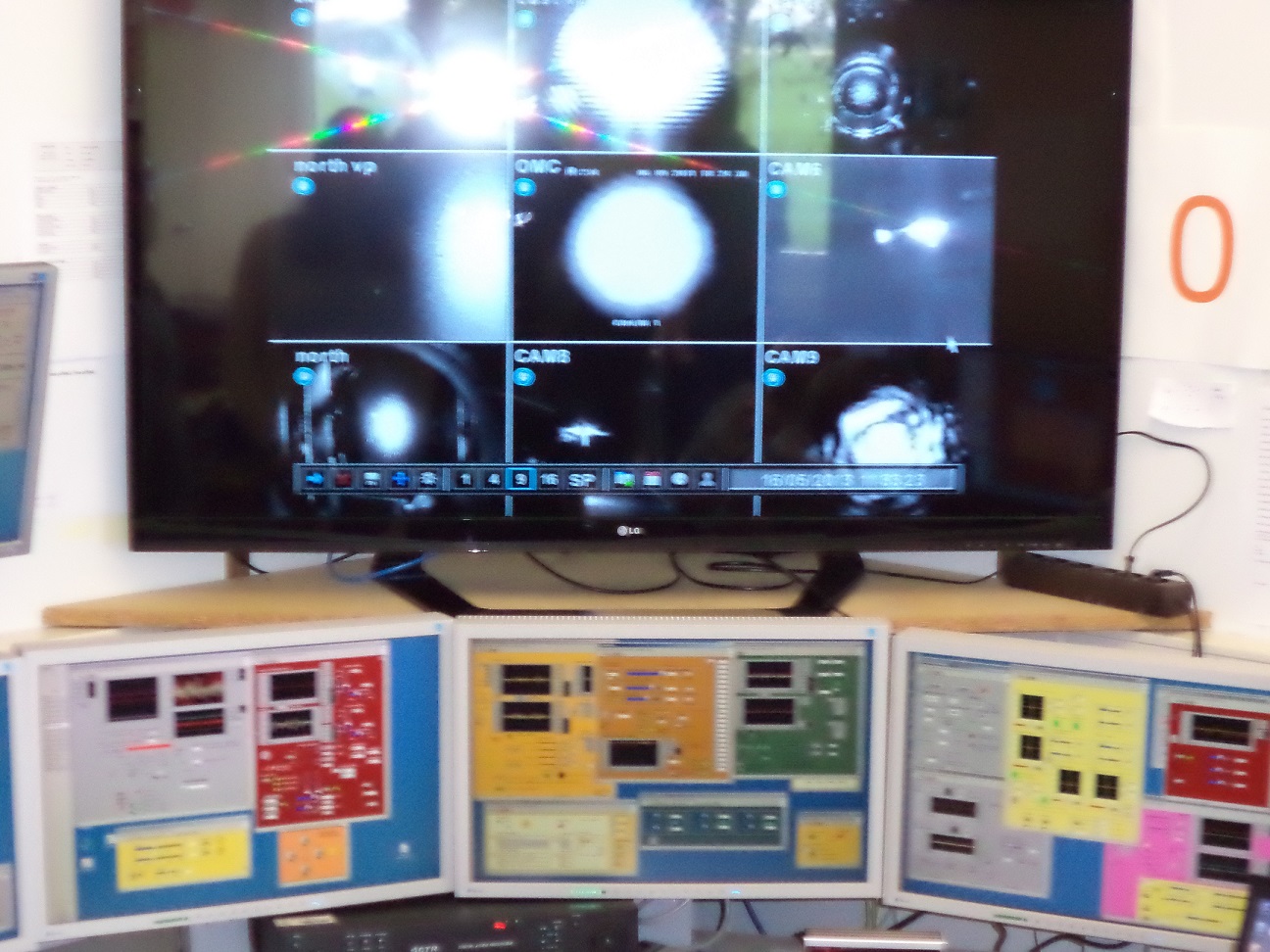}\quad \includegraphics[width=0.224\textwidth]{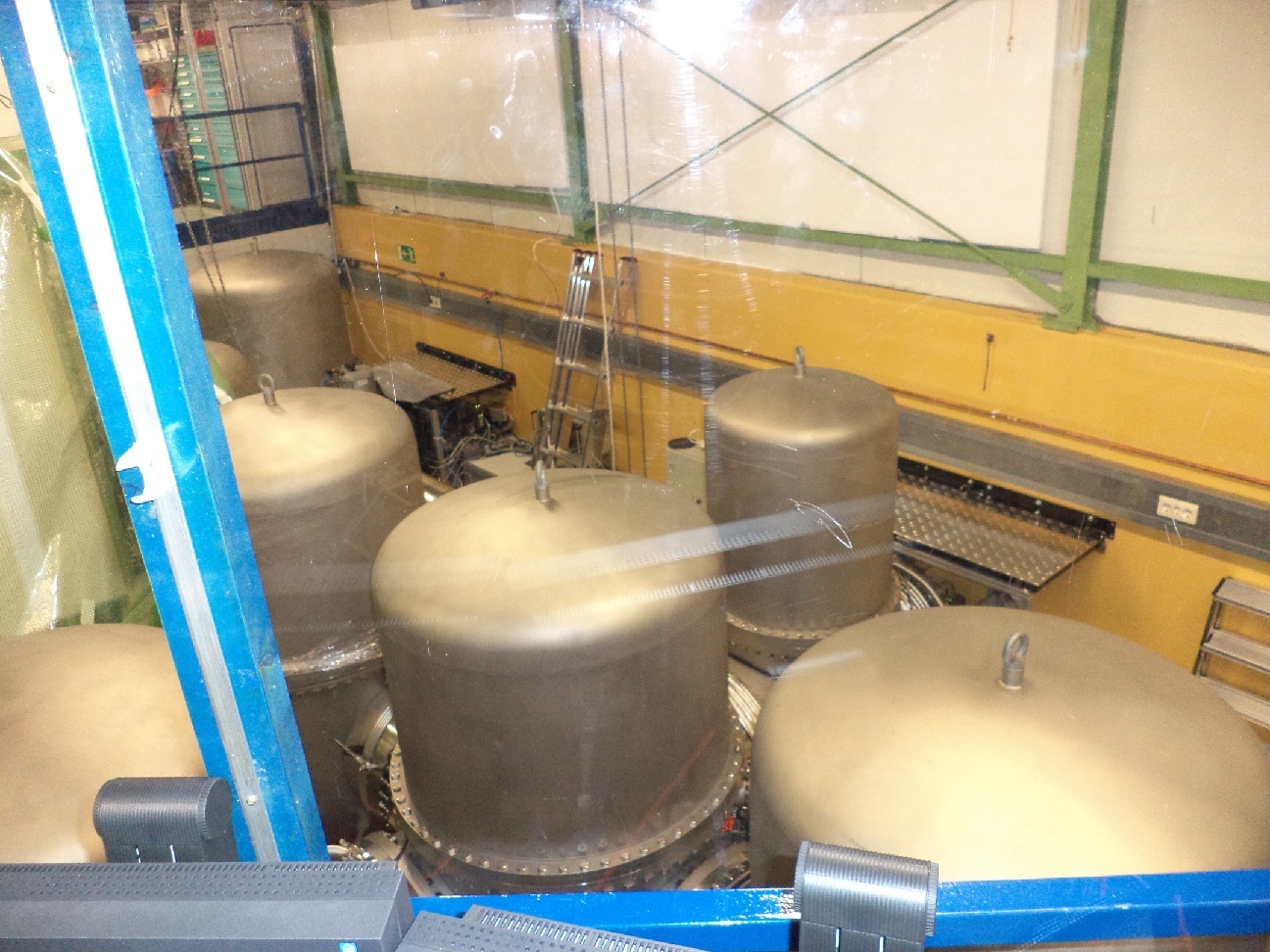}\quad\includegraphics[width=0.224\textwidth]{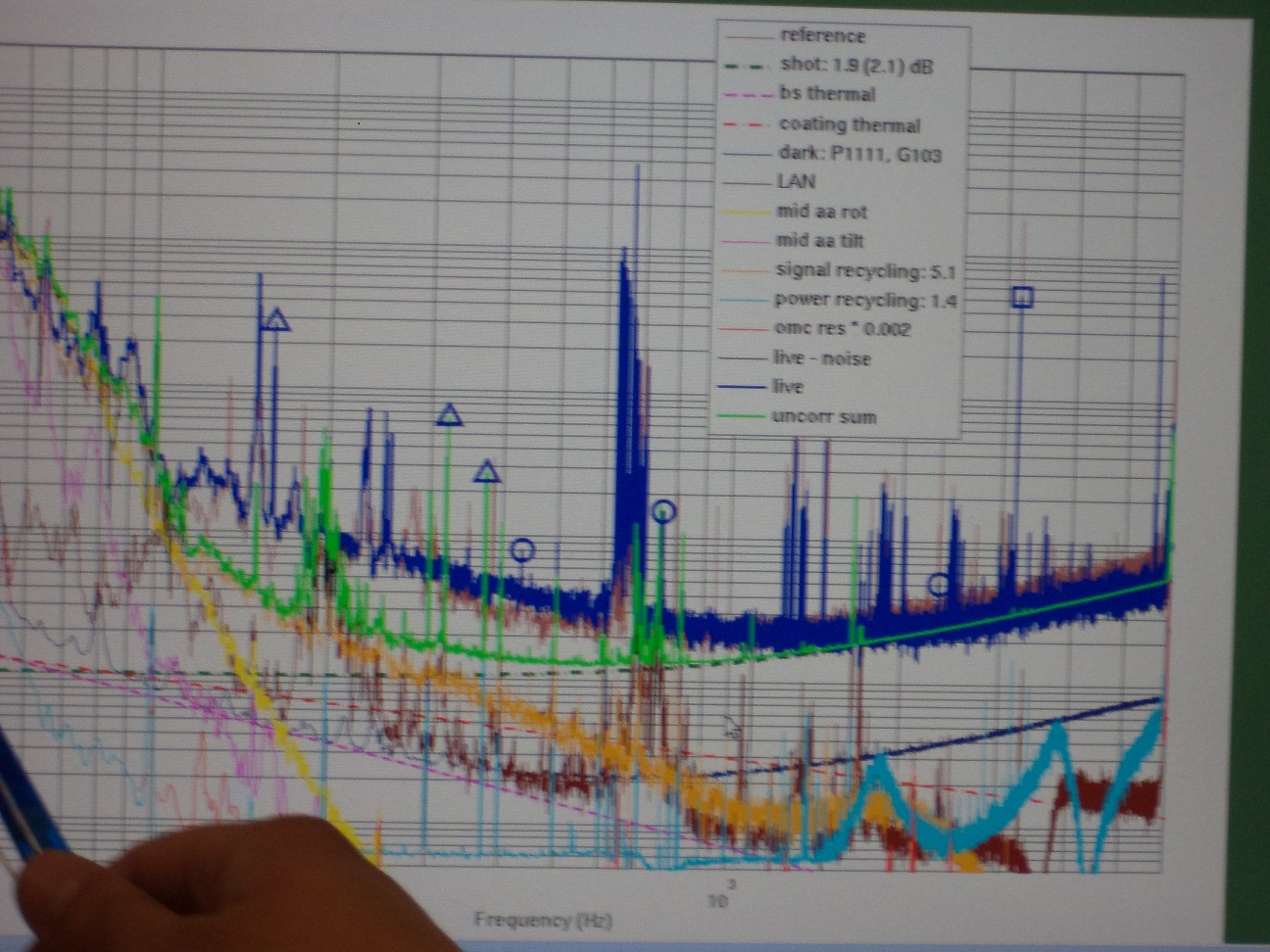}}
\end{center}
\caption[GEO 600 interferometer: 6 pictures]{\label{GEO600}These pictures show, from top-left to down-right, one of the two arms (600 meters) of the GEO600 laser-interferometer located in Hannover, Germany. An on-site control room equipped with monitors providing, among other things, informations about the laser beam, that propagates in the two perpendicular tubes. Moreover we observe in the third upper photgraph the vacuum chambers where the beam splitter is located (Figure \ref{Interferometer-23.04.17}) followed by a picture with a strain sensitivity [$1/\sqrt{Hz}$] - frequency plot providing information about the noise budget affecting the detector. The author would like to thank the \href{http://www.aei.mpg.de/165375/AEI_Hannover}{Albert-Einstein-Institut in Hannover}, together with the \href{http://www.aei.mpg.de/2162/de}{Max Planck Institute for Gravitational Physics} as well as the whole on site Max Planck team for the kind hospitality during the visit.}
\end{figure}
These interferometers, which are used to detect coincident signals, are run by the LIGO collaboration mentioned above in the context of the first direct detection of gravitational radiation. The VIRGO interferometer, which is a French-Italian joint project, is located near Pisa in Italy and has arms of 3 km length. Beside the large gravitational waves observatories there are two smaller ones located in Europe and Japan. The GEO600 facility, which is located near Hannover in Germany and has arms of 600 meters length, is a German-British project and works in close collaboration with the LIGO scientific community (Figure \ref{GEO600}). The TAMA detector is situated in Tokyo and posses arms of 300 meters length. These two smaller detectors are employed as back-up detectors and above all to develop and test new measurement techniques that can be used to upgrade the more advanced detectors (LIGO, VIRGO). Additional ground-based detectors are currently being built (KAGRA) or being developed (INDIGO) and are planned to be operational in the near future. In December 2015 the European Space Agency (ESA) launched the LISA Pathfinder spacecraft designed to prepare the scientific and technical grounds for a space-borne gravitational waves interferometer \cite{eLISA, ELISA2,LISA1}. Following the successful flight of LISA Pathfinder, the LISA Consortium now proposes a 4 year mission in response to ESA's call for missions for L3 \cite{LISA1,LISA2}. The observatory will be based on three arms with six active laser links, between three identical spacecraft in a triangular formation separated by 2.5 million km. It is supposed that this increase in sensitivity will allow to discern wave signals from millions of additional sources located in our Galaxy and beyond. Space-based interferometers are likely to have million kilometer long arms and thus be sensitive to the milli-Hz band $(\sim 10^{-3}$ Hz) \cite{YunesSiemens,Will1, eLISA,ELISA2,ELISA3,ELISA4}. Different types of interferometers are then sensitive to different types of gravitational-waves sources. In the context of binary coalescences, ground based interferometers are sensitive to late inspirals and mergers of neutron stars stellar massive black holes \cite{YunesSiemens,Will1,LIGO2,LIGO3}, while space based detectors will be sensitive to supermassive black-hole binaries with masses around $10^5 M_{\odot}$ \cite{eLISA,LISA1,ELISA2,ELISA3,ELISA4}. LISA is an all-sky monitor and will offer a wide view of a dynamic cosmos using gravitational waves as new and unique messengers to unveil the mysteries of the universe and gravity. It will be able to probe the entire Universe, from its smallest scales near the horizons of black holes, all the way to cosmological scales. In this regard the LISA mission will scan the entire sky as it follows behind the Earth in its orbit, obtaining both polarisations of the gravitational radiation simultaneously, and will measure source parameters with astrophysically relevant sensitivity in a band from below $10^{-4}$Hz to above $10^{-1}$Hz.
\vspace{0.2cm}

\begin{figure}[h]
\begin{center}
\includegraphics[width=10cm,height=5cm]{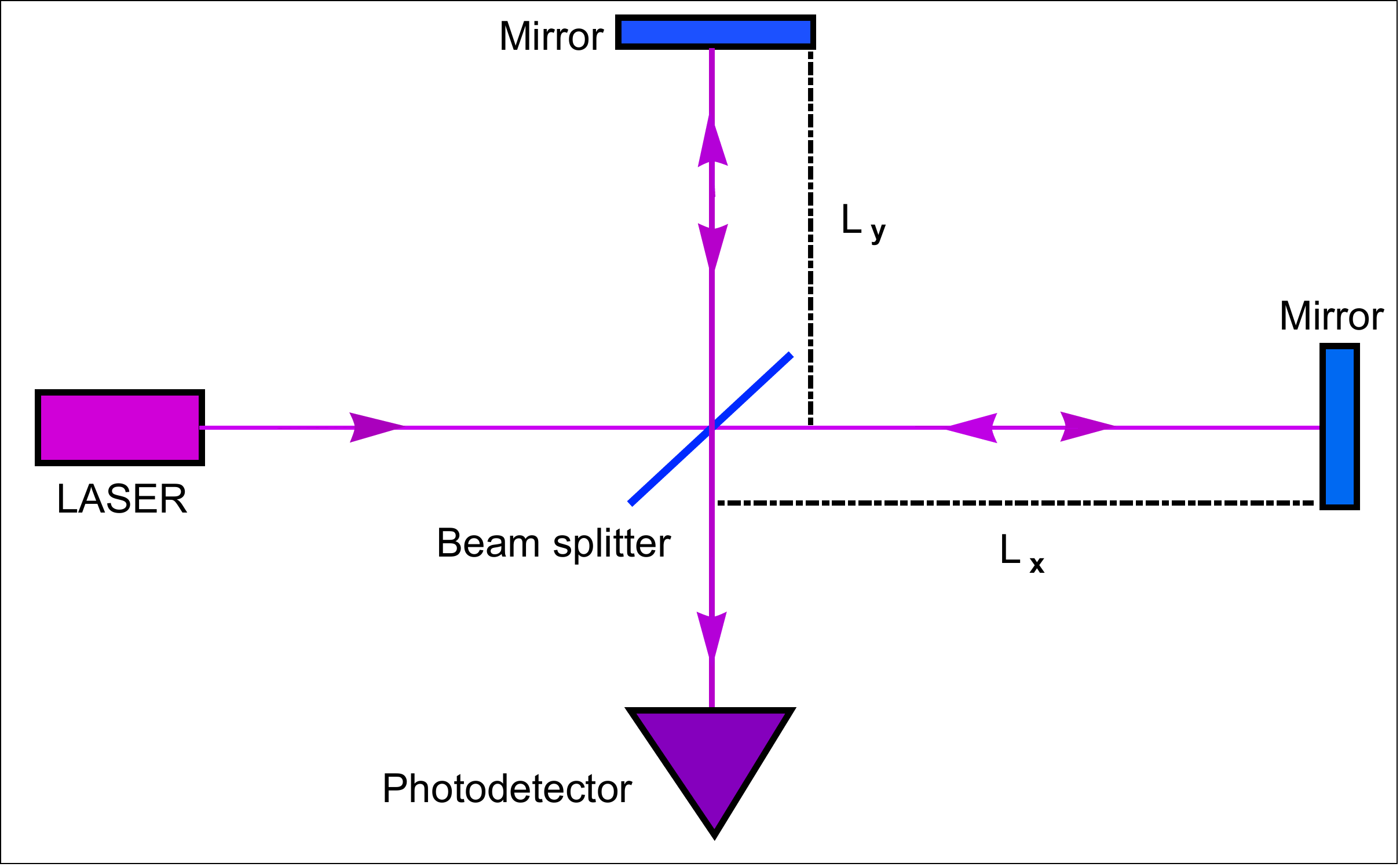}
\end{center}
\caption[Michelson-type interferometer]{\label{Interferometer-23.04.17}This figure illustrates the layout of a Michelson-type interferometer. The monochromatic light, emitted by a laser source, is sent to a beam-splitter which separates the light, with equal probability, into two beams travelling along the two arms $L_x$ and $L_y$. The ends of the two arms are equipped with two totally reflecting mirrors and after travelling once back and forth the two beams recombine at the beam-splitter where part of the light is conducted towards a photodetector.}
\end{figure}

\begin{figure}[h]
\begin{center}
\includegraphics[width=0.325\textwidth]{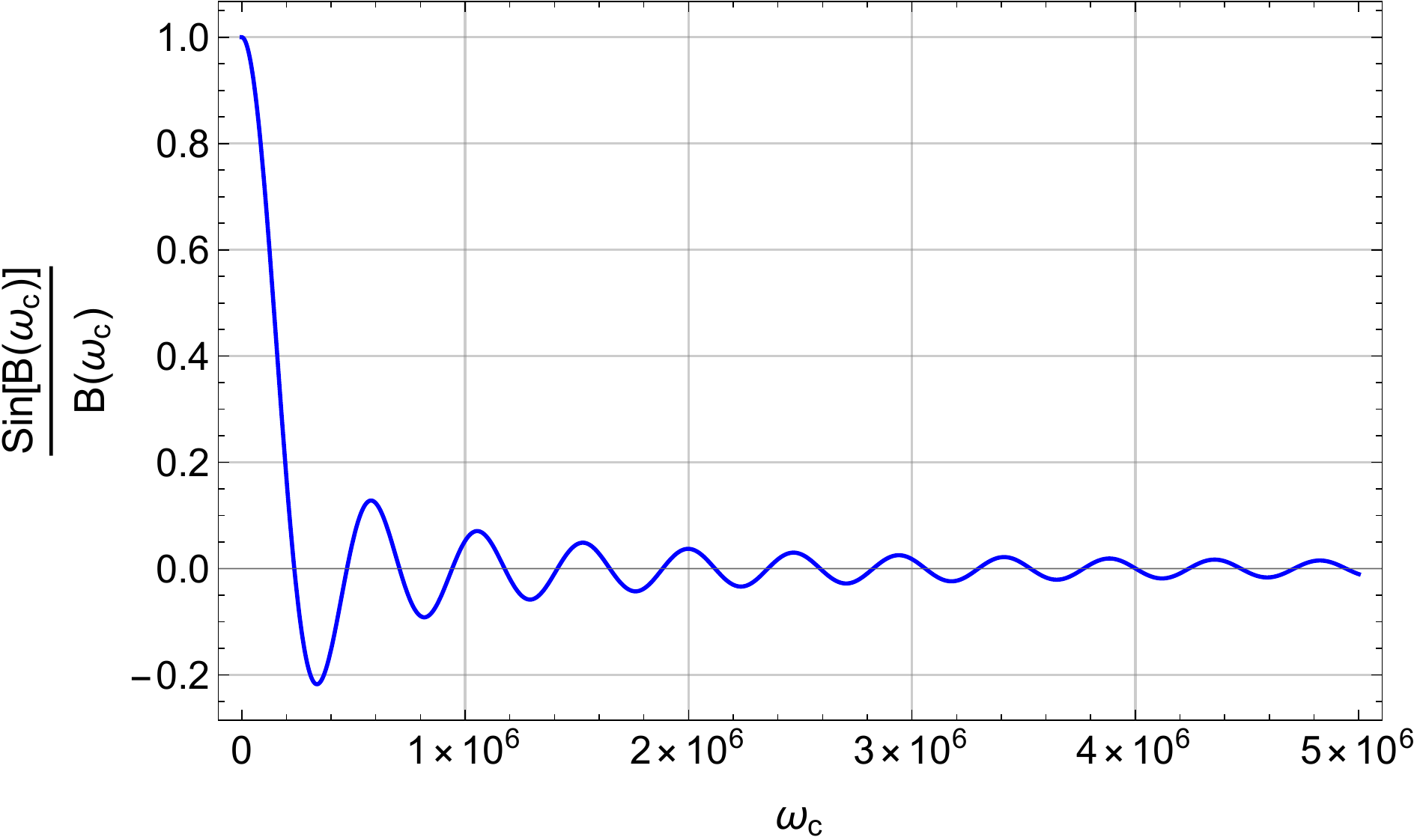}\quad \ \ \includegraphics[width=0.325\textwidth]{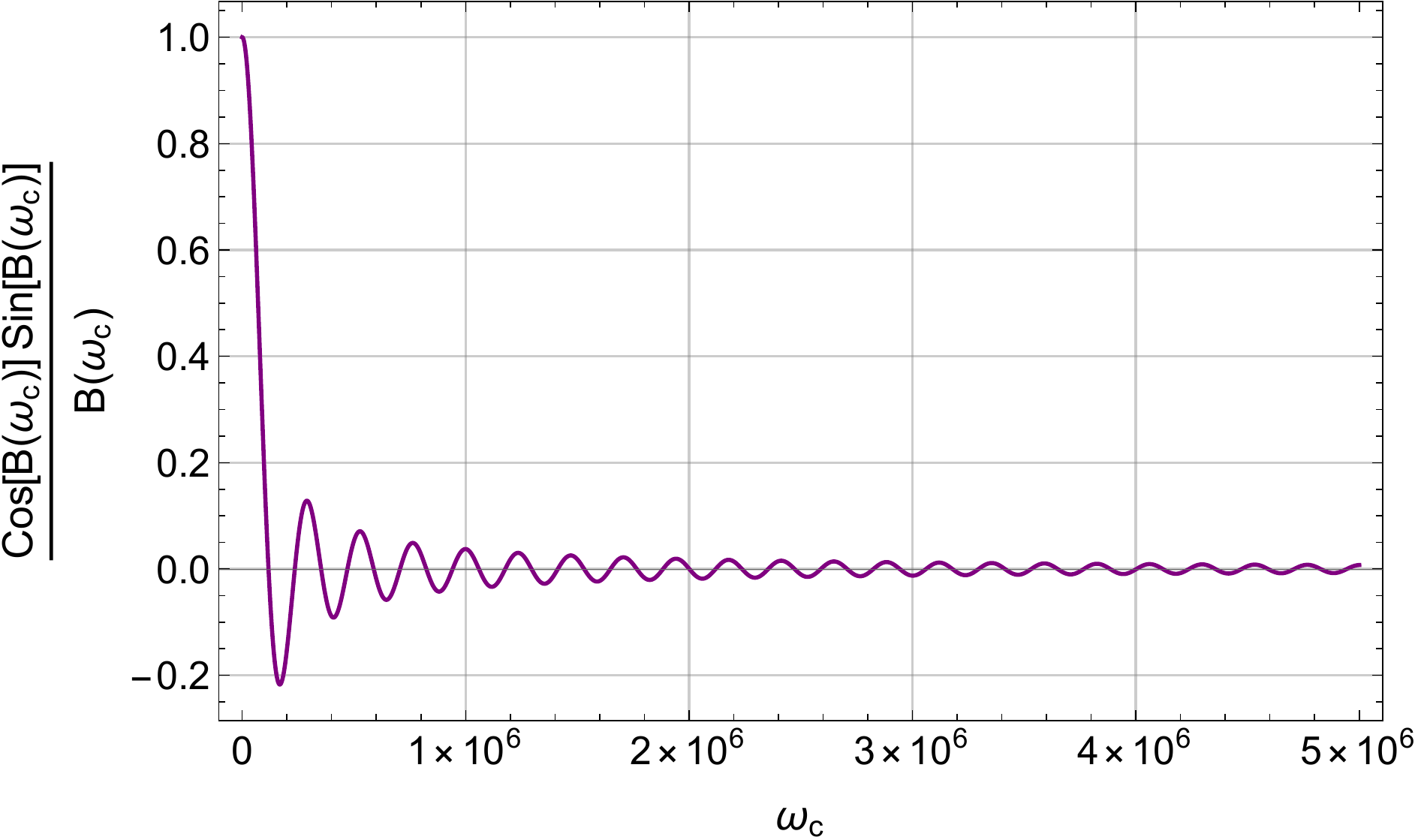}
\end{center}
\caption[]{\label{Sinc-04.05.17}The left figure illustrates the function $\frac{\sin[B(\omega_c)]}{B(\omega_c)}$, while the right figure displays the function $\frac{\cos[B(\omega_c)]\ \sin[B(\omega_c)]}{B(\omega_c)}$, where $B(\omega_c)=\frac{\omega_cL_x}{c}$ is a dimensionless quantity. Here $\omega_c$ is the characteristic frequency of the incoming gravitational radiation, $L_x$ is the length of the $x$ arm of the laser interferometer and $c$ is the speed of light. These functions appear in the time interval (equation \eqref{timex-25.04.17}) of a photon travelling from the beam-splitter to the mirror and back (Figure \ref{Interferometer-23.04.17}). The length of the $x$ arm is $L_x=4$ km and $\omega_c$ is measured in Hz. We observe that for low frequencies the effect due to the incoming gravitational waves is strongest.}
\end{figure}

In order to understand the fundamental functioning of these detectors we will discuss the well known Michelson interferometer, which is an extraordinarily accurate instrument to measure changes in the travel time of light in its arms. In Figure \ref{Interferometer-23.04.17} we present the layout of a simple Michelson-type interferometer. It consists of a monochromatic light source generated by a laser, whose light is sent to a beam-splitter. The latter separates the light, with equal probability amplitudes, into two beams travelling in the two arms with respective lengthes $L_x$ and $L_y$. The ends of the two arms are equipped with totally reflecting mirrors and after travelling once back and forth the two beams recombine at the beam-splitter. While part of the beam goes back towards the laser another part of the light is captured by a photodetector, that measures the beam's intensity. The monochromatic laser light is denoted $\omega_M$ and we obtain for the power of the two electric fields originating from the two beams travelling respectively in the $L_x$ and $L_y$ arms and recombining at some given time $t$ at the beam-splitter,
\begin{equation}
\label{Power1-24.04.17}
P_{out}\,=\,P_0\sin^2[k_M(L_x-L_y)].
\end{equation}
The amplitude $P_0\sim E_0^2$ is proportional to the square of the amplitude of the two electric fields ($E_1=-(E_0/2) e^{-i \omega_M t+2ik_ML_x}$and $E_2=+(E_0/2) e^{-i \omega_M t+2ik_ML_y}$) and $k_M=\omega_M/c$ is the wavenumber of the laser light. The phase of the electric field is conserved, while the field acquire overall factors from reflections and transmissions at the mirrors. Further computational details can be withdrawn from the appendix-section \ref{AppendixGRad} related to this section.
 We observe from equation \eqref{Power1-24.04.17} that any variation in the length of the two arms will be translated into a variation in the power at the photdetector. In what follows we will employ this general idea in the context of the detection of gravitational waves using separately the TT-frame and the proper detector frame, both introduced previously in this chapter. The origin of the coordinate system will be chosen in order to coincide with the position of the beam-splitter and the position of the two totally reflecting mirrors will have the coordinates $(L_x,0)$ and $(0,L_y)$ respectively. In the TT-gauge description the physical effect of the gravitational waves is manifested in the way that it affects the propagation of light between these two points \cite{Maggiore1, Buonanno1,Schutz1,Riles1}. Provided that we consider only the $+$-polarisation of a gravitational wave, travelling in the $z$-direction, we obtain for the spacetime interval in this gauge,
\begin{equation}
\label{intervalplus-24.04.17}
ds^2\,=\,-c^2dt^2+[1+h_+(t)]dx^2+[1-h_+(t)]dy^2+dz^2,
\end{equation}
where we remind that $h_+$ is the gravitational wave polarisation defined by equation \eqref{polla2-20.04.17}. Here we choose for reasons of simplicity and in analogy to equation \eqref{Polarisation-24.04.17} $P_+=\cos(\omega_ct)$, where $\omega_c$ is the characteristic frequency of the gravitational waves. In general, the lengths of the arms $L_x$ and $L_y$ will be chosen to be as close as possible ($L_x=L_y=L$) in order to cancel common noise in both of the arms. There are multiple noise sources like seismic noise originating from the Earth's ground continual motion or thermal noise which induce vibrations both in the mirrors and in the suspensions. We will briefly come back to this rather complex issue towards the end of this discussion by reviewing the so called shot noise which originates from the fact that the laser light comes in discrete quantas. In the particular context of equation \eqref{intervalplus-24.04.17} the total power observed at the photodetector (Figure \ref{Interferometer-23.04.17}) is modulated by the gravitational wave signal as,
\begin{equation}
P(t)\,=\,P_0/2\ \big(1-\cos[2\phi_0+\Delta\phi_I(t)]\big),
\end{equation}
where $\Delta\phi_I$ is the total phase difference in the Michelson interferometer induced by the gravitational radiation and $\phi_0=k_M(L_x-L_y)$ is a parameter that the experimenter can adjust, choosing the best working point for the interferometer \cite{Maggiore1}. Additional computational details are provided in the appendix-section \ref{AppendixGRad}. In the limit $\omega_cL/c\ll 1$ we immediately obtain for the phase shift, $\Delta\phi_I(t)\approx h(t-L/c)k_M L/2$ and this translates into a variation in the lengths of the arms, $\frac{2\Delta (L_x-L_y)}{L}\approx h(t-L/c)$, where $h(t-L/c)=A_o\cos[\omega_c(t-L/c)]$. The proper detector frame has the advantage of being very intuitive, since in a first approximation we can use the language of flat spacetime, $ds^2=-c^2dt^2+d\textbf{x}^2+\mathcal{O}(|\textbf{x}^2|/\mathcal{C}^2)$, where $|\textbf{x}|/\mathcal{C}\ll 1$ and $\mathcal{C}^{-2}\sim|R_{\mu\nu\rho,\sigma}|$ is the curvature radius. In the case where the observer does not experience rotations or accelerations one can derive an expression for the metric, that is accurate to second order in $x$,
\begin{equation}
\label{PDF-03.05.17}
ds^2\,=\,-c^2dt^2(1+R_{0a0b}x^ax^b)-c dtdx^a\Big(\frac{4}{3}R_{0bac}x^bx^c\Big)+dx^adx^b\Big(\delta_{ab}-\frac{1}{3}R_{acbd}x^cx^d\Big)+\mathcal{O}\Big(\frac{|\textbf{x}|^3}{\mathcal{C}^3}\Big),
\end{equation}
where we remind that $R_{abcd}$ is the Riemann curvature tensor \cite{Buonanno1,MisnerThroneWheeler,Maggiore1,Zimmermann1}. The even more accurate form of the line element with rotations and accelerations, for an Earthbound detector, is presented in equation \eqref{PDFRA-03.05.17} in the appendix-section \ref{AppendixGRad} related to this chapter. The interaction of the mirrors with incoming gravitational radiation is described by the equation of geodesic deviation outlined in equation \eqref{GeoForce-10.04.17}. If again we consider only the $+$-polarisation we obtain, $\ddot{\epsilon}_x=\ddot{h}_+\epsilon_x$ for the mirror on the $x$ arm, while $\epsilon_y(t)$ remains zero at all times provided that $\epsilon_y(t)|_{t=0}=\dot{\epsilon}_y(t)|_{t=0}=0$. It should be noticed, that although we are working here in the proper detector frame, the TT-gauge can be used because in linearised theory the Riemann tensor is invariant under coordinate transformations. Thus we are free to compute it in the frame that we wish and in particular we can use the form of $h_{\alpha\beta}$ in the TT-gauge. This equation can be solved perturbatively and we obtain to zeroth order in the amplitude, $\epsilon_x=L_x$. In this regard, for a careful choice of integration constants the solution to first order in $A_o$ becomes, $\epsilon_x(t)=L_x+(A_oL_x/2)\cos(\omega_ct)$. For a photon that starts at the beam-splitter at a time $t_0$ and travels along the positive $x$-axis, reaches the mirror and travels back towards the beam-splitter and attains it at a time $t_2$ we get,
\begin{equation}
\label{timesimple-03.05.17}
t_2-t_0\,=\,\frac{2L_x}{c}+\frac{A_oL_x}{c}\cos[\omega_c(t_0+L_x/c)].
\end{equation} 
Additional computational details are displayed in the appendix-section \ref{AppendixGRad} related to the present chapter. The result outlined in equation \ref{timesimple-03.05.17} should be compared to the outome presented in equation \eqref{timex-25.04.17}. We observe, that besides some minor notational details, the main difference is that in equation \ref{timesimple-03.05.17} the function $\sin(\omega_cL_x/c)/(\omega_cL_x/c)$ has been replaced by one. Sine one is the first term in a Taylor expansion in $\omega_cL_x/c$ of this function, it is interesting to work out the time interval $t_2-t_0$ to a higher order of accuracy by using the metric presented in equation \eqref{PDF-03.05.17}. For the propagation of a photon along a trajectory where $y=z=0$ and therefore with $dy=dz=0$, the line element reduces to,
\begin{equation}
ds^2\,=\,-c^2dt^2(1+R_{0101}x^2)+dx^2,
\end{equation}
where we used the Riemann tensor properties \cite{PoissonWill,Inverno,LandauLifshitz}. The position $x(t)$, at next-to-leading order, of a photon travelling along the $x$ arm is obtained by integrating the relation, $dx=\pm cdt[1+\omega_c^2/(4c^2)x^2(t)A_o\cos(\omega_ct)$. With this we obtain for a round-trip of the photon, starting at the beam-splitter to the mirror and back,
\begin{equation}
\label{timeadvanced-04.05.17}
t_2-t_0\,=\,\frac{2L_x}{c}+\frac{A_oL_x}{c}\cos[\omega_c(t_0+L_x/c)]\Big[1-\frac{1}{6}\ \Big(\frac{\omega_c L_x}{c}\Big)^2\Big],
\end{equation}
where we remind that $\omega_c$ is the characteristic frequency of the incoming gravitational radiation. It should be noticed that in the last pair of brackets in equation \eqref{timeadvanced-04.05.17} we recognize the first two terms of the Taylor expansion of the function $\frac{\sin(\omega_c L_x/c)}{\omega_cL_x/c}$ encountered in equation \eqref{timex-25.04.17} and illustrated in Figure \ref{Sinc-04.05.17}. This shows that the analysis performed in the proper detector frame consistently reproduces the leading and next-to-leading terms of the TT-gauge result \cite{Maggiore1}. Moreover we observe that, while the description in the proper detector frame is more intuitive, the TT-gauge description is more accurate in the sense that it allows us to derive the precise closed form of the dependence of $\omega_cL_x/c$. 
We will close this discussion by mentioning some of the main noise sources that affect the highly subtle interferometric measurements. It is rather obvious that that there are many different noise sources. One of them comes from the fact that the Earth's ground is in continual motion, with amplitudes of the order of a few microns. Human activities such as traffic or trains as well as local phenomena such as winds produce a noise background of the order of 1-10 Hz. In addition there is a micro-seismic background, which affects the gravitational wave detector mostly in the form of surface waves that shake the suspension mechanisms and eventually the laser light reflecting mirrors. This is the main reason why ground-based interferometers can hardly detect gravitational radiation below the $\sim 10$ Hz region \cite{Maggiore1}. Another important source is Newtonian noise, also known as gravity gradient noise, which is due to mass density fluctuations induced by micro-seismic noise. These fluctuations produce inhomogeneities of the gravitational field of the Earth, which couple directly to the test masses (mirrors). Thermal noise, which is another noise source, induces vibrations both in the suspensions and in the mirrors. A beam of photons that impinges on a mirror and is reflected exerts a pressure on the mirror itself. If the radiation pressure were constant, it could simply be compensated by the mechanism that holds the mirrors in place. Unfortunately the number of photons arriving on the mirror fluctuates and therefore the radiation pressure fluctuates too generating in this way a stochastic force that shakes the mirrors. The sensitivity at which a gravitational wave detector must function in order to have good chances of detection is extremely ambitious. We saw in Table \ref{Amplitudes-20.04.17} that the amplitude of a gravitational wave signal produced by a stellar black hole binary system, located at a distance of approximately $1.8\times 10^{16}$ km (distance Earth-Double Pulsar) 
is of the order $A_o\sim 10^{-19}$. Using the relation mentioned previously, we find, for the displacement of the mirror of the interferometer,
\begin{equation}
\label{displacement-03.05.17}
\Delta L\sim 10^{-15} \text{m},
\end{equation}
where the length of the arms is set to $L=4$ km. The variation in the length of the arms produced by the gravitational radiation emitted by this particular binary system is of the order of the diameter of a free proton \cite{Maggiore1,Buonanno1,Schutz1,Riles1}. In the context of the Double Pulsar, which is less massive than a stellar binary black hole system, the amplitude (Table \ref{Amplitudes-20.04.17}) and the resulting mirror-displacement are even three orders of magnitude smaller than the displacement outlined in equation \eqref{displacement-03.05.17}.

\section{Conclusion}
\label{Conclusion}
In this review manuscript we outlined the conceptual foundations of gravitational radiation emission. Following Albert Einstein's first attempts we derived the linearised Einstein field equations and worked out the corresponding wave equation. Furthermore we determined the gravitational potentials in the far away wave zone using the relaxed Einstein equations \cite{PoissonWill,PatiWill1,WillWiseman}. We closed this review on gravitational waves by discussing the radiative losses of gravitating systems and finally took a look on the outstanding experimental achievements performed in the context of laser interferometer gravitational waves detectors. However the final answer of whether Einstein's theory of gravity constitutes the ultimate truth or whether it has to be extended or even replaced by some totally different approach, will definitely be provided by space technology that has, at least to a large extend, not even been developed yet. We previously saw that only gravitational waves sample the fully nonlinear and highly dynamical part of gravity and in this regard they definitely are the most promising observational signals for constraining gravitational theories in a regime where Einstein's field equations have not been validated so far. In fact the astrophysical sources generating the gravitational radiation can be such that the two dimensionless parameters, the compactness parameter $C=\frac{G m_c}{c^2 r_c}$ and the velocity parameter $\mathcal{V}=\frac{v_c}{c}$, approach both a value close to order unity. We remind that the compactness parameter for the Earth-Sun binary system and its corresponding characteristic velocity are respectively, $C_{ES}\sim 10^{-8}$ and $\mathcal{V}_{ES}\sim 10^{-4}$. We saw that the most relativistic binary pulsar system known to date, the Double Pulsar (PSRJ0737-3039A/B), precisely studied by radio astronomy observations, posses values of the order, $C_{DP}\sim 4\cdot 10^{-4}$ and $\mathcal{V}_{DP}\sim 2\cdot 10^{-2}$ \cite{YunesSiemens,Stairs1,Burgay1,DoublePulsar}. The gravitational radiation, emitted by such massive sources ($m_{SMBH}\sim 10^{6}\cdot M_\odot$) close to the inspiral phase, would eventually allow us to verify Einstein's theory to an order of precision that has never be obtained before. As we have already hinted previously this will require space technologies that will permit to detect a proper wave signal (little noise) together with a solid observation of the radiation emitting source itself by some kind of technologically advanced satellites. Only the two ingredients together will eventually permit us to deceiver the true nature of gravity \cite{YunesSiemens}. We recall that there is a considerable difference between ground-based and space-based detectors due to the different frequency ranges these detectors are sensitive to. Over the past years major scientific and financial efforts have been made on a global scale in order to directly measure the gravitational radiation. Indeed, many highly sensitive ground-based detectors (LIGO, VIRGO, GEO600, TAMA300, KAGRA), equipped with cutting-edge technology, are currently being operated with the aim to disclose perturbations in the gravitational field produced by far distant and violent astrophysical processes \cite{LIGO0,LIGO1, LIGO2,LIGO3,LIGO4, LIGO5, Virgo1, GEO1, Tama1, KAGRA1}. Ground based-interferometers are bound to the surface and curvature of the Earth and thus they have kilometer long arms and are sensitive to the deca Hz and the hecta Hz band $(\sim$ 10-100 Hz). In the quest of increasing the sensitivity we remind that in December 2015 the European Space Agency (ESA) launched the LISA (Laser Interferometer Space Antenna) Pathfinder spacecraft designed to prepare the scientific and technical grounds for a space-borne gravitational waves interferometer \cite{eLISA,LISA1}. It is supposed that this increase in sensitivity will allow to discern wave signals from millions of additional sources located in our Galaxy and beyond. Space-based interferometers are likely to have million kilometer long arms and thus be sensitive to the milli-Hz band $(\sim 10^{-3}$ Hz) \cite{YunesSiemens,Will1, eLISA,ELISA2,ELISA3,ELISA4}. Following the successful flight of the LISA Pathfinder project, the LISA Consortium now proposes a 4 year mission in response to ESA's call for missions for L3 \cite{LISA2}. The observatory will be based on three arms equipped with six active laser links, between three identical spacecrafts in a triangular formation separated by 2.5 million kilometers. In the context of binary coalescences, ground based interferometers are sensitive to late inspirals and mergers of neutron stars stellar massive black holes \cite{YunesSiemens,Will1,LIGO2,LIGO3}, while space based detectors will be sensitive to supermassive black-hole binaries with masses around $10^5 M_{\odot}$ \cite{eLISA,LISA1,ELISA2,ELISA3,ELISA4}. Bianary system inspirals are made of a stellar-mass compact object in a generic decaying orbit around a supermassive black hole. These inspirals produce millions of cycles of gravitational waves in the sensitivity band of space-based detectors \cite{Inspiral1,Inspiral2,Inspiral3,Inspiral4}. LISA is an all-sky monitor and will offer a wide view of a dynamic cosmos using gravitational waves as new and unique messengers to unveil the mysteries of the universe and gravity. In this regard it will be able to probe the entire universe, from its smallest scales near the horizons of black holes, all the way down to cosmological scales. Thus the LISA mission will be able to scan the entire sky as it follows behind the Earth in its orbit, obtaining both polarisations of the gravitational radiation simultaneously, and will measure source parameters with astrophysically relevant sensitivity in a band from below $10^{-4}$Hz to above $10^{-1}$Hz. Meanwhile earth bound measurements, like those obtained by radio astronomy surveys \cite{Taylor1,Burgay1,Stairs1,Stairs2,DoublePulsar,Wex1} and especially the recent detection of the gravitational wave signals GW150914, GW151226, GW170104, GW170814 constitute a strong additional step towards the final truth \cite{LIGO1,LIGO2,LIGO3,LIGO4,LIGO5,LIGO6,LIGO7}. It will however be the simultaneous interplay of all the different experimental devices as well as the complementarity between analytic post-Newtonian calculations and numerical relativity simulations that will eventually allow for the disclosure of the remaining astrophysical and cosmological mysteries intimately related to the concept of gravity itself.

\begin{appendix}
\section{Gravitational Radiation}
\label{AppendixGRad}
In this appendix-subsection we provide some additional technical derivation details regarding the results presented in the main part of this review article.
\subsubsection{The linearised Einstein field equations}
By using the decomposition of the metric into a Minkowski metric contribution plus a small perturbation inside the general transformation behaviour of the metric, outlined in equation \eqref{gtrafo-06.04.17}, we obtain,
\begin{equation}
\eta_{\alpha\beta}+d'_{\alpha\beta}\,=\,(1-\partial_\alpha t^\rho)\ (1-\partial_\beta t^\sigma)\ (\eta_{\rho\sigma}+d_{\rho\sigma})\,=\,\eta_{\rho\sigma}-\partial_\sigma t_\rho-\partial_\rho t_\sigma+d_{\rho\sigma}+\mathcal{O}((\partial t)^2, d\partial t),
\end{equation}
where we recall that the derivatives of the spacetime translation components are at most of the same order of smallness as the gravitational potentials, $|\partial_\alpha t_\beta|\leq |d_{\alpha\beta}|$. We obtain for the Lorentz transformation of the linearised metric,
\begin{equation}
g'_{\alpha\beta}(x')\,=\,\Lambda^{\ \rho}_{\alpha} \Lambda^{\ \sigma}_{\beta} g_{\rho\sigma}(x)\,=\,\Lambda^{\ \rho}_{\alpha} \Lambda^{\ \sigma}_{\beta}(\eta_{\rho\sigma}+d_{\rho\sigma}(x))\,=\,\eta_{\alpha\beta}+\Lambda^{\ \rho}_{\alpha} \Lambda^{\ \sigma}_{\beta} d_{\rho\sigma}(x)\,=\,\eta_{\alpha\beta}+d'_{\alpha\beta}(x),
\end{equation}
where we remind that $d'(x')=\Lambda^{\ \rho}_\alpha \Lambda^{\ \sigma}_\beta d_{\rho\sigma}(x)$ is a tensor under Lorentz transformations \cite{LandauLifshitz,MisnerThroneWheeler,PoissonWill,Maggiore1,Schutz0A}. It is a lengthy but straightforward exercise to linearise the Christoffel symbols and we obtain,
\begin{equation}
\begin{split}
\Gamma^\rho_{\mu\nu}\,=\,\frac{1}{2}g^{\rho\lambda} \big(\partial_\mu g_{\nu\lambda}+\partial_\nu g_{\lambda\mu}-\partial_\lambda g_{\mu\nu}\big)\,
=&\,\frac{1}{2}\big(\eta^{\rho\lambda}+d^{\rho\lambda}\big)\ \big(\partial_\mu d_{\nu\lambda}+\partial_\nu d_{\lambda\mu}-\partial_\lambda d_{\mu\nu}\big)+\mathcal{O}(d^2)\\
=&\, \frac{1}{2}\eta^{\rho\lambda}\big(\partial_\mu d_{\nu\lambda}+\partial_\nu d_{\lambda\mu}-\partial_\lambda d_{\mu\nu}\big)+\mathcal{O}(d^2)
\end{split}
\end{equation}
By using this result we can work out the linearised Riemann tensor,
\begin{equation}
\begin{split}
R^\mu_{\ \beta \rho \sigma}\,=&\,\partial_\rho \Gamma^\mu_{\sigma\beta}-\partial_\sigma \Gamma^\mu_{\rho\beta}+\Gamma^\mu_{\lambda\rho}\Gamma^\lambda_{\beta\sigma}-\Gamma^\mu_{\lambda\sigma}\Gamma^\lambda_{\beta\rho}\\
=&\,\frac{1}{2}\eta^{\mu\lambda}\big(\partial_\rho\partial_\beta d_{\lambda\sigma}+\partial_\sigma \partial_\lambda d_{\rho\beta}-\partial_\rho \partial_\lambda d_{\sigma \beta}-\partial_\sigma \partial_\beta d_{\lambda\rho}\big)+\mathcal{O}(d^2).
\end{split}
\end{equation}
Using the Minkowski metric we can reformulate the linearised Riemann tensor in order to obtain,
\begin{equation}
\begin{split}
R_{\alpha\beta\rho\sigma}\,=\,\eta_{\alpha\mu}R^\mu_{\ \beta\rho\sigma}\,=&\,\frac{1}{2}\delta^\lambda_{\ \alpha}\big(\partial_\rho\partial_\beta d_{\lambda\sigma}+\partial_\sigma \partial_\lambda d_{\rho\beta}-\partial_\rho \partial_\lambda d_{\sigma\beta}-\partial_\sigma \partial_\beta d_{\lambda\rho}\big)+\mathcal{O}(d^2)\\
=&\, \frac{1}{2}\big(\partial_\rho \partial_\beta d_{\alpha\sigma}+\partial_\sigma \partial_\alpha d_{\rho\beta}-\partial_\rho \partial_\alpha d_{\sigma\beta}-\partial_\sigma \partial_\beta d_{\alpha\rho}\big)+\mathcal{O}(d^2).
\end{split}
\end{equation}
With this we can work out the linearised Ricci tensor and we obtain,
\begin{equation}
\begin{split}
R_{\alpha\beta}\,=\,R^\rho_{\ \alpha\rho\beta}\,=\,\eta^{\rho\mu} R_{\mu\alpha\rho\beta}\,=&\,\eta^{\rho\mu}\frac{1}{2}\big(\partial_\rho\partial_\alpha d_{\mu\beta}+\partial_\beta \partial_\mu d_{\rho\alpha}-\partial_\rho\partial_\mu d_{\alpha\beta}-\partial_\alpha\partial_\beta d_{\mu\rho}\big)+\mathcal{O}(d^2)\\
=&\, \frac{1}{2}\big(\partial_\rho\partial_\alpha d^\rho_{\ \beta}+\partial_\beta\partial_\mu d^\mu_{\ \alpha}-\Box d_{\alpha\beta}-\partial_\alpha \partial_\beta d\big)+\mathcal{O}(d^2),
\end{split}
\end{equation}
where we recall that $\Box=\partial_\rho\partial^\rho$ and $d=d^{\rho}_{\ \rho}$. The Ricci scalar is obtained from the contraction of the Minkowski metric with the Ricci tensor and we get,
\begin{equation}
\begin{split}
R\,=\,\eta^{\alpha\beta}R_{\alpha\beta}\,=&\,\frac{1}{2}\eta^{\alpha\beta}\big(\partial_\rho\partial_\alpha d^\rho_{\ \beta}+\partial_\beta \partial_\mu d^\mu_{\ \alpha}-\Box d_{\alpha\beta}-\partial_\alpha\partial_\beta d\big)+\mathcal{O}(d^2)\\
=&\,\frac{1}{2}\big(\partial_\rho \partial_\alpha d^{\rho\alpha}+\partial_\beta\partial_\mu d^{\mu\beta}-2\Box d\big)+\mathcal{O}(d^2)\\
=&\,\partial_\mu\partial_\nu d^{\mu\nu}-\Box d+\mathcal{O}(d^2)
\end{split}
\end{equation}
The Einstein tensor was introduced in equation \eqref{EFE-24.03.17} and we can gather the results worked out previously in order to determine the linearised version of it,
\begin{equation}
\begin{split}
G_{\alpha\beta}\,=&\,R_{\alpha\beta}-\frac{1}{2}Rg_{\alpha\beta}\\
=&\,\frac{1}{2}\big(\partial_\rho \partial_\alpha d^\rho_{\ \beta}+\partial_\beta \partial_\mu d^\mu_{\ \alpha} -\Box d_{\alpha\beta}-\partial_\alpha\partial_\beta d\big)-\frac{1}{2}\big(\partial_\mu \partial_\nu d^{\mu\nu}-\Box d\big)\eta_{\alpha\beta}+\mathcal{O}(d^2)
\end{split}
\end{equation}
We mentioned in the main part of this manuscript that the linearised Einstein tensor $G^{(1)}_{\alpha\beta}$ is invariant under the transformations outlined in equation \eqref{trafocoordinates-07.04.17} and equation \eqref{trafodeviation-07.04.17}. We have for the linearised Ricci-tensor,
\begin{equation}
\begin{split}
R^{(1)}_{\alpha\beta}\rightarrow R'^{(1)}_{\alpha\beta}\,=&\,\frac{1}{2}\big\{\partial_\rho\partial_\alpha d'^\rho_{\ \beta}+\partial_\beta \partial_\mu d'^\mu_{\ \alpha}-\Box d'_{\alpha\beta}-\partial_\alpha\partial_\beta d'\big\}\\
=&\,\frac{1}{2}\big\{\partial_\rho \partial_\alpha d^\rho_{\ \beta}-\partial_\rho\partial_\alpha \partial^\rho t_\beta-\partial_\rho\partial_\alpha\partial_\beta t^\rho+\partial_\beta\partial_\mu d^\mu_{\ \alpha}-\partial_\beta \partial_\mu \partial^\mu t_\alpha-\partial_\beta\partial_\mu \partial_\alpha t^\mu\\
&\,\quad-\Box d_{\alpha\beta}+\partial_\alpha \Box t_\beta+\partial_\beta \Box t_\alpha-\partial_\alpha \partial_\beta d+\partial_\alpha \partial_\beta \partial^\nu t_\nu+\partial_\alpha\partial_\beta\partial^\mu t_\mu\big\}\\
=&\,\frac{1}{2}\big\{\partial_\rho\partial_\alpha d^\rho_{\ \beta}+\partial_\beta \partial_\mu d^\mu_{\ \alpha}-\Box d_{\alpha\beta}-\partial_\alpha\partial_\beta d\big\}\,=R^{(1)}_{\alpha\beta},
\end{split}
\end{equation}
where $R_{\alpha\beta}=R^{(1)}_{\alpha\beta}+\mathcal{O}(d^2)$. The linearised Ricci scalar is of course also invariant under the transformations mentioned in equation \eqref{trafocoordinates-07.04.17} and equation \eqref{trafodeviation-07.04.17} and we have,
\begin{equation}
\begin{split}
R^{(1)}\rightarrow R'^{(1)}\,=&\,\partial_\mu \partial_\nu d'^{\mu\nu}-\Box d'\\
=&\,\partial_\mu\partial_\nu d^{\mu\nu} -\Box \partial_\nu t^\nu-\Box \partial_\mu t^\mu-\Box h+\Box \partial_\rho t^\rho+\Box \partial_\sigma t^\sigma\,=\,R^{(1)},
\end{split}
\end{equation}
where $R=R^{(1)}+\mathcal{O}(d^2)$. It is possible to simplify the Einstein tensor even further as we saw in equation \eqref{SWave-06.04.17} in the main part of this review,
\begin{equation}
\begin{split}
G_{\alpha\beta}\,=&\,\frac{1}{2}\big[\partial^\rho\partial_\alpha\big(\bar{d}_{\rho\beta}-\frac{1}{2}\eta_{\rho\beta}\bar{d}\big)+\partial_\beta\partial^\mu \big(\bar{d}_{\mu\alpha}-\frac{1}{2}\eta_{\mu\alpha}\bar{d}\big)-\Box\big(\bar{d}_{\alpha\beta}-\frac{1}{2}\eta_{\alpha\beta}\bar{d}\big)\\
&\,+\partial_\alpha\partial_\beta \bar{d}-\eta_{\alpha\beta}\partial_\mu\partial_\nu\big(\bar{d}^{\mu\nu}-\frac{1}{2}\eta^{\mu\nu}\bar{d}\big)-\eta_{\alpha\beta} \Box \bar{d}\big]+\mathcal{O}(d^2)\\
=&\,\frac{1}{2}\big[\partial^\rho\partial_\alpha \bar{d}_{\rho\beta}+\partial_\beta \partial^\mu \bar{d}_{\mu\alpha}-\Box \bar{d}_{\alpha\beta}-\eta_{\alpha\beta} \partial_\mu \partial_\nu \bar{d}^{\mu\nu}\big]+\mathcal{O}(d^2)\,=\,\frac{1}{2}\Box \bar{d}_{\alpha\beta}+\mathcal{O}(d^2)
\end{split}
\end{equation}
where we used the inverse relation, $d_{\alpha\beta}=\bar{d}_{\alpha\beta}-\frac{1}{2}\eta_{\alpha\beta} \bar{d}$ of the trace reverse tensor defined by $\bar{d}_{\alpha\beta}=d_{\alpha\beta}-\frac{1}{2}\eta_{\alpha\beta} d$. In this particular context we observe that $\bar{d}\equiv \eta_{\alpha\beta}d^{\alpha\beta}=d-2d=-d$. In the ultimate line of computation we employed the harmonic gauge condition defined by $\partial^\alpha \bar{d}_{\alpha\beta}=0$.

\subsubsection{Gravitational potentials in the far away wave zone}
We will provide some computational details about the derivation rule introduced in the main part,
\begin{equation}
\partial_a h^{\alpha\beta}\,=\,\frac{\partial h^{\alpha\beta}}{\partial \tau}\frac{\partial\tau}{\partial x^a}\,=\,-\frac{N_a}{c}\partial_\tau h^{\alpha\beta},
\end{equation}
where we remind that $\tau=t-R/c$ is the retarded time and $N_a=x_a/R$. In this particular context it should be noticed that the harmonic gauge conditions translate into,
\begin{equation}
\partial_\beta h^{0\beta}\,=\,0\ \Rightarrow\ \dot{h}^{00}-\dot{h}^{0b} N_b\,=\,0, \quad \ \partial_\beta h^{\alpha\beta}\,=\,0\ \Rightarrow\ \dot{h}^{a0}-\dot{h}^{ab} N_b\,=\,0,
\end{equation}
where the overhead dot stands for the retarded time derivative. By introducing the potentials into the harmonic gauge condition we obtain,
\begin{equation}
\dot{h}^{00}-N_b \dot{h}^{0b}\,=\,0\Rightarrow \dot{X}-N_b(\dot{Y} N^b+\dot{Y}^b_T)\,=\,0\Rightarrow X\,=\,Y,
\end{equation}
where we used $N^bN_b=1$, $N_bY^b_T=0$ and the constant of integration was set to zero. Similarly we have have,
\begin{equation}
\dot{h}^{0a}-N_b\dot{h}^{ab}\,=\,0\Rightarrow \dot{Y}^a-N_b \dot{Z}^{ab}\,=\,0\Rightarrow \dot{Y} N^a+\dot{Y}^a_T-\frac{N^a}{3}\dot{A}-\frac{2}{3}N^a\dot{B}-\dot{Z}^a_T\,=\,0,
\end{equation} 
where in addition to the relations already outlined for the derivation of the previous equation, we employed here the constraint $N_b Z^{ab}_{TT}=0$ presented in equation \eqref{constraints-06.04.17}. From this we obtain a scalar and a vectorial constraint,
\begin{equation}
Y\,=\,\frac{A}{3}+\frac{2}{3}B,\quad \ Y^a_T\,=\,Z^a_T,
\end{equation}
where again we set the constants of integration to zero. From the transformation behaviour of the metric deviation under small translations, presented in equation \eqref{trafocoordinates-07.04.17}, and the relation between the metric deviation and the gravitational potentials we can work out the transformation behaviour of the gravitational potentials, 
\begin{equation}
\label{AppHG-07.04.17}
\begin{split}
h'^{\alpha\beta}\,=\,d'^{\alpha\beta}-\frac{d'}{2}\eta^{\alpha\beta}\,=&\,h^{\alpha\beta}-\frac{h}{2}\eta^{\alpha\beta}-\partial^\alpha t^\beta-\partial^\beta t^\alpha+\frac{h}{2}\eta^{\alpha\beta}+\partial_\mu t^\mu \eta^{\alpha\beta}\\ 
=&\,h^{\alpha\beta}-\partial^\alpha t^\beta-\partial^\beta t^\alpha+\eta^{\alpha\beta}\partial_\mu t^\mu.
\end{split}
\end{equation}
The harmonic gauge condition, presented in equation \eqref{HG-07.04.17}, is preserved whenever the small translation vectors posses harmonic components, $\partial_\beta h'^{\alpha\beta}=0\ \Rightarrow\ \partial_\beta h^{\alpha\beta}-\Box t^\alpha=0$. We aim to provide some additional details about the changes in the components of the gravitational potentials under the gauge transformation outlined in equation \eqref{AppHG-07.04.17}. We have for the time-time component of the gravitational potentials,
\begin{equation}
h'^{00}\,=\,h^{00}-2\partial^0t^0+(\partial_0t^0+\partial_bt^b)\eta^{00}\,=\,h^{00}+c^{-1}\dot{t}^0+c^{-1}\dot{t}^bN_b,
\end{equation}
where we used $\partial_bt^b=-c^{-1}\dot{t}^bN_b$. By using the potentials outlined in equation \eqref{IntermediatPot-07.04.17} as well as equation \eqref{Vectors-07.04.17} we eventually obtain,
\begin{equation}
\label{IG-07.04.17A}
A'+2B'\,=\,A+2B+3\dot{\gamma}+3\dot{\kappa},
\end{equation}
where we used $n_b\dot{\kappa}^b_T=0$. In a similar way we have for the time-space components of the gravitational potentials,
\begin{equation}
h'^{0a}\,=\,h^{0a}-\partial^0t^a-\partial^at^0+(\partial_0t^0+\partial_bt^b)\eta^{0a}\,=\,h^{0a}+c^{-1}\dot{t}^a+c^{-1}\dot{t}^0N^a,
\end{equation}
where we used $\partial^at^0=-c^{-1}\dot{t}^0N^a$. By using the potentials outlined in equation \eqref{IntermediatPot-07.04.17} as well as equation \eqref{Vectors-07.04.17} we eventually obtain,
\begin{equation}
\label{IG-07.04.17B}
(A'+2B')N^a+3Y'^a_T=(A+2B)N^a+3Y^a_T+3\dot{\kappa}N^a+3\dot{\kappa}^a_T+3\dot{\gamma}N^a.
\end{equation}
By combining the results presented in equation \eqref{IG-07.04.17A} and equation \eqref{IG-07.04.17B} we obtain the following transformation behaviours for the components,
\begin{equation}
\label{IGC-08.04.17}
\begin{split}
A\rightarrow A'\,=\,A+3\dot{\gamma}-\dot{\kappa},\quad \ B\rightarrow B'\,=\,B+2\dot{\kappa},\quad \ Y^a_T\rightarrow Y'^a_T\,=\,Y^a_T+\dot{\kappa}_T^a.
\end{split}
\end{equation}
Finally we can work out for the space-space components of the gravitational potentials the transformation behaviour and we obtain,
\begin{equation}
h'^{ab}\,=\,h^{ab}-\partial^at^b-\partial^b t^a+(\partial_0t^0+\partial_ct^c)\eta^{ab}\,=\,h^{ab}+c^{-1}\dot{t}^bN^a+c^{-1}\dot{t}^aN^b+c^{-1}(\dot{t}^0-\dot{t}^cN_c)\delta^{ab},
\end{equation}
where we used $\partial_c t^b=\dot{t}^b(-c^{-1}N_c)$. Here again we obtain by using the potentials outlined in equation \eqref{IntermediatPot-07.04.17} as well as equation \eqref{Vectors-07.04.17},
\begin{equation}
\begin{split}
\frac{\delta^{ab}}{3}A'+\big(N^aN^b-\frac{\delta^{ab}}{3}\big)B'+&N^aY'^b_T+N^bY'^a_T+Z'^{ab}_{TT}\,
=\,\frac{\delta^{ab}}{3}A+\big(N^aN^b-\frac{\delta^{ab}}{3}\big)B+N^aY^b_T+N^bY^a_T\\&+Z^{ab}_{TT}+N^b(\dot{\kappa}N^a+\dot{\kappa}_T^a)+N^a(\dot{\kappa}N^b+\dot{\kappa}_T^b)+[\dot{\gamma}-N_c(\dot{\kappa}N^c+\dot{\kappa}^c_T)]\delta^{ab}
\end{split}
\end{equation}
By combining these results together with those outlined under equation \eqref{IGC-08.04.17} we obtain the following transformation behaviour for the TT-component,
$Z^{ab}_{TT}\rightarrow Z^{ab}_{TT}$. In order to find the relation outlined in equation \eqref{GeoForce-10.04.17}, we need to derive the geodesic deviation equation \cite{Weinberg2}. We obtain for the difference of two nearby geodesics separated by a small spacetime vector $\epsilon^\alpha$,
\begin{gather}
\frac{d^2(x^\alpha+\epsilon^\alpha)}{d\lambda^2}+\Gamma^\alpha_{\beta\rho}(x+\epsilon)\ \frac{d (x^\beta +\epsilon^\beta)}{d\lambda}\frac{d(x^\rho+\epsilon^\rho)}{d\lambda}-\frac{d^2 x^\alpha}{d\lambda^2}-\Gamma^\alpha_{\beta\rho}(x)\ \frac{dx^\beta}{d\lambda}\frac{dx^\rho}{d\lambda}\,=\,0\\
\Rightarrow\frac{d^2 \epsilon^\alpha}{d\lambda^2}+\epsilon^\sigma \partial_\sigma \Gamma^\alpha_{\beta\rho} u^\beta u^\rho+2\Gamma^\alpha_{\beta\rho} u^\beta \frac{d \epsilon^\rho}{d\lambda}+\mathcal{O}(\epsilon^2)\,=\,0,
\end{gather}
where we used $\Gamma^\alpha_{\beta\rho}(x+\epsilon)=\Gamma^\alpha_{\beta\rho}(x)+\epsilon^\sigma\partial_\sigma \Gamma^\alpha_{\beta\rho}(x)+\mathcal{O}(\epsilon^2)$ and we remind that $u^\alpha=\frac{dx^\alpha}{d\lambda}$ is the four-velocity. Moreover we have that,
\begin{equation}
\begin{split}
\frac{D^2 \epsilon^2}{D\tau^2}\,=&\,\frac{d}{d\tau}\Big[\frac{d\epsilon^\alpha}{d\tau}+\Gamma^\alpha_{\mu\nu} \epsilon^\mu u^\nu\Big]+\Gamma^\alpha_{\beta\rho}\Big[\frac{d\epsilon^\beta}{d\tau}+\Gamma^\beta_{\gamma\eta} \epsilon^\gamma u^\eta\Big] u^\rho\\
=&\,\frac{d^2\epsilon^\alpha}{d\tau^2}+\frac{d\Gamma^\alpha_{\mu\nu}}{d\tau}\epsilon^\mu u^\nu +\Gamma^\alpha_{\mu\nu} \frac{d\epsilon^\mu}{d\tau}u^\nu+\Gamma^\alpha_{\mu\nu} \epsilon^\mu \frac{d u^\nu}{d\tau}+\Gamma^\alpha_{\beta\rho} \frac{d\epsilon^\beta}{d\tau} u^\rho+\Gamma^\alpha_{\beta\rho}\Gamma^\beta_{\gamma\eta}\epsilon^\gamma u^\eta u^\rho\\
=&\,\frac{d^2 \epsilon^\alpha}{d\tau^2}+\partial_\sigma \Gamma^\alpha_{\mu\nu} \epsilon^\mu u^\nu u^\sigma +2\Gamma^\alpha_{\beta\rho} \frac{d\epsilon^\beta}{d\tau} u^\rho+\Gamma^\alpha_{\mu\nu}\epsilon^\mu \frac{d^2x^\nu}{d\tau^2}+\Gamma^\alpha_{\beta\rho}\Gamma^\beta_{\gamma\eta} \epsilon^\gamma u^\eta u^\rho\\
=&\,u^\sigma \partial_\sigma \Gamma^\alpha_{\mu\nu}\epsilon^\mu u^\nu-\epsilon^\kappa \partial_\kappa \Gamma^\alpha_{\beta\rho} u^\beta u^\rho-\Gamma^\alpha_{\mu\nu} \epsilon^\mu \Gamma^\nu_{\kappa\rho} u^\kappa u^\rho +\Gamma^\alpha_{\beta\rho} \Gamma^\beta_{\gamma\eta} \epsilon^\gamma u^\eta u^\rho\\
=&\,-\big[\partial_\kappa \Gamma^\alpha_{\beta\rho}-\partial_\beta \Gamma^\alpha_{\kappa\rho}+\Gamma^\alpha_{\kappa\nu}\Gamma^\nu_{\beta\rho}-\Gamma^\alpha_{\nu\beta}\Gamma^\nu_{\kappa\rho}\big]\ \epsilon^\kappa u^\beta u^\rho\\
=&\,-R^\alpha_{\ \rho\kappa \beta}\ \epsilon^\kappa u^\beta u^\rho,
\end{split}
\end{equation}
where we used the covariant derivative of a generic vector field $V^\alpha(x)$ along the curve $x^\alpha(\lambda)$, $\frac{DV^\alpha}{D\lambda}=\frac{d V^\alpha}{d\lambda}+\Gamma^\alpha_{\nu\rho} V^\nu \frac{dx^\rho}{d\lambda}$ as well as the geodesic deviation equation \cite{MisnerThroneWheeler,PoissonWill,Maggiore1,Buonanno1,Schutz1}. The linearised Riemann tensor reads,
\begin{equation}
R_{0a0b}\,=\,\frac{1}{2}\big[\partial_0\partial_ad_{0b}+\partial_b\partial_0d_{0a}-\partial^2_0 d_{ab}-\partial_b\partial_a d_{00}\big]\,=\,-\frac{1}{2}\partial_0^2 h_{ab}\,=\,\frac{G}{2c^4 R}\ddot{Z}_{ab}^{TT},
\end{equation}
where we remind that $d_{\alpha\beta}=h_{\alpha\beta}-1/2\ h\eta_{\alpha\beta}$ and that the gravitational potentials for a far away wave zone field point are, $h^{00}=\frac{4 G M}{c^2 R}, \ h^{0a}=0, \ h^{ab}=\frac{G}{c^4 R} Z^{ab}_{TT}(\tau,\textbf{N})$. We will proove some of the properties of the transverse projector outlined in the main part,
\begin{gather}
P^i_{\ i}\,=\,\delta^i_{\ i}-N^iN_i\,=\,3-1\,=\,2,\\
P^i_{\ j} N_i\,=\,\delta^i_{\ j}N_i-N^iN_jN_i\,=\,N_j-N_j\,=\,0,\\
P^i_{\ k}P^k_{\ j}\,=\,\delta^i_{\ k}\delta^k_{\ j}-N^iN_j-N^iN_j+N^iN_j\,=\,\delta^i_{\ j}-N^iN_j\,=\,P^i_{\ j},
\end{gather}
where we have that $N^iN_i=1$ \cite{PoissonWill,Maggiore1,Buonanno1, Schutz1}. From these properties we can immediately deduce for the transverse-tracefree operators that,
\begin{equation}
\label{TT-16.04.17}
\begin{split}
(TT)^{ij}_{\ \ kl}N^k\,=&\,P^i_{\ k}P^j_{\ l}N^k-\frac{1}{2}P^{ij}P_{kl}N^k\,=\,0,\\
(TT)^{ij}_{\ \ kl}\delta^{kl}\,=&\,(P^i_{\ k}P^j_{\ l}-\frac{1}{2}P^{ij}P_{kl})\delta^{kl}\,=\,P^{ij}-P^{ij}\,=\,0,\\
(TT)^{ij}_{\ \ kl}\ (TT)^{kl}_{\ \ mn}\,=&\,(P^i_{\ k}P^j_{\ l}-\frac{1}{2}P^{ij}P_{kl})\ (P^k_{\ m}P^l_{\ n}-\frac{1}{2}P^{kl}P_{mn})\\
=&\,P^i_{\ m}P^j_{\ n}-\frac{1}{2}P^{ij}P_{mn}-\frac{1}{2}P^{ij}P_{mn}+\frac{1}{2}P^{ij}P_{mn}\,=\,(TT)^{ij}_{\ \ mn}.
\end{split}
\end{equation}
The polarizations of a generic tensor $Z^{ab}$ can be extracted by using the relations outlined in equation \eqref{Polar-10.04.17} together with the polar angles presented in equation \eqref{PolarAngles-10.04.17} and we obtain,
\begin{equation}
\begin{split}
Z_+\,=&\,-\frac{1}{4}\ \sin^2\theta\ (Z^{xx}+Z^{yy}-2Z^{zz})+\frac{1}{4}\ (1+\cos^2\theta)\cos2\varphi\ (Z^{xx}-Z^{yy})\\
&\,+\frac{1}{2}\ (1+\cos^2\theta)\ \sin2\varphi\ Z^{xy}-\frac{1}{2}\ \sin2\theta\ (\cos\varphi\ Z^{xz}+\sin\varphi\ Z^{yz}),\\
Z_\times\,=&\,-\frac{1}{2}\ \cos\theta\ [\sin2\varphi\ (Z^{xx}-Z^{yy})-2\cos2\varphi\ Z^{xy}]+\sin\theta(\sin\varphi\ Z^{xz}-\cos\varphi\ Z^{yz}).
\end{split}
\end{equation}
By way of example we aim to explicitly work out the first component of the matrix $Z^{ab}$ presented in equation \eqref{Components-11.04.17} using the relation $Z^{ab}_{TT}=Z_+(m^am^b-k^ak^b)+Z_\times(m^ak^b+k^am^b)$ together with the polar angles displayed in equation \eqref{PolarAngles-10.04.17} and we obtain,
\begin{equation}
\begin{split}
Z_{TT}^{xx}\,=&\,Z_+(m^xm^x-k^xk^x)+2Z_\times m^xk^x\\
=&\,Z_+(\cos^2\theta\cos^2\varphi-\sin^2\varphi)-2Z_\times\cos\theta\cos\varphi\sin\varphi\\
=&\,-\frac{1}{2}\ \big[\sin^2\theta-(1+\cos^2\theta)\cos2\varphi\big]\ Z_+-\cos\theta\ \sin2\varphi\ Z_\times.
\end{split}
\end{equation}
The remaining components of the tensor $Z^{ab}$ can be determined in a very similar way \cite{Schutz0C,PoissonWill,Maggiore1,Riles1}. Choosing $\theta=0$ and $\phi=0$ we observe that this component reduces to $Z^{xx}_{TT}(\theta=0,\phi=0)=Z_+$ the $+$-polarisation. The remaining components, in this particular case, are outlined in equation \eqref{Tensor-11.04.17} related to this appendix-section. We aim to provide some additional details regarding the derivation of the relations presented under equation \eqref{Ellipses-11.04.17}. We have for the ring of test masses, before the arrival of the gravitational radiation propagating in the $z$-direction, the following equation, $x_0^2+y_0^2=1$. Considering only the $+$-mode or only the $\times$-mode we obtain, from equation \eqref{Displ-11.04.17}, respectively,
\begin{equation}
\begin{split}
x_0\,=&\,\frac{x}{1+\Delta_+},\quad \quad \quad \quad \ \ x_0\,=\,\frac{x-\Delta_\times y}{1-\Delta_\times^2},\\
y_0\,=&\,\frac{y}{1-\Delta_+},\quad \quad \quad \quad \ \  y_0\,=\,\frac{y-\Delta_\times x}{1-\Delta_\times^2},
\end{split}
\end{equation}
where we remind that $\Delta_+=\frac{G\ Z_+(t)}{2c^4\ R}$ and $\Delta_\times=\frac{G\ Z_\times(t)}{2c^4\ R}$ are time-dependent quantities varying between their minimum and maximum values. We will provide some additional computational steps for the derivation of the result outlined in equation \eqref{BSGP-12.04.17}. 
in this regard we remind from equation \eqref{FAWZFP-28.03.17} that to leading order the gravitational potentials for a far away wave zone field point are proportional to the second temporal derivative of the radiative quadrupole moment. Taking the second derivatives of this result leads to,
\begin{equation}
\label{SecondD-17.04.17}
h^{ab}\,=\,\frac{2G}{c^4 R}\ \ddot{I}^{ab}\,=\,\frac{4 G \eta m}{c^4 R}\ \Big[v^av^b-\frac{Gm}{r} e_r^ae_r^b\Big],
\end{equation}
where $\textbf{v}=\textbf{v}_1-\textbf{v}_2$ is the relative velocity vector and where we used Newton's second law for a fictitious particle with reduced mass $\mu=m_1m_2/(m_1+m_2)$. Moreover we recall that $I^{ab}$ is the Newtonian quadrupole moment and $\eta=m_1m_2/m^2$ is a dimensionless parameter. We recall from that $\textbf{e}_r=[\cos\phi,\sin\phi,0]$ and $\textbf{e}_\phi=[-\sin\phi,\cos\phi,0]$ are the time dependent unit polar vectors expressed in a cartesian basis ($\textbf{e}_x$, $\textbf{e}_y$) \cite{PoissonWill}. We will evaluate first,
\begin{equation}
\begin{split}
v^av^b\,=&\,[\dot{r}\textbf{e}_r^a+r\dot{\phi}\textbf{e}_\phi^a]\ [\dot{r}\textbf{e}_r^b+r\dot{\phi}\textbf{e}_\phi^b]\\
=&\,\frac{G m}{p}\big[e^2\sin^2\phi\ \textbf{e}_r^a\textbf{e}_r^b+e\sin\phi\ (1+e\cos\phi)\ [\textbf{e}_r^a\textbf{e}_\phi^b+\textbf{e}_r^b\textbf{e}_\phi^a]+(1+ e\cos\phi)^2\ \textbf{e}_\phi^a\textbf{e}_\phi^b\big],
\end{split}
\end{equation}
where we employed the relations from equation \eqref{Deriv-12.04.17}, $\textbf{r}=r\ \textbf{e}_r$, $\textbf{v}=\dot{r}\ \textbf{e}_r+r\dot{\phi}\ \textbf{e}_\phi$ and $\dot{r}=\sqrt{\frac{G m}{p}}\ e \ \sin\phi$. The second contribution $\frac{Gm}{r}\ \textbf{e}_r^a\textbf{e}_\phi^b=\frac{Gm}{p}\ (1+e\cos\phi)\  \textbf{e}_r^a\textbf{e}_\phi^b$ can be rephrased in terms of the {semi-latus rectum} \cite{PoissonWill}. The combination of these two intermediate results leads to the relation given in equation \eqref{BSGP-12.04.17}. The polarisations for circular orbits mentioned in equation \eqref{Circular-13.04.17} are derived from the more general case given by,
\begin{equation}
\begin{split}
P_+\,=&\,-(1+\cos^2 i)\ \big[\cos(2\phi+2\omega)+\frac{5}{4}e\cos(\phi+2\omega)+\frac{1}{4}e \cos(3\phi+2\omega)+\frac{1}{2}e^2\cos2\omega\big]\\
&\,+\frac{1}{2}e\sin^2 i(\cos\phi+e),\\
P_\times\,=&\,-2\cos i\big[\sin(2\phi+2\omega)+\frac{5}{4}e\sin(\phi+2\omega)+\frac{1}{4}e\sin(3\phi+2\omega)+\frac{1}{2}e^2\sin2\omega\big],
\end{split}
\end{equation}
where we remind that $e$ is the eccentricity and $\omega$ is the angle between the pericenter and the line of nodes outlined in Figure \ref{DetAd}. These relations were derived by employing equation \eqref{polllla-13.04.17} together with equation \eqref{polla-13.04.17} within equation \eqref{BSGP-12.04.17}. 
\subsubsection{Radiative losses in gravitating systems}
We will provide some further details about the derivation of the result presented in equation \eqref{DerivativePot-15.04.17}. We will analyse temporal and spatial differentiation of the gravitational potentials separately, 
\begin{equation}
\begin{split}
c\partial_0 h^{\alpha\beta}\,=&\,\epsilon\ \dot{w}^{\alpha\beta}_1+\epsilon^2\ \dot{w}^{\alpha\beta}_2+\cdots,\\
c\partial_ah^{\alpha\beta}\,=&\,c\lambda_c\big[\big(\partial_a R^{-1}\big)\ w^{\alpha\beta}_1+R^{-1}\ \partial_aw^{\alpha\beta}_1\big]+c\lambda^2_c\big[\big(\partial_aR^{-2}\big)\ w^{\alpha\beta}_2+R^{-2}\ \partial_aw^{\alpha\beta}_2\big]+\cdots,\\
=&\,-\epsilon\ N_a\ \dot{w}_1^{\alpha\beta}-\epsilon^2\ \big[N_a\ \dot{w}^{\alpha\beta}_2+(c/\lambda_c)\ N_aw^{\alpha\beta}_1-(c/\lambda_c)\ P_a^{\ b}\ \frac{\partial w_1^{\alpha\beta}}{\partial N^b}\big]+\cdots,
\end{split}
\end{equation}
where we have, $\partial_aR^{-1}=-R^{-2}\partial_a R=-R^{-2}N_a$, $\partial_aw^{\alpha\beta}_1=-c^{-1}\dot{w}^{\alpha\beta}_1N_a+R^{-1}\ P_a^{\ b} \partial_{N_b}w^{\alpha\beta}_1$, $\partial_aR^{-2}=-2R^{-3}N_a$ and $\epsilon^2 \partial_aw^{\alpha\beta}_2=-\epsilon^2c^{-1}\dot{w}^{\alpha\beta}_2 N_a+\mathcal{O}(\epsilon^3)$. For the last derivative we also used the result $\partial_a N_b=R^{-1}\ P_{ab}$, where we remind that $P_{ab}=\delta_{ab}-N_aN_b$ is the transverse projector \cite{MisnerThroneWheeler,PoissonWill}. In order to obtain the result displayed in equation \eqref{TT-15.04.17} we need to use equation \eqref{DerivativePot-15.04.17} within the Landau-Lifshitz pseudotensor given by equation \eqref{LL-15.04.17} and we obtain,
\begin{equation}
\begin{split}
g_{\lambda\mu}g^{\nu\rho}\ \partial_\nu h^{\alpha\lambda}\ \partial_\rho h^{\beta\mu}\,=&\,+c^{-2}\ \eta_{lm}\ \eta^{\nu\rho}s_\nu s_\rho\ \dot{h}_{TT}^{al}\dot{h}_{TT}^{bm}+\mathcal{O}(\epsilon^2)\,=\,0+\mathcal{O}(\epsilon^2),\\
\frac{1}{2}\ g_{\lambda\mu}g^{\alpha\beta}\ \partial_\rho h^{\lambda\nu}\ \partial_\nu h^{\rho\mu}\,=&\,+\frac{c^{-2}}{2}\ \eta^{\alpha\beta}\eta_{lm}\ N_r\dot{h}_{TT}^{ln}\ N_n\dot{h}_{TT}^{rm}+\mathcal{O}(\epsilon^2)\,=\,0+\mathcal{O}(\epsilon^2),\\
-g_{\mu\nu}g^{\lambda\alpha}\ \partial_\rho h^{\beta\nu}\ \partial_\lambda h^{\rho\mu}\,=&\,-c^{-2}\ \eta_{m n}\eta^{la}\ N_l\dot{h}_{TT}^{b n} \ N_r\dot{h}_{TT}^{rm}+\mathcal{O}(\epsilon^2)\,=\,0+\mathcal{O}(\epsilon^2),\\
-g_{\mu\nu}g^{\lambda\beta}\ \partial_\rho h^{\alpha\nu}\ \partial_\lambda h^{\rho\mu}\,=&\,-c^{-2}\ \eta_{mn}\eta^{lb}\ N_l\dot{h}_{TT}^{a n}\ N_r\dot{h}_{TT}^{rm}+\mathcal{O}(\epsilon^2)\,=\,0+\mathcal{O}(\epsilon^2),\\
\frac{4}{8}\ g^{\alpha\lambda}g^{\beta\mu}\ g_{\nu\rho}g_{\sigma\tau}\ \partial_\lambda h^{\nu\tau}\ \partial_\mu h^{\rho\sigma}\,=&\,+\frac{c^{-2}}{2}\ \eta^{\alpha \lambda}\eta^{\beta \mu}\ \eta_{nr}\eta_{st}\ s_\lambda\dot{h}_{TT}^{nt}\ s_\mu \dot{h}_{TT}^{rs}+\mathcal{O}(\epsilon^2)\\
=&\,+\frac{c^{-2}}{2}\ \dot{h}_{TT}^{nt}\ \dot{h}^{TT}_{nt}\ s^\alpha s^\beta+\mathcal{O}(\epsilon^2),\\
-\frac{2}{8}\ g^{\alpha\beta}g^{\lambda\mu}\ g_{\nu\rho}g_{\sigma\tau}\ \partial_\lambda h^{\nu\tau}\ \partial_\mu h^{\rho\sigma}\,=&\,-\frac{c^{-2}}{4}\eta^{\alpha\beta}\eta^{\lambda\mu}s_\lambda s_\mu\ \eta_{nr}\eta_{st}\ \dot{h}_{TT}^{nt}\ \dot{h}_{TT}^{rs}+\mathcal{O}(\epsilon^2)\,=\,0+\mathcal{O}(\epsilon^2),\\
-\frac{2}{8}\ g^{\alpha\lambda}g^{\beta\mu}\ g_{\rho\sigma}g_{\nu\tau}\ \partial_\lambda h^{\nu\tau}\ \partial_\mu h^{\rho\sigma}\,=&\,-\frac{c^{-2}}{4}\ s^\alpha s^\beta\ \delta_{nt}\dot{h}^{nt}_{TT}\ \delta_{rs}\dot{h}^{rs}_{TT}+\mathcal{O}(\epsilon^2)\,=\,0+\mathcal{O}(\epsilon^2),\\
\frac{1}{8}\ g^{\alpha\beta}g^{\lambda\mu}\ g_{\rho\sigma}g_{\nu\tau}\ \partial_\lambda h^{\nu\tau}\ \partial_\mu h^{\rho\sigma}\,=&\,+\frac{c^{-2}}{8}\ \eta^{\alpha\beta}\ \eta^{\lambda\mu}s_\lambda s_\mu\ \delta_{nt}\dot{h}^{nt}_{TT}\ \delta_{rs}\dot{h}^{rs}_{TT}\,=\,0+\mathcal{O}(\epsilon^2).
\end{split}
\end{equation}
Here we used the following properties, $h^{00}=\mathcal{O}(\epsilon^2)$, $h^{0a}=\mathcal{O}(\epsilon^2)$ and $h^{ab}=h^{ab}_{TT}+\mathcal{O}(\epsilon^2)$, $N_ah^{ab}_{TT}=0$, $\eta^{\alpha\beta}s_\alpha s_\beta=0$ and $\delta_{ab}h^{ab}_{TT}=0$. The angular momentum flux relation can be formulated in terms of transverse traceless potentials,
\begin{equation}
\label{AMf-16.04.17}
\mathcal{A}^{ab}\,=\,\frac{c^3}{16\pi G}\int d\Omega\ R^2\ \big[h^{aq}_{TT}\dot{h}^{bq}_{TT}-h^{bq}_{TT}\dot{h}^{aq}_{TT}-(1/2)\dot{h}^{qp}_{TT}(x^a\partial^b-x^b\partial^a)h_{qp}^{TT}\big],
\end{equation}
where the overdot means differentiation with respect to the retarded time \cite{PoissonWill}. We will provide some additional computational steps in the derivation of the quadrupole formula presented in equation \eqref{QF-16.04.17}. We know from equation \eqref{Fluxes-16.04.17} together with equation \eqref{N-16.04.17} that,
\begin{equation}
\mathcal{P}\,=\,\frac{c^3}{32\pi G}\int d\Omega\ R^2\ [\dot{h}^{ab}_{TT}\dot{h}^{TT}_{ab}]\,=\,\frac{G}{8c^5\pi}\int d\Omega\  (TT)^{kl}_{\ \ ij}\dddot{I}^{\langle ij\rangle}\dddot{I}_{\langle kl\rangle}\,=\,\frac{G}{5c^5}\dddot{I}^{\langle kl\rangle}\dddot{I}_{\langle kl\rangle},
\end{equation}
where we used $(TT)^{ab}_{\ \ ij}(TT)^{kl}_{\ \ ab}=(TT)^{kl}_{\ \ ij}$ (equation \eqref{TT-16.04.17}) together with $\int d\Omega \ (TT)^{kl}_{\ \ ij}=\frac{2\pi}{15}(11\delta_i^{ \ k}\delta_j^{\  l}-4\delta_{ij}\delta^{kl}+\delta_i^{ \ l}\delta_j^{\ k}$) and $I^{\langle ij\rangle}\delta_{ij}=0$ (TT-gauge) \cite{Maggiore1}. In what follows we will present the major computational steps necessary to derive the result outlined in equation \eqref{AverageEmission-17.04.17} for the average energy emission due to gravitational waves. In order to compute the third derivative of the Newtonian quadrupole moments we can use the result displayed in equation \eqref{SecondD-17.04.17} and we obtain,
\begin{equation}
\begin{split}
\frac{\partial^3}{\partial_\tau^3} I^{ab}\,=&\,2\eta m\ \frac{\partial}{\partial\tau} \big[v^av^b-\frac{G m}{r}\ e_r^ae_r^b\big]\\
=&\,2\eta m \Big[\Big(-G\frac{m}{r^2}\frac{r^a}{r}\Big) v^b+v^a\Big(-G\frac{m}{r^2}\frac{r^b}{r}\Big)-\frac{Gm}{r^3}\Big(-3\frac{\dot{r}}{r} r^ar^b+v^ar^b+r^av^b\Big)\Big]\\
=&\,-2\eta m \frac{Gm}{r^2}\ \Big[\frac{2}{r}\big(r^av^b+v^ar^b\big)-\frac{3\dot{r}}{r^2} r^ar^b\Big]\\
=&\,-2\eta m \ \frac{Gm}{r^2}\ \big[\dot{r}e^a_re^b_r+2r\dot{\phi}\ (e^a_re_\phi^b+e^b_re^a_\phi)\big],
\end{split}
\end{equation}
where we remind that the unit vectors $\textbf{e}_r$ and $\textbf{e}_\phi$ lie within the orbital plane of the binary system. The quadrupole moment was defined in equation \eqref{N-16.04.17} and becomes,
\begin{equation}
\begin{split}
\dddot{I}^{\langle ab\rangle}\,=\,\dddot{I}^{ab}-\frac{\delta^{ab}}{3}\dddot{I}^{pp}\,
=&\,-2\eta m \frac{(Gm)^{3/2}}{p^{5/2}}\big[1+e \cos(\phi)\big]^2\ \big[e \sin(\phi) e^a_r e^b_r\\
&\,+2\big(1+e\cos(\phi)\big)\ \big(e^a_r e_\phi^b+e_\phi^ae_r^b\big)-\frac{\delta^{ab}}{3} e \sin(\phi)\big],
\end{split}
\end{equation}
where $I^{\langle ab\rangle}\,=\,\int d\textbf{x}\ \rho(\tau,\textbf{x})\ \big(x^ax^b-\frac{x^px_p}{3} \delta^{ab}\big)$ is the quadrupole moment, $\rho(\tau,\textbf{x})=T^{00}\ c^{-2}$ is the source term and $\dddot{I}^{pp}=-2\eta m \frac{(Gm)^{3/2}}{p^{5/2}}\big[1+e \cos(\phi)\big]^2\ e\sin(\phi)$. The product of the third temporal derivatives of the quadrupole moments is,
\begin{equation}
\begin{split}
\dddot{I}^{\langle ab\rangle} \dddot{I}^{\langle ab\rangle}\,=&\,4\eta^2 m^2 \frac{(Gm)^3}{p^5}\ \big[1+e \cos(\phi)\big]^4\\
&\quad\ \big[e^2 \sin^2(\phi)-(2/3)\ e^2\sin^2(\phi)+8 \big(1+e\cos(\phi)\big)^2+(1/3)\ e^2 \sin^2(\phi)\big]+o\\
=&\,32\eta^2 m^2 \frac{(Gm)^3}{p^5}\ \big[1+e \cos(\phi)\big]^4\ \Big[\frac{e^2}{12}+\frac{11}{12}e^2 \cos^2(\phi)+1+2e\cos(\phi)\Big]+o,
\end{split}
\end{equation}
where we used the well known trigonometric relation $\sin^2(\phi)=1-\cos^2(\phi)$ and $o$ denotes terms that are odd and finally disappear in the orbital average integration. From equation \eqref{QF-16.04.17} we eventually obtain for the quadrupole formula,
\begin{equation}
\mathcal{P}(\phi)\,=\,\frac{G}{5c^5}\ \dddot{I}^{\langle ab\rangle} \dddot{I}^{\langle ab\rangle}\,=\,\frac{32}{5c^5}\ \eta^2 \ \frac{G^4m^5}{p^5}\ \big[1+e \cos(\phi)\big]^4\ \Big[\frac{e^2}{12}+\frac{11}{12}e^2 \cos^2(\phi)+1+2e\cos(\phi)\Big]
\end{equation}
For binary-systems evolving on elliptic trajectories the released energy-flux should be averaged over a complete orbital cycle and we get,
\begin{equation}
\begin{split}
\langle\mathcal{P}\rangle\,=\,\frac{1}{P}\int_0^P dt\ \mathcal{P}(t)\,=&\,\frac{(1-e^2)^{3/2}}{2\pi}\int_0^{2\pi}d\phi \ \big(1+e\cos(\phi)\big)^{-2}\ \mathcal{P}(\phi)\\
=&\,\frac{32}{5c^5}\ \eta^2 \ \frac{G^4m^5}{p^5}\ (1-e^2)^{3/2}\ \Big[1+\frac{73}{24}e^2+\frac{37}{96}e^4\Big],
\end{split}
\end{equation}
where we used the relation between the orbital period and the semi-major axis (Kepler's third law), $P=2\pi \sqrt{a^3/(Gm)}$, $p=a(1-e^2)$ and $\dot{\phi}=\sqrt{\frac{Gm}{p^3}}\ \big[1+e\cos(\phi)\big]^2$. For the derivation of the decrease rate of the semi-major axis outlined in equation \eqref{SMAD-17.04.17} we used the chain derivative rule $dE/dt=d E/da\ da/dt$ together with the total energy relation $E=-\eta Gm^2/(2a)$ \cite{Maggiore1} of the fictitious body with reduced mass $\mu=m_1m_2/m$ orbiting the center-of-mass of the binary system on an elliptic orbit. We aim to provide some further computational details about the derivation of the result presented in equation \eqref{PeriodVar-18.04.17}. As a first step towards this particular result we work out the correlation between the orbital period and the total orbital energy and we obtain, $P\,=\,Gm\ \mu\pi \ \sqrt{\frac{\mu}{2}}\ (-E)^{-3/2}$. Differentiation with respect to time immediately leads to the left-hand side of the relation presented in equation \eqref{PeriodVar-18.04.17}. The final result (right-hand side) is obtained by incorporating equation \eqref{AverageEmission-17.04.17}. The relation presented in equation \eqref{Periastronshift-18.04.17} is obtained from the following reasoning. Because of the binary system's permanent energy loss due to the emission of gravitational radiation, the frequency $\nu$ cannot be constant. For small variations we can expand the frequency around some reference value at $t=t_0$, $\nu(t)=\nu(t_0)+\dot{\nu}(t_0)t+\frac{1}{2}\ddot{\nu}(t_0)t^2+\cdots$. From this it is straightforward to derive the accumulated orbital phase, $\Phi(T)=2\pi \int_0^Tdt\ \nu(t)=2\pi (\nu T+\dot{\nu}/2\ T^2+\cdots)$. The $n$-th time of periastron passage, $T_n$, takes place when $\Phi(T_n)=2\pi n$. Thus the cumulative difference between the periastron passages $T_n$ and the values $n\cdot P$ is given by, 
\begin{equation}
\label{HulseTaylor-21.04.17}
T_n-P\cdot n\,=\,\frac{\dot{P}}{2P}\ T_n^2+\mathcal{O}(T^3_n),
\end{equation}
where we remind that $\nu=1/P$ \cite{Maggiore1}.
By combining this relation with equation \eqref{PeriodVar-18.04.17} and neglecting terms of the order $\mathcal{O}(T^3_n)$, we eventually obtain the result outlined in equation \eqref{Periastronshift-18.04.17}. In the context of the Michelson-type interferometer discussed in the main part, we will provide some additional details about the derivation of the result displayed in equation \eqref{Power1-24.04.17}. The sum of the two electric fields, $E_1=-(E_0/2) e^{-i \omega_M t+2ik_ML_x}$and $E_2=+(E_0/2) e^{-i \omega_M t+2ik_ML_y}$ that recombine at some time $t$ at the beam splitter are,
\begin{equation}
E_{out}\,=\,-iE_0e^{-i\omega_Mt+ik_M(L_x+L_y)}\sin[k_M(L_x-L_y)],
\end{equation}
where we remind that $k_M=\omega_M/c$ is the wavenumber of the monochromatic laser light with frequency $\omega_M$ (Figure \ref{Interferometer-23.04.17}). The power measured at the photodetector is proportional $P_{out}\sim|E_{out}|^2$ to the complex conjugate of the sum of the two electric fields at some time $t$ at the beam-splitter. We will briefly review the main computational steps that lead to the result outlined under equation \eqref{Power1-24.04.17}. From the spacetime interval given in equation \eqref{intervalplus-24.04.17} we obtain for the light (photons travelling along null geodesics $ds^2=0$) in the $x$-arm to first order in $A_o$, $dx=\pm cdt[1-h_+/2]$, where the plus sign holds for the travel from the beam-splitter to the mirror and the minus sign for the return trip. We assume that a photon leaves the beam-splitter at a given time $t^x_0$ and reaches the mirror at the coordinate position $(L_x,0)$ at some time $t^x_1$. With this we obtain after integrating the $x$-arm interval relation in the plus direction, $L_x=c(t^x_1-t^x_0)-c/2\int_{t_0^x}^{t^x_1}dt'\ h_+(t')$. A similar relation, $L_x=c(t^x_2-t^x_1)-c/2\int_{t_1^x}^{t_2^x}dt'\ h_+(t')$ is obtained for the reflection of the photon, where $t^x_2$ is the moment in time where the photon returns to the beam-splitter. Summation of these two relations as well as the replacement in the resulting integral of $t^x_2$ by $t^x_0+2L_x/c$ leads to,
\begin{equation}
\label{timex-25.04.17}
t_2^x-t_0^x\,=\,\frac{2L_x}{c}+\frac{A_oL_x}{c}\cos[\omega_c(t_0+L_x/c)]\frac{\sin(\omega_c L_x/c)}{\omega_cL_x/c},
\end{equation}
where we remind that $A_o$ is the amplitude of the incoming gravitational radiation. A similar relation can be obtained for the $y$-arm, where $L_x$ is replaced by $L_y$ and the second term on the right-hand side picks up a minus sign. Replacing $t_2$ by some given value of the observation time $t$ and taking into account that in order to come back at the beam-splitter at the time $t$, the light that went through the $x$-arm must have started its round-trip travel at a time $t^x_0$ obtained by inverting equation \eqref{timex-25.04.17} to first order in $A_o$. This means that $A_o\cos[\omega_c(t_0+L_x/c)]$ is replaced by $A_o\cos[\omega_c(t-2L_x/c+L_x/c)]$ and we obtain, $t_0^x=t-2L_x/c-L_x/c\ A_o\cos[\omega_c(t-L_x/c)]\frac{\sin(\omega_c L_x/c)}{\omega_cL_x/c}$. A similar relation can be obtained for the $y$-arm and we will choose the lengths of the two arms as close as possible, $2L_x=2L+(L_x-L_y)$ and $2L_y=2L-(L_x-L_y)$. From this we obtain for the two electric fields $E^x(t)=-(E_0/2)\ e^{-i\omega_M(t-2L/c)+i\phi_0+i\Delta\phi_x(t)}$ and $E^y(t)=+(E_0/2)\ e^{-i\omega_M(t-2L/c)-i\phi_0+i\Delta\Phi_y(t)}$, where $\phi_0=k_M(L_x-L_y)$ is a parameter that the experimenter can adjust, choosing the best working point for the interferometer \cite{Maggiore1}. The total phase difference induced by the gravitational radiation in the Michelson interferometer is $\Delta\phi_I=\Delta\phi_x-\Delta\phi_y=2\Delta\phi_x$, where $\Delta\phi_x(t)=A_ok_ML\frac{\sin(\omega_c L/c)}{\omega_cL/c}\cos[\omega_c(t-L/c)]$. The total electric field at the output is, $E_{tot}=E^x(t)+E^y(t)=-iE_0e^{-i\omega_M(t-L/c)}\sin[\phi_0+\frac{\Delta\phi_I}{2}]$.

We will present the line element for an observer in the proper detector frame with rotations and accelerations up to second order in $x$,
\begin{equation}
\begin{split}
\label{PDFRA-03.05.17}
ds^2\,=\,&-c^2dt^2\Big(1+\frac{2\textbf{a}\cdot \textbf{x}}{c^2}+\frac{(\textbf{a}\cdot\textbf{x})^2}{c^4}-\frac{(\boldsymbol{\Omega}\times\textbf{x})^2}{c^2}+R_{0a0b}x^ax^b\Big)\\
&+2cdtdx^a\Big(\epsilon_{abc}\frac{\Omega^bx^c}{c}-\frac{2}{3}R_{0bac}x^bx^c\Big)+dx^adx^b\Big(\delta_{ab}-\frac{1}{3}R_{acbd}x^cx^d\Big)+\mathcal{O}\Big(\frac{|\textbf{x}|^3}{\mathcal{C}^3}\Big),
\end{split}
\end{equation}
where $\textbf{a}$ is the acceleration of the laboratory with respect to a local free falling frame and $\boldsymbol{\Omega}$ is the angular velocity of the laboratory with respect to local gyroscopes \cite{Maggiore1,Zimmermann1}. The term $2\textbf{a}\cdot \textbf{x}/c^2$ gives the inertial acceleration, while the term $(\textbf{a}\cdot\textbf{x}/c^2)^2$ is a gravitational redshift. The term $(\boldsymbol{\Omega}\times \textbf{x}/c)^2$ gives a Lorentz time dilatation due to the angular velocity of the laboratory and the term $\epsilon_{abc}\Omega^bx^c$ is known as the Sagnac effect. It should be noticed that for $\textbf{a}=\boldsymbol{\Omega}=0$ we recover the metric outlined in equation \eqref{PDF-03.05.17}. Since spacetime is flat, a photon that starts at the beam-splitter at a time $t_0$, moving along the positive $x$-axis, follows the trajectory $x(t)=c(t-t_0)$ and reaches the mirror at a time $t_1$ given by $\epsilon_x(t_1)=c(t-t_1)$. Using the solution of the geodesic deviation equation to first order in the amplitude $A_o$ we finally obtain, $c(t_1-t_0)=L_x+A_oL_x/2\cos[\omega_c(t_0+L_x/c)]$. The round-trip (beam-splitter - mirror - beam-splitter) is twice as large and we obtain the result presented in equation \eqref{timesimple-03.05.17} \cite{Maggiore1}. We know that we have,
\begin{equation}
\label{interm-04.05.17}
dx=\pm cdt[1+\omega_c^2/(4c^2)x^2(t)A_o\cos(\omega_ct).
\end{equation}
We consider a photon that leaves the beam-splitter at some time $t_0$ and propagates along the positive $x$ direction. To lowest order in $A_o$ we have the trivial result $x(t)=c(t-t_0)$ and we can insert this result into equation \eqref{interm-04.05.17} in order to obtain the solution to first order in the amplitude \cite{Maggiore1}.  

\end{appendix}

\end{document}